\documentclass[11pt]{article}
\usepackage{amssymb}
\usepackage{amsthm}
\usepackage{amsmath}
\usepackage[dvips]{epsfig}
\usepackage{graphicx}
\usepackage{color}
\usepackage{url}

\newtheorem{theorem}{Theorem}
\newtheorem{lemma}{Lemma}

\newtheorem{remark}{Remark}

\newcommand{\R}{\mathbb{R}}
\newcommand{\C}{\mathbb{C}}
\newcommand{\Z}{\mathbb{Z}}
\newcommand{\N}{\mathbb{N}}

\begin{document}

\title{Periodic Travelling Waves in Dimer Granular Chains}

\author{Matthew Betti and Dmitry E. Pelinovsky \\
{\small Department of Mathematics and Statistics, McMaster
University, Hamilton, Ontario, Canada, L8S 4K1 }}

\date{\today}
\maketitle

\begin{abstract}
We study bifurcations of periodic travelling waves in granular dimer chains
from the anti-continuum limit, when the mass ratio between the light and heavy beads
is zero. We show that every limiting periodic wave is uniquely continued with
respect to the mass ratio parameter and the periodic waves with the wavelength larger than
a certain critical value are spectrally stable. Numerical computations are developed
to study how this solution family is continued to the limit of equal mass ratio between
the beads, where periodic travelling waves of granular monomer chains exist.
\end{abstract}

\tableofcontents

\section{Introduction}\label{sec:intro}

Wave propagation in granular crystals has been studied quite intensively in the past
ten years. Granular crystals are thought to be closely-packed chains of elastically interacting
particles, which obey the Fermi-Pasta-Ulam (FPU) lattice equations with Hertzian interaction forces.
Experimental work with granular crystals and
their numerous applications \cite{Daraio,Sen} stimulated theoretical and mathematical
research on the granular chains of particles.

Existence of solitary waves in granular chains was considered with a number of
analytical and numerical techniques. In his two-page note, MacKay \cite{MacKay} showed
how to adopt the technique of Friesecke and Wattis \cite{Wattis} to the proof of existence
of solitary waves. English and Pego \cite{Pego} used these results to prove the double-exponential
decay of spatial tails of solitary waves. Numerical convergence to the solitary wave solutions was studied
by Ahnert and Pikovsky \cite{Pikovsky}. Stefanov and Kevrekidis \cite{Stefanov}
reviewed the variational technique of \cite{Wattis} and proved that the solitary
waves are bell-shaped (single-humped).

Recently, the interest to granular crystals has shifted towards periodic travelling waves
as well as travelling waves in heterogeneous (dimer) chains, as more relevant for
physical experiments \cite{Harbola,Porter}. Periodic wave solutions of the
differential advance-delay equation were considered by James
in the context of Newton's cradle \cite{James1} and homogeneous granular crystals \cite{James2}.
In particular, numerical approximations in \cite{James2} suggested that
periodic waves with wavelength larger than a critical value are spectrally unstable.
Convergence to solitary waves in the limit of infinite wavelengths was also illustrated
numerically and asymptotically in \cite{James2}. More recent work \cite{James3}
showed non-existence of time-periodic breathers in homogeneous granular crystals and
existence of these breathers in Newton's cradle, where a discrete $p$-Schr\"{o}dinger
equation provides a robust approximation.

Periodic waves in a chain of finitely many beads closed in a periodic loop
were approximated by Starosvetsky {\em et al.} in monomers \cite{Star1} and
dimers \cite{Star2} by using numerical techniques based
on Poincar\'{e} maps. Interesting enough, solitary waves were found in the limit
of zero mass ratio between lighter and heavy beads in \cite{Star2}. It is explained
in \cite{Star2} that these solitary waves are in resonance with linear waves and hence
they do not persist with respect to the mass ratio parameter. Numerical results
of \cite{Star2} indicate the existence of a countable set of the mass ratio parameter
values, for which solitary waves should exist, but no rigorous studies
of this problem have been developed so far. Recent work \cite{Star3} contains
numerical results on existence of periodic travelling waves in granular dimer chains.

In our present work, we rely on the anti-continuum limit of the FPU lattice,
which was recently explored in the context of existence and stability of discrete
multi-site breathers
by Yoshimura \cite{Yoshimura}. By using a variant of the Implicit Function Theorem,
we prove that every limiting periodic wave is uniquely continued with
respect to the mass ratio parameter. By the perturbation theory arguments
(which are similar to the recent work in \cite{PelSak}),
we also show that the periodic waves with the wavelength larger than
a certain critical value are spectrally stable. Our results are different from
the asymptotic calculations in \cite{Star2}, where a different limiting
solution is considered in the anti-continuum limit.

The family of periodic nonlinear waves bifurcating from the anti-continuum
limit are shown numerically to extend all way to the limit
of equal masses for the dimer beads. The periodic travelling waves
of the homogeneous (monomer) chains considered in \cite{James2}
are different from the periodic waves extended here from the anti-continuum limit.
In other words, the periodic waves in dimers do not satisfy the reductions
to the periodic waves in monomers even if the mass ratio is one. Similar
travelling waves consisting of binary oscillations in monomer chains
were considered a while ago with the center manifold reduction methods
\cite{IoossJames}.

The paper is organized as follows.
Section 2 introduces the model and sets up the scene for the search of
periodic travelling waves. Continuation from the anti-continuum limit
is developed in Section 3. Section 4 gives perturbative results that characterize
Floquet multipliers in the spectral stability problem associated with the
periodic waves near the anti-continuum limit.
Numerical results are collected together in Section 5.
Section 6 concludes the paper.

\section{Formalism}

\subsection{The model}

We consider an infinite granular chain of spherical beads of alternating masses
(a so-called {\em dimer}), which obey Newton's equations of motion,
\begin{eqnarray}
\label{FPU-before}
\left\{ \begin{array}{l}
m \ddot{x}_n = V'(y_n-x_n)-V'(x_n - y_{n-1}), \\
M \ddot{y}_n = V'(x_{n+1}-y_n)- V'(y_n - x_n),  \end{array} \right. \quad n \in \mathbb{Z},
\end{eqnarray}
where $m$ and $M$ are masses of light and heavy beads with coordinates $\{ x_n \}_{n \in \Z}$
and $\{ y_n \}_{n \in \Z}$, respectively, whereas $V$ is the interaction potential.
The potential $V$ represents the Hertzian contact forces for perfect spheres and is given by
\begin{equation}
 \label{eq:potential}
 V(x)=\frac{1}{1+\alpha}|x|^{1+\alpha}H(-x),
\end{equation}
where $\alpha=\frac{3}{2}$ and $H$ is the Heaviside step function with $H(x) = 1$ for $x > 0$ and
$H(x) = 0$ for $x \leq 0$. The mass ratio is modeled by the parameter $\varepsilon^2 := \frac{m}{M}$.
Using the substitution,
\begin{equation}
n \in \Z : \quad x_n(t) = u_{2n-1}(\tau), \quad
y_n(t) = \varepsilon w_{2n}(\tau), \quad t = \sqrt{m} \tau,
\end{equation}
we rewrite the system of Newton's equations (\ref{FPU-before}) in the equivalent form:
\begin{equation}
\label{eq:ADD}
\left\{ \begin{array}{l}
\ddot{u}_{2n-1} = V'(\varepsilon w_{2n} - u_{2n-1}) - V'(u_{2n-1} - \varepsilon w_{2 n - 2}), \\
\ddot{w}_{2n} = \varepsilon V'(u_{2 n+1}- \varepsilon w_{2n}) -
\varepsilon V'(\varepsilon w_{2n} - u_{2 n-1}),
\end{array} \right. \quad n \in \mathbb{Z}.
\end{equation}
The value $\varepsilon = 0$ correspond to the anti-continuum limit, when the heavy particles do
not move.

At the limit of equal mass ratio $\varepsilon = 1$, we note the reduction,
\begin{equation}
\label{scalar-constraint}
n \in \Z : \quad u_{2n-1}(\tau) = U_{2n-1}(\tau), \quad w_{2n}(\tau) = U_{2n}(\tau),
\end{equation}
for which the system of two granular chains (\ref{eq:ADD}) reduces to
the scalar granular chain (a so-called {\em monomer}),
\begin{equation}
\label{eq:Mono}
\ddot{U}_n = V'(U_{n+1} - U_n) - V'(U_n - U_{n-1}), \quad n \in \Z.
\end{equation}

The system of dimer equations (\ref{eq:ADD}) has two symmetries.
One symmetry is the translational invariance of solutions with respect to
$\tau$, that is, if $\{ u_{2n-1}(\tau), w_{2n}(\tau) \}_{n \in \Z}$ is a
solution of (\ref{eq:ADD}), then
\begin{equation}
\label{symmetry1}
\{ u_{2n-1}(\tau + b), w_{2n}(\tau + b) \}_{n \in \Z}
\end{equation}
is also a solution of (\ref{eq:ADD}) for any $b \in \R$.
The other symmetry is a uniform shift of coordinates $\{ u_{2n-1}, w_{2n} \}_{n \in \Z}$
in the direction of $(\varepsilon, 1)$, that is, if
$\{ u_{2n-1}(\tau), w_{2n}(\tau) \}_{n \in \Z}$ is a
solution of (\ref{eq:ADD}), then
\begin{equation}
\label{symmetry2}
\{ u_{2n-1}(\tau) + a \varepsilon, w_{2n}(\tau) + a \}_{n \in \Z}
\end{equation}
is also a solution of (\ref{eq:ADD}) for any $a \in \R$.

The system of dimer equations (\ref{eq:ADD}) has the symplectic structure
\begin{equation}
\label{symplectic}
\dot{u}_{2n-1} = \frac{\partial H}{\partial p_{2n-1}}, \quad
\dot{p}_{2n-1} = -\frac{\partial H}{\partial u_{2n-1}}, \quad
\dot{w}_{2n} = \frac{\partial H}{\partial q_{2n}}, \quad
\dot{q}_{2n} = -\frac{\partial H}{\partial w_{2n}}, \quad n \in \Z,
\end{equation}
where the Hamiltonian function is
\begin{equation}
\label{Hamiltonian}
H = \frac{1}{2} \sum_{n \in \Z} \left( p_{2n-1}^2 + q_{2n}^2 \right) +
\sum_{n \in \Z} V(\varepsilon w_{2n} - u_{2n-1}) + \sum_{n \in \Z} V(u_{2n-1} - \varepsilon w_{2n-2}),
\end{equation}
written in canonical variables
$\left\{ u_{2n-1}, p_{2n-1} = \dot{u}_{2n-1}, w_{2n}, q_{2n} = \dot{w}_{2n} \right\}_{n \in \Z}$.

\subsection{Periodic traveling waves}

We shall consider $2\pi$-periodic solutions of the dimer system (\ref{eq:ADD}),
\begin{equation}
\label{periodicity}
u_{2n-1}(\tau) = u_{2n-1}(\tau + 2\pi), \quad w_{2n}(\tau) = w_{2n}(\tau + 2\pi),
\quad \tau \in \R, \quad n \in \Z.
\end{equation}
Travelling waves correspond to the special solution
to the dimer system (\ref{eq:ADD}), which satisfies the following reduction,
\begin{equation}
\label{traveling}
u_{2n+1}(\tau) = u_{2n-1}(\tau + 2 q), \quad w_{2n+2}(\tau) = w_{2n}(\tau + 2q),
\quad \tau \in \R, \quad n \in \Z,
\end{equation}
where $q \in [0,\pi]$ is a free parameter. We note that the constraints
(\ref{periodicity}) and (\ref{traveling}) imply that there exists $2\pi$-periodic
functions $u_*$ and $w_*$ such that
\begin{equation}
\label{advance-delay}
u_{2n-1}(\tau) = u_*(\tau + 2 q n), \quad w_{2n}(\tau) = w_*(\tau + 2 q n),
\quad \tau \in \R, \quad n \in \Z.
\end{equation}
In this context, $q$ is inverse proportional to the wavelength of the periodic wave
over the chain $n \in \Z$.
The functions $u_*$ and $w_*$ satisfy the following
system of differential advance-delay equations:
\begin{equation}
\label{eq:advance-delay}
\left\{ \begin{array}{l}
\ddot{u}_*(\tau) = V'(\varepsilon w_*(\tau) - u_*(\tau)) - V'(u_*(\tau) -
\varepsilon w_*(\tau - 2 q)), \\
\ddot{w}_*(\tau) = \varepsilon V'(u_*(\tau + 2 q) - \varepsilon w_*(\tau)) -
\varepsilon V'(\varepsilon w_*(\tau) - u_*(\tau)),
\end{array} \right. \quad \tau \in \mathbb{R}.
\end{equation}

\begin{remark}
A more general traveling periodic wave can be sought in the form
$$
u_{2n-1}(\tau) = u_*(c \tau + 2 q n), \quad
w_{2n}(\tau) = w_*(c \tau + 2 q n), \quad
\tau \in \R, \quad n \in \Z,
$$
where $c > 0$ is an arbitrary parameter. However, the parameter $c$ can be normalized to one
thanks to invariance of the system of dimer equations (\ref{eq:ADD}) with respect to
a scaling transformation.
\end{remark}

\begin{remark}
For particular values $q = \frac{\pi m}{N}$, where $m$ and $N$ are positive integers such that
$1 \leq m \leq N$, periodic travelling waves satisfy a system of $2 m N$ second-order
differential equations that follows from the system of lattice differential equations (\ref{eq:ADD})
subject to the periodic conditions:
\begin{equation}
\label{boundary-conditions}
u_{-1} = u_{2 m N-1}, \quad u_{2 m N+1} = u_1, \quad w_0 = w_{2 m N}, \quad w_{2 m N+2} = w_2.
\end{equation}
This reduction is useful for analysis of stability of periodic travelling waves and
for numerical approximations.
\end{remark}

\subsection{Anti-continuum limit}

Let $\varphi$ be a solution of the nonlinear oscillator equation,
\begin{equation}
\ddot{\varphi} = V'(-\varphi) - V'(\varphi) \quad \Rightarrow \quad
\ddot{\varphi} + |\varphi|^{\alpha - 1} \varphi = 0. \label{oscillator}
\end{equation}
Because $\alpha = \frac{3}{2}$, bootstrapping arguments show that if there exists
a classical $2\pi$-periodic solution of the differential equation (\ref{oscillator}),
then $\varphi \in C^3_{\rm per}(0,2\pi)$.

The nonlinear oscillator equation (\ref{oscillator}) has the first integral,
\begin{equation}
E = \frac{1}{2} \dot{\varphi}^2 + \frac{1}{1 + \alpha} |\varphi|^{\alpha + 1}. \label{energy}
\end{equation}
The phase portrait of the nonlinear oscillator (\ref{oscillator}) on
the $(\varphi,\dot{\varphi})$-plane consists of a family of closed
orbits around the only equilibrium point $(0,0)$. Each orbit corresponds to
the $T$-periodic solution for $\varphi$, where $T$ is determined uniquely by
energy $E$. It is well-known \cite{James2,Yoshimura} that, for $\alpha > 1$,
the period $T$ is a monotonically decreasing
function of $E$ such that $T \to \infty$ as $E \to 0$ and $T \to 0$ as
$E \to \infty$. Therefore, there exists a unique $E_0 \in \R_+$ such that
$T = 2\pi$ for this $E = E_0$. We also know that the nonlinear oscillator (\ref{oscillator})
is non-degenerate in the sense that $T'(E_0) \neq 0$ (to be more precise,
$T'(E_0) < 0$).

In what follows, we only consider $2\pi$-periodic
functions $\varphi$ which are defined by (\ref{energy}) for $E = E_0$.
For uniqueness arguments, we shall consider initial conditions
$\varphi(0) = 0$ and $\dot{\varphi}(0) > 0$, which determine uniquely
one of the two odd $2\pi$-periodic functions $\varphi$.

The limiting $2\pi$-periodic travelling wave solution at $\varepsilon = 0$
should satisfy the constraints (\ref{traveling}), which we do by
choosing for any fixed $q \in [0,\pi]$,
\begin{equation}
\label{limit-solution}
\varepsilon = 0: \quad u_{2n-1}(\tau) = \varphi(\tau + 2q n), \quad w_{2n}(\tau) = 0,
\quad \tau \in \R,  \quad n \in \Z.
\end{equation}
To prove the persistence of this limiting solution in powers of $\varepsilon$
within the granular dimer chain (\ref{eq:ADD}), we shall work in the Sobolev spaces
of odd $2\pi$-periodic functions for $\{ u_{2n-1}\}_{n \in \Z}$,
\begin{equation}
\label{space-1}
H_u^k = \left\{ u \in H^k_{\rm per}(0,2\pi) : \quad u(-\tau) = -u(\tau), \;\; \tau \in \R \right\},
\quad k \in \N_0,
\end{equation}
and in the Sobolev spaces of $2\pi$-periodic functions with zero mean for $\{ w_{2n}\}_{n \in \Z}$,
\begin{equation}
\label{space-2}
H_w^k = \left\{ w \in H^k_{\rm per}(0,2\pi) : \quad \int_0^{2\pi} w(\tau) d \tau = 0 \right\},
\quad k \in \N_0.
\end{equation}
The constraints in (\ref{space-1}) and (\ref{space-2}) reflects the presence of two symmetries
(\ref{symmetry1}) and (\ref{symmetry2}). The two symmetries generate a two-dimensional kernel of
the linearized operators. Under the constraints in (\ref{space-1})
and (\ref{space-2}), the kernel of the linearized operators is trivial, zero-dimensional.

It will be clear from analysis that the vector space $H_w^k$ defined by (\ref{space-2})
is not precise enough to prove the persistence of travelling wave solutions
satisfying the constraints (\ref{traveling}). Instead of this space,
for any fixed $q \in [0,\pi]$, we introduce the vector space $\tilde{H}_w^k$ by
\begin{equation}
\label{space-3}
\tilde{H}_w^k = \left\{ w \in H^k_{\rm per}(0,2\pi) : \quad w(\tau) = -w(-\tau-2q) \right\},
\quad k \in \N_0.
\end{equation}
We note that $\tilde{H}_w^k \subset H_w^k$, because if the constraint $w(\tau) = -w(-\tau-2q)$
is satisfied, then the $2\pi$-periodic function $w$ has zero mean.

\subsection{Special periodic traveling waves}

Before developing persistence analysis, we shall point out three remarkable explicit
periodic travelling solutions of the granular dimer chain (\ref{eq:ADD}) for $q = 0$,
$q = \frac{\pi}{2}$ and $q = \pi$. For $q = \frac{\pi}{2}$, we have an exact solution
\begin{equation}
\label{exact-1}
q = \frac{\pi}{2} : \quad u_{2n-1}(\tau) = \varphi(\tau + \pi n), \quad w_{2n}(\tau) = 0.
\end{equation}
This solution preserves the constraint $V'(u_{2n+1}) = V'(-u_{2n-1})$ in equations
(\ref{eq:ADD}) thanks to the symmetry $\varphi(\tau - \pi) = \varphi(\tau + \pi) = -\varphi(\tau)$
on the $2\pi$-periodic solution of the nonlinear oscillator equation (\ref{oscillator}).

For either $q = 0$ or $q = \pi$, we obtain another exact solution,
\begin{equation}
\label{exact-2}
q = \{0,\pi\} : \quad u_{2n-1}(\tau) = \frac{\varphi(\tau)}{(1+\varepsilon^2)^3}, \quad
w_{2n}(\tau) = \frac{- \varepsilon \varphi(\tau)}{(1+\varepsilon^2)^3},
\end{equation}

By construction, these solutions (\ref{exact-1}) and (\ref{exact-2})
persist for any $\varepsilon \geq 0$. We shall investigate if the continuations
are unique near $\varepsilon = 0$ for these special values of $q$ and
if there is a unique continuation of the general limiting solution (\ref{limit-solution})
in $\varepsilon$ for any other fixed value of $q \in [0,\pi]$.

Furthermore, we note that the exact solution (\ref{exact-2}) for $q = \pi$ at $\varepsilon = 1$
satisfies the constraint (\ref{scalar-constraint}) with $U_{2n-1}(\tau) = - U_{2n}(\tau) =
U_{2n}(\tau - \pi)$. This reduction indicates that the function (\ref{exact-2})
for $\varepsilon = 1$ satisfies the granular monomer chain (\ref{eq:Mono}) and coincides
with the solution considered by James \cite{James2}. On the other hand,
the exact solutions (\ref{exact-1}) for $q = \frac{\pi}{2}$
and (\ref{exact-2}) for $q = 0$ do not
produce any solutions of the monomer chain at $\varepsilon = 1$. This indicates that
there exists generally two distinct solutions at $\varepsilon = 1$, one is continued
from $\varepsilon = 0$ and the other one is constructed from the solution of the
monomer chain (\ref{eq:Mono}) in \cite{James2}.

\section{Persistence of periodic traveling waves near $\varepsilon = 0$}

\subsection{Main result}

We consider the system of differential advance-delay equations (\ref{eq:advance-delay}).
The limiting solution (\ref{limit-solution}) becomes now
\begin{equation}
\label{limit-solution-advanced-delay}
\varepsilon = 0: \quad u_*(\tau) = \varphi(\tau), \quad w_*(\tau) = 0,
\quad \tau \in \R,
\end{equation}
where $\varphi$ is a unique odd $2\pi$-periodic solution of the nonlinear oscillator
equation (\ref{oscillator}) with $\dot{\varphi}(0) > 0$.
We now formulate the main result of this section.

\begin{theorem}
Fix $q \in [0,\pi]$. There is a unique $C^1$ continuation of
$2\pi$-periodic traveling wave (\ref{limit-solution-advanced-delay})
in $\varepsilon$, that is, there is a $\varepsilon_0 > 0$ such that
for every $\varepsilon \in (0,\varepsilon_0)$, there are $C > 0$ and
a unique $2\pi$-periodic solution $(u_*,w_*) \in H^2_u \times \tilde{H}^2_w$
of the system of differential advance-delay equations (\ref{eq:advance-delay}) such that
\begin{equation}
\label{bound-limit-solution}
\| u_* - \varphi \|_{H^2_{\rm per}} \leq C \varepsilon^2, \quad
\| w_* \|_{H^2_{\rm per}} \leq C \varepsilon.
\end{equation}
\label{theorem-1}
\end{theorem}

\begin{remark}
By Theorem \ref{theorem-1}, the limiting solution (\ref{limit-solution-advanced-delay})
for $q \in \left\{ 0, \frac{\pi}{2}, \pi \right\}$ is uniquely continued in $\varepsilon$.
These continuations coincide with the exact solutions (\ref{exact-1}) and (\ref{exact-2}).
\end{remark}

\subsection{Formal expansions in powers of $\varepsilon$}

Let us first consider formal expansions in powers of $\varepsilon$ to understand
the persistence analysis from $\varepsilon = 0$. Expanding the solution
of the system of differential advance-delay equations (\ref{eq:advance-delay}), we write
\begin{equation}
\label{expansion}
u_*(\tau) = \varphi(\tau) + \varepsilon^2 u_*^{(2)}(\tau) + {\rm o}(\varepsilon^2),
\quad w_*(\tau) = \varepsilon w_*^{(1)}(\tau) + {\rm o}(\varepsilon^2),
\end{equation}
and obtain the linear inhomogeneous equations
\begin{equation}
\label{inhomogeneous-1}
\ddot{w}^{(1)}_*(\tau) = F_w^{(1)}(\tau) := V'(\varphi(\tau + 2q)) - V'(-\varphi(\tau))
\end{equation}
and
\begin{equation}
\label{inhomogeneous-2}
\ddot{u}^{(2)}_*(\tau) + \alpha |\varphi(\tau)|^{\alpha - 1} u_*^{(2)}(\tau) = F_u^{(2)}(\tau) :=
V''(-\varphi(\tau)) w_*^{(1)}(\tau) + V''(\varphi(\tau)) w_*^{(1)}(\tau - 2q).
\end{equation}
Because $V$ is $C^2$ but not $C^3$, we have to truncate
the formal expansion (\ref{expansion}) at ${\rm o}(\varepsilon^2)$
to indicate that there are obstacles to continue the power series beyond
terms of the $\mathcal{O}(\varepsilon^2)$ order.

Let us consider two differential operators
\begin{eqnarray}
\label{operator-1}
L_0 = \frac{d^2}{d \tau^2} & : & \quad H^2_{\rm per}(0,2\pi) \to L^2_{\rm per}(0,2\pi), \\
\label{operator-2}
L = \frac{d^2}{d \tau^2} + \alpha |\varphi(\tau)|^{\alpha - 1} & : & \quad
H^2_{\rm per}(0,2\pi) \to L^2_{\rm per}(0,2\pi),
\end{eqnarray}
As a consequence of two symmetries, these operators are not invertible because
they admit one-dimensional kernels,
\begin{equation}
\label{kernels}
{\rm Ker}(L_0) = {\rm span}\{1\}, \quad {\rm Ker}(L) = {\rm span}\{ \dot{\varphi} \}.
\end{equation}
Note that the kernel of $L$ is one-dimensional under the constraint $T'(E_0) \neq 0$
(see Lemma 3 in \cite{James2} for a review of this standard result).

To find uniquely solutions of the inhomogeneous equations (\ref{inhomogeneous-1}) and
(\ref{inhomogeneous-2}) in function spaces $H^2_w$ and $H^2_u$ respectively,
see (\ref{space-1}) and (\ref{space-2}) for definition of function spaces, the
source terms must satisfy the Fredholm conditions
$$
\langle 1, F_w^{(1)} \rangle_{L^2_{\rm per}} = 0 \quad \mbox{\rm and} \quad
\langle \dot{\varphi}, F_u^{(2)}\rangle_{L^2_{\rm per}} = 0.
$$
The first Fredholm condition is satisfied,
$$
\int_0^{2\pi} \left[ V'(\varphi(\tau + 2q)) - V'(-\varphi(\tau)) \right] d \tau =
\int_0^{2\pi} V'(\varphi(\tau + 2q)) d \tau - \int_0^{2\pi} V'(-\varphi(\tau)) d \tau = 0,
$$
because the mean value of a periodic function is independent on the limits of integration
and the function $\varphi$ is odd in $\tau$. Since $F_w^{(1)} \in L^2_w$, there is a unique
solution $w^{(1)} \in H^2_w$ of the linear inhomogeneous equation (\ref{inhomogeneous-1}).

The second Fredholm condition is satisfied,
$$
\int_0^{2\pi}\dot{\varphi}(\tau) \left[ V''(-\varphi(\tau)) w_*^{(1)}(\tau) +
V''(\varphi(\tau)) w_*^{(1)}(\tau - 2q) \right] d \tau = 0,
$$
if the function $F_u^{(2)}$ is odd in $\tau$. If this is the case, then $F_u^{(2)} \in L^2_u$
and there is a unique solution $u^{(2)} \in H^2_u$ of the linear inhomogeneous equation (\ref{inhomogeneous-2}).
To show that $F_u^{(2)}$ is odd in $\tau$, we will prove that $w_*^{(1)}$ satisfies the reduction
\begin{equation}
\label{reduction-w}
w_*^{(1)}(\tau) = -w_*^{(1)}(-\tau-2q), \quad \Rightarrow \quad F_u^{(2)}(-\tau) = - F_u^{(2)}(\tau),
\quad \tau \in \R.
\end{equation}
It follows from the linear inhomogeneous equation (\ref{inhomogeneous-1}) that
$$
\ddot{w}^{(1)}_*(\tau) + \ddot{w}^{(1)}_*(-\tau -2q) =
V'(\varphi(\tau + 2q)) - V'(-\varphi(\tau)) + V'(\varphi(-\tau)) - V'(-\varphi(-\tau-2q)) = 0,
$$
where the last equality appears because $\varphi$ is odd in $\tau$. Integrating this equation
twice and using the fact that $w^{(1)}_* \in H^2_w$, we obtain reduction (\ref{reduction-w}).
Note that the reduction (\ref{reduction-w}) implies that $w^{(1)}_* \in \tilde{H}^2_w$,
where $\tilde{H}^2_w \subset H^2_w$ is given by (\ref{space-3}).

\subsection{Proof of Theorem \ref{theorem-1}}

To prove Theorem \ref{theorem-1}, we shall consider the vector fields of the system
of differential advance-delay equations (\ref{eq:advance-delay}),
\begin{equation}
\label{vector-field}
\left\{ \begin{array}{l}
F_u(u(\tau),w(\tau),\varepsilon) := V'(\varepsilon w(\tau) - u(\tau)) - V'(u(\tau) - \varepsilon w(\tau - 2 q)), \\
F_w(u(\tau),w(\tau),\varepsilon) := \varepsilon V'(u(\tau + 2 q) - \varepsilon w(\tau)) -
\varepsilon V'(\varepsilon w(\tau) - u(\tau)),
\end{array} \right. \quad \tau \in \mathbb{R}.
\end{equation}
We are looking for a strong solution $(u_*,w_*)$ of the system (\ref{eq:advance-delay})
satisfying the reduction,
\begin{equation}
\label{reduction-u-w}
u_*(-\tau) = -u_*(\tau), \quad w_*(\tau) = -w_*(-\tau-2q), \quad \tau \in \R,
\end{equation}
that is, $u_* \in H^2_u(\R)$ and $w_* \in \tilde{H}^2_w(\R)$.

If $(u,w) \in H^2_u \times \tilde{H}^2_w$,
then $F_u$ is odd in $\tau$. Furthermore, since $V$ is $C^2$, then $F_u$ is a $C^1$ map
from $H^2_u \times \tilde{H}^2_w \times \R$ to $L^2_u$ and its Jacobian
at $\varepsilon = 0$ is given by
\begin{equation}
\label{Jacobian-1}
D_u F_u(u,w,0) = V''(- u) - V''(u) = -\alpha |u|^{\alpha - 1},
\quad D_w F_u(u,w,0) = 0.
\end{equation}
On the other hand, under the constraints (\ref{reduction-u-w}),
we have $F_w \in L^2_w$, because
\begin{eqnarray*}
\int_0^{2\pi} F_w(u(\tau),w(\tau),\varepsilon) d \tau & = &
\varepsilon \int_0^{2\pi} V'(u(\tau + 2 q) + \varepsilon w(-\tau-2q)) d\tau \\
& \phantom{t} & \phantom{texttest} -
\varepsilon \int_0^{2\pi} V'(\varepsilon w(\tau) + u(-\tau)) d \tau = 0.
\end{eqnarray*}
Moreover, under the constraints (\ref{reduction-u-w}), we actually have $F_w \in \tilde{L}^2_w$ because
\begin{eqnarray*}
& \phantom{t} & F_w(u(\tau),w(\tau),\varepsilon) + F_w(u(-\tau - 2q),w(-\tau - 2q),\varepsilon) \\
& \phantom{t} & \phantom{texttest} = \varepsilon V'(u(\tau + 2 q) - \varepsilon w(\tau)) -
\varepsilon V'(\varepsilon w(\tau) - u(\tau)) \\
& \phantom{t} & \phantom{texttexttext}
+ \varepsilon V'(u(-\tau) - \varepsilon w(-\tau-2q)) -
\varepsilon V'(\varepsilon w(-\tau-2q) - u(-\tau-2q)) \\
& \phantom{t} & \phantom{texttest} = 0.
\end{eqnarray*}
Since $V$ is $C^2$, then $F_w$ is a $C^1$ map
from $H^2_u \times \tilde{H}^2_w$ to $\tilde{L}^2_w$ and
its Jacobian at $\varepsilon = 0$ is given by
\begin{equation}
\label{Jacobian-2}
D_u F_w(u,w,0) = 0, \quad D_w F_w(u,w,0) = 0.
\end{equation}

Let us now define the nonlinear operator
\begin{equation}
\label{operator-dimer}
\left\{ \begin{array}{l}
f_u(u,w,\varepsilon) := \frac{d^2 u}{d \tau^2} - F_u(u,w,\varepsilon), \\
f_w(u,w,\varepsilon) := \frac{d^2 w}{d \tau^2} - F_w(u,w,\varepsilon).
\end{array} \right.
\end{equation}
We have $(f_u,f_w) : H^2_u \times \tilde{H}^2_w \times \R \to L^2_u \times \tilde{L}^2_w$
are $C^1$ near the point $(\varphi,0,0) \in H^2_u \times \tilde{H}^2_w \times \R$.
To apply the Implicit Function Theorem near this point, we need
$(f_u,f_w) = 0$ at $(u,w,\varepsilon) = (\varphi,0,0)$ and the invertibility
of the Jacobian operator $(f_u,f_w)$ with respect to $(u,w)$ near $(\varphi,0,0)$.

It follows from (\ref{Jacobian-1}) and (\ref{Jacobian-2}) that the Jacobian operator
of $(f_u,f_w)$ at $(\varphi,0,0)$ is given by the diagonal
matrix of operators $L$ and $L_0$ defined by (\ref{operator-1}) and (\ref{operator-2}).
The kernels of these operators in (\ref{kernels}) are zero-dimensional in
the constrained vector spaces (\ref{space-1}) and (\ref{space-2}) (we actually use space (\ref{space-3})
in place of space (\ref{space-2})).

By the Implicit Function Theorem, there exists a
$C^1$ continuation of the limiting solution (\ref{limit-solution-advanced-delay})
with respect to $\varepsilon$ as the $2\pi$-periodic solutions
$(u_*,w_*) \in H^2_u \times \tilde{H}^2_w$ of the system
of differential advance-delay equations (\ref{eq:advance-delay}) near $\varepsilon = 0$.
From the explicit expression (\ref{vector-field}), we can see
that $\| w_* \|_{H^2_{\rm per}} = \mathcal{O}(\varepsilon)$ whereas
$\| u_* - \varphi \|_{H^2_{\rm per}} = \mathcal{O}(\varepsilon^2)$ as $\varepsilon \to 0$.
The proof of Theorem \ref{theorem-1} is complete.

\section{Spectral stability of periodic traveling waves near $\varepsilon = 0$}

\subsection{Linearization at the periodic traveling waves}

We shall consider the dimer chain equations (\ref{eq:ADD}),
which admit for small $\varepsilon > 0$ the periodic traveling waves
in the form (\ref{advance-delay}), where $(u_*, w_*)$ is defined
by Theorem \ref{theorem-1}. Linearizing the system of nonlinear equations
(\ref{eq:ADD}) at the periodic traveling waves (\ref{advance-delay}),
we obtain the system of linearized dimer equations
for small perturbations,
\begin{equation}
 \label{eq:ADD-linear}
\left\{ \begin{array}{l}
\ddot{u}_{2n-1} = V''(\varepsilon w_*(\tau + 2qn) - u_*(\tau + 2qn))
(\varepsilon w_{2n} - u_{2n-1}) \\ \phantom{texttexttexttexttext}
- V''(u_*(\tau + 2qn) - \varepsilon w_*(\tau + 2qn - 2q)) (u_{2n-1} - \varepsilon w_{2 n - 2}), \\
\ddot{w}_{2n} = \varepsilon V''(u_*(\tau + 2qn + 2 q) - \varepsilon w_*(\tau + 2qn))
(u_{2 n+1}- \varepsilon w_{2n}) \\ \phantom{texttexttexttexttext}
- \varepsilon V''(\varepsilon w_*(\tau + 2qn) - u_*(\tau + 2qn)) (\varepsilon w_{2n} - u_{2 n-1}),
\end{array} \right.
\end{equation}
where $n \in \Z$. A technical complication
is that $V''$ is continuous but not continuous differentiable. This will complicate our analysis
of perturbation expansions for small $\varepsilon > 0$. Note that the technical complications
does not occur for exact solutions (\ref{exact-1}) and (\ref{exact-2}).
Indeed, for exact solution (\ref{exact-1}) with $q = \frac{\pi}{2}$, the linearized system
(\ref{eq:ADD-linear}) is rewritten explicitly as
\begin{equation}
 \label{eq:ADD-linear-1}
\left\{ \begin{array}{l}
\ddot{u}_{2n-1} + \alpha |\varphi|^{\alpha - 1} u_{2n-1} = \varepsilon
\left( V''(- \varphi) w_{2n} + V''(\varphi) w_{2 n - 2} \right), \\
\ddot{w}_{2n} + 2 \varepsilon^2 V''(-\varphi) w_{2n} = \varepsilon
V''(-\varphi) (u_{2 n+1} + u_{2 n-1}).
\end{array} \right.
\end{equation}
For exact solution (\ref{exact-2}) with $q = 0$ or $q = \pi$, the linearized system
(\ref{eq:ADD-linear}) is rewritten explicitly as
\begin{equation}
 \label{eq:ADD-linear-2}
\left\{ \begin{array}{l}
\ddot{u}_{2n-1} + \frac{\alpha}{1 + \varepsilon^2} |\varphi|^{\alpha - 1} u_{2n-1} =
\frac{\varepsilon}{1 + \varepsilon^2}
\left( V''(- \varphi) w_{2n} + V''(\varphi) w_{2 n - 2} \right), \\
\ddot{w}_{2n} + \frac{\alpha \varepsilon^2}{1 + \varepsilon^2} |\varphi|^{\alpha - 1} w_{2n}
= \frac{\varepsilon}{1 + \varepsilon^2} \left( V''(\varphi) u_{2 n+1} + V''(-\varphi) u_{2 n-1}  \right).
\end{array} \right.
\end{equation}
In both cases, the linearized systems (\ref{eq:ADD-linear-1}) and (\ref{eq:ADD-linear-2})
are analytic in $\varepsilon$ near $\varepsilon = 0$.

The system of linearized equations (\ref{eq:ADD-linear}) has the same symplectic structure
(\ref{symplectic}), but the Hamiltonian is now given by
\begin{eqnarray}
\nonumber
H & = & \frac{1}{2} \sum_{n \in \Z} \left( p_{2n-1}^2 + q_{2n}^2 \right) +
\frac{1}{2} \sum_{n \in \Z} V''(\varepsilon w_*(\tau + 2q n) - u_*(\tau + 2 q n))
(\varepsilon w_{2n} - u_{2n-1})^2 \\
\label{Hamiltonian-quadratic} & \phantom{t} & \phantom{texttext}
+ \frac{1}{2} \sum_{n \in \Z} V''(u_*(\tau + 2 q n) - \varepsilon w_*(\tau + 2q n - 2q))
(u_{2n-1} - \varepsilon w_{2n-2})^2.
\end{eqnarray}
The Hamiltonian $H$ is quadratic in canonical variables
$\left\{ u_{2n-1}, p_{2n-1} = \dot{u}_{2n-1}, w_{2n}, q_{2n} = \dot{w}_{2n} \right\}_{n \in \Z}$.

\subsection{Main result}

Because coefficients of the linearized dimer equations (\ref{eq:ADD-linear})
are $2\pi$-periodic in $\tau$, we shall look for
an infinite-dimensional analogue of the Floquet theorem that states that all solutions
of the linear system with $2\pi$-periodic coefficients satisfies the reduction
\begin{equation}
\label{floquet-reduction}
{\bf u}(\tau + 2\pi) = \mathcal{M} {\bf u}(\tau), \quad \tau \in \R,
\end{equation}
where ${\bf u} := \left[ \cdots, w_{2n-2}, u_{2n-1}, w_{2n}, u_{2n+1}, \cdots \right]$
and $\mathcal{M}$ is the monodromy operator.

\begin{remark}
Let $q = \frac{\pi m}{N}$ for some positive integers $m$ and $N$ such that
$1 \leq m \leq N$. In this case, the system of dimer equations
(\ref{eq:ADD}) can be closed into a chain of $2 m N$ second-order differential
equations subject to the periodic boundary conditions (\ref{boundary-conditions}). Similarly,
the linearized system (\ref{eq:ADD-linear}) can also be closed as a system
of $2 m N$ second-order equations
and the monodromy operator $\mathcal{M}$ becomes an infinite diagonal composition
of a $4 m N$-by-$4 m N$ Floquet matrix, each matrix has $4 m N$ eigenvalues
called the Floquet multipliers.
\end{remark}

We can find eigenvalues of the monodromy operator ${\cal M}$
by looking for the set of eigenvectors in the form,
\begin{equation}
\label{floquet-eigenvectors}
u_{2n-1}(\tau) = U_{2n-1}(\tau) e^{\lambda \tau}, \quad u_{2n}(\tau) = W_{2n}(\tau) e^{\lambda \tau},
\quad \tau \in \R,
\end{equation}
where $(U_{2n-1},W_{2n})$ are $2\pi$-periodic functions and
the admissible values of $\lambda$ are found
from the existence of these $2\pi$-periodic functions.
The admissible values of $\lambda$ are called the
{\em characteristic exponents} and they define the Floquet multipliers $\mu$ by the standard
formula $\mu = e^{2\pi \lambda}$.

Eigenvectors (\ref{floquet-eigenvectors}) are defined as
$2\pi$-periodic solutions of the linear eigenvalue problem,
\begin{equation}
 \label{eq:ADD-eigenvalue}
\left\{ \begin{array}{l}
\ddot{U}_{2n-1} + 2\lambda \dot{U}_{2n-1} + \lambda^2 U_{2n-1}
= V''(\varepsilon w_*(\tau + 2qn) - u_*(\tau + 2qn)) (\varepsilon W_{2n} - U_{2n-1})
 \\ \phantom{texttexttexttexttext}
- V''(u_*(\tau + 2qn) - \varepsilon w_*(\tau + 2qn - 2q)) (U_{2n-1} - \varepsilon W_{2 n - 2}), \\
\ddot{W}_{2n} + 2\lambda \dot{W}_{2n} + \lambda^2 W_{2n} =
\varepsilon V''(u_*(\tau + 2qn + 2 q) - \varepsilon w_*(\tau + 2qn)) (U_{2 n+1}- \varepsilon W_{2n}) \\
\phantom{texttexttexttexttext}
- \varepsilon V''(\varepsilon w_*(\tau + 2qn) - u_*(\tau + 2qn)) (\varepsilon W_{2n} - U_{2 n-1}).
\end{array} \right.
\end{equation}

The Krein signature, which plays an important role in the studies of spectral stability
of periodic solutions (see Section4 in \cite{Aubry}), is defined as the sign of the $2$-form
associated with the symplectic structure (\ref{symplectic}):
\begin{equation}
\label{Krein}
\sigma = i \sum_{n \in \mathbb{Z}} \left[ u_{2n-1} \bar{p}_{2n-1} - \bar{u}_{2n-1} p_{2n-1}
+ w_{2n} \bar{q}_{2n} - \bar{w}_{2n} q_{2n} \right],
\end{equation}
where $\left\{ u_{2n-1}, p_{2n-1} = \dot{u}_{2n-1}, w_{2n}, q_{2n} = \dot{w}_{2n} \right\}_{n \in \Z}$
is an eigenvector (\ref{floquet-eigenvectors}) associated with an eigenvalue $\lambda \in i \R_+$.
Note that by the symmetry of the linear eigenvalue problem (\ref{eq:ADD-eigenvalue}), it follows that
if $\lambda$ is an eigenvalue, then $\bar{\lambda}$ is also an eigenvalue, whereas
the $2$-form $\sigma$ is constant with respect to $\tau \in \R$.

If $\varepsilon = 0$, the monodromy operator $\mathcal{M}$ in (\ref{floquet-reduction})
is block-diagonal and consists of an infinite set of $2$-by-$2$ Jordan blocks, because
the dimer system (\ref{eq:ADD}) is decoupled into a countable set of uncoupled
second-order differential equations. As a result, the linear eigenvalue problem
(\ref{eq:ADD-eigenvalue}) with the limiting solution (\ref{limit-solution}) admits
an infinite set of $2\pi$-periodic solutions for $\lambda = 0$,
\begin{equation}
\label{floquet-limit}
\varepsilon = 0 : \quad U^{(0)}_{2n-1} = c_{2n-1} \dot{\varphi}(\tau + 2 q n), \quad
W^{(0)}_{2n} = a_{2n}, \quad n \in \Z,
\end{equation}
where $\{ c_{2n-1}, a_{2n} \}_{n \in \Z}$ are arbitrary coefficients. Besides
eigenvectors (\ref{floquet-limit}), there exists another countable set of
generalized eigenvectors for each of the uncoupled second-order differential equations,
which contribute to $2$-by-$2$ Jordan blocks. Each block corresponds
to the double Floquet multiplier $\mu = 1$ or the double characteristic exponent $\lambda = 0$.
When $\varepsilon \neq 0$ but $\varepsilon \ll 1$, the characteristic exponent $\lambda = 0$
of a high algebraic multiplicity splits. We shall study the splitting of the characteristic
exponents $\lambda$ by the perturbation arguments.

We now formulate the main result of this section.

\begin{theorem}
Fix $q = \frac{\pi m}{N}$ for some positive integers $m$ and $N$ such that
$1 \leq m \leq N$. Let $(u_*,w_*) \in H^2_u \times \tilde{H}^2_w$ be
defined by Theorem \ref{theorem-1} for sufficiently small positive $\varepsilon$.
Consider the linear eigenvalue problem (\ref{eq:ADD-eigenvalue})
subject to $2 m N$-periodic boundary conditions (\ref{boundary-conditions}).
There is a $\varepsilon_0 > 0$
such that, for every $\varepsilon \in (0,\varepsilon_0)$,
there exists $q_0(\varepsilon) \in \left(0,\frac{\pi}{2}\right)$
such that for every $q \in (0,q_0(\varepsilon))$ or $q \in (\pi - q_0(\varepsilon),\pi]$,
no values of $\lambda$ with ${\rm Re}(\lambda) \neq 0$ exist, whereas
for every $q \in (q_0(\varepsilon),\pi - q_0(\varepsilon))$,
there exist some values of $\lambda$ with ${\rm Re}(\lambda) > 0$.
\label{theorem-2}
\end{theorem}

\begin{remark}
By Theorem \ref{theorem-2}, periodic traveling waves
are spectrally stable for $q \in (0,q_0(\varepsilon))$ and
$q \in (\pi - q_0(\varepsilon),\pi]$ and unstable for
$q \in (q_0(\varepsilon),\pi - q_0(\varepsilon))$.
Therefore, the linearized system (\ref{eq:ADD-linear-1})
for the exact solution (\ref{exact-1}) with $q = \frac{\pi}{2}$
subject to $4$-periodic boundary conditions
($m = 1$ and $N = 2$) is unstable for small $\varepsilon > 0$,
where the linearized system (\ref{eq:ADD-linear-2})
for the exact solution (\ref{exact-2}) with $q = \pi$
subject to $2$-periodic boundary conditions
($m = 1$ and $N = 1$) is stable for small $\varepsilon > 0$.
\end{remark}

\begin{remark}
The result of Theorem \ref{theorem-2} is expected to
hold for all values of $q$ in $[0,\pi]$ but the spectrum
of the linear eigenvalue problem (\ref{eq:ADD-eigenvalue})
for the characteristic  exponent $\lambda$ becomes continuous
and connected to zero. An infinite-dimensional analogue of the
perturbation theory is required to study eigenvalues of the monodromy
operator $\mathcal{M}$ in this case.
\end{remark}

\begin{remark}
The case $q = 0$ is degenerate for an application of the perturbation theory. Nevertheless,
we show numerically that the linearized system (\ref{eq:ADD-linear-2}) for
the exact solution (\ref{exact-2}) with $q = 0$ ($m = 1$ and $N \to \infty$)
is stable for small $\varepsilon > 0$ and all characteristic exponents are
at least double for any $\varepsilon > 0$.
\end{remark}

\subsection{Formal perturbation expansions}

We would normally expect splitting $\lambda = \mathcal{O}(\varepsilon^{1/2})$ if the
limiting linear eigenvalue problem at $\varepsilon = 0$ is diagonally decomposed into
$2$-by-$2$ Jordan blocks \cite{PelSak}. However, in the linearized dimer problem
(\ref{eq:ADD-eigenvalue}), this splitting occurs in a higher
order, that is, $\lambda = \mathcal{O}(\varepsilon)$,
because the coupling between the particles of equal masses
shows up at the $\mathcal{O}(\varepsilon^2)$ order of the perturbation theory.
Regular perturbation computations in $\mathcal{O}(\varepsilon^2)$
would require $V''$ to be at least $C^1$, which we do not have. In the computations below,
we neglect this discrepancy, which is valid at least for $q = \frac{\pi}{2}$ and $q = \pi$.
For other values of $q$, the formal perturbation expansion is justified with
the renormalization technique (Section 4.6).

We expand $2\pi$-periodic solutions of the linear eigenvalue problem (\ref{eq:ADD-eigenvalue})
into power series of $\varepsilon$:
\begin{equation}
\label{expansions-eigenvalue}
\lambda = \varepsilon \lambda^{(1)} + \varepsilon^2 \lambda^{(2)} + {\rm o}(\varepsilon^2)
\end{equation}
and
\begin{eqnarray}
\label{expansions-eigenvectors}
\left\{ \begin{array}{l}
U_{2n-1} = U_{2n-1}^{(0)} + \varepsilon U_{2n-1}^{(1)} + \varepsilon^2 U_{2n-1}^{(2)} + {\rm o}(\varepsilon^2), \\
W_{2n} = W_{2n}^{(0)} + \varepsilon W_{2n}^{(1)} + \varepsilon^2 W_{2n}^{(2)} + {\rm o}(\varepsilon^2), \end{array} \right.
\end{eqnarray}
where the zeroth-order terms are given by (\ref{floquet-limit}).
To determine corrections of the power series expansions uniquely, we shall require that
\begin{equation}
\langle \dot{\varphi}, U_{2n-1}^{(j)} \rangle_{L^2_{\rm per}} =
\langle 1, W_{2n}^{(j)} \rangle_{L^2_{\rm per}} = 0, \quad n \in \Z, \quad j = 1,2.
\end{equation}
Indeed, if $U_{2n-1}^{(j)}$ contains a component, which is parallel to $\dot{\varphi}$, then
the corresponding term only changes the value of $c_{2n-1}$
in the eigenvector (\ref{floquet-limit}), which is yet to be determined.
Similarly, if a $2\pi$-periodic function $W_{2n}^{(j)}$ has a nonzero mean value,
then the mean value of $W_{2n}^{(j)}$
only changes the value of $a_{2n}$ in the eigenvector (\ref{floquet-limit}),
which is yet to be determined.

The linear equations (\ref{eq:ADD-eigenvalue}) are satisfied at the $\mathcal{O}(\varepsilon^0)$
order. Collecting terms at the $\mathcal{O}(\varepsilon)$ order, we obtain
\begin{equation}
 \label{eq:ADD-order-1}
\left\{ \begin{array}{l}
\ddot{U}^{(1)}_{2n-1} + \alpha |\varphi(\tau + 2 q n)|^{\alpha - 1} U^{(1)}_{2n-1}
= - 2\lambda^{(1)} \dot{U}^{(0)}_{2n-1} \\
\phantom{texttexttexttexttexttext} + V''(-\varphi(\tau + 2 q n)) W_{2n}^{(0)} +
V''(\varphi(\tau + 2 q n)) W^{(0)}_{2 n - 2}, \\
\ddot{W}^{(1)}_{2n} = - 2\lambda^{(1)} \dot{W}^{(0)}_{2n}
+ V''(\varphi(\tau + 2 q n + 2 q)) U^{(0)}_{2 n+1} + V''(-\varphi(\tau + 2 q n)) U^{(0)}_{2 n-1}.
\end{array} \right.
\end{equation}
Let us define solutions of the following linear inhomogeneous equations:
\begin{eqnarray}
\label{eq-1}
\ddot{v} + \alpha |\varphi|^{\alpha - 1} v = -2 \ddot{\varphi}, \\
\label{eq-2}
\ddot{y}_{\pm} + \alpha |\varphi|^{\alpha - 1} y_{\pm} = V''(\pm \varphi), \\
\label{eq-3}
\ddot{z}_{\pm} = V''(\pm \varphi) \dot{\varphi}.
\end{eqnarray}
If we can find uniquely $2\pi$-periodic solutions of these equations
such that
$$
\langle \dot{\varphi}, v \rangle_{L^2_{\rm per}} =
\langle \dot{\varphi}, y_{\pm} \rangle_{L^2_{\rm per}} =
\langle 1, z_{\pm} \rangle_{L^2_{\rm per}} = 0,
$$
then the perturbation
equations (\ref{eq:ADD-order-1}) at the $\mathcal{O}(\varepsilon)$ order are satisfied with
\begin{eqnarray}
\label{floquet-order-1}
\left\{ \begin{array}{l} U_{2n-1}^{(1)} =
c_{2n-1} \lambda^{(1)} v(\tau + 2 q n) + a_{2n} y_-(\tau + 2 q n) + a_{2n-2} y_+(\tau + 2 q n), \\
W_{2n}^{(1)} = c_{2n+1} z_+(\tau + 2 q n + 2q) + c_{2n-1} z_-(\tau + 2 q n).
\end{array} \right.
\end{eqnarray}

The linear equations (\ref{eq:ADD-eigenvalue}) are now satisfied up to
the $\mathcal{O}(\varepsilon)$ order.
Collecting terms at the $\mathcal{O}(\varepsilon^2)$ order, we obtain
\begin{equation}
 \label{eq:ADD-order-2}
\left\{ \begin{array}{l}
\ddot{U}^{(2)}_{2n-1} + \alpha |\varphi(\tau + 2 q n)|^{\alpha - 1} U^{(2)}_{2n-1}
= - 2\lambda^{(1)} \dot{U}^{(1)}_{2n-1} - 2\lambda^{(2)} \dot{U}^{(0)}_{2n-1}
- (\lambda^{(1)})^2 U^{(0)}_{2n-1} \\
\phantom{texttext} + V''(-\varphi(\tau + 2 q n)) W_{2n}^{(1)} + V''(\varphi(\tau + 2 q n)) W^{(1)}_{2 n - 2}\\
\phantom{texttext}
- V'''(-\varphi(\tau + 2 q n)) (w_*^{(1)}(\tau + 2 q n) - u_*^{(2)}(\tau + 2 q n)) U_{2n-1}^{(0)}\\
\phantom{texttext}
- V'''(\varphi(\tau + 2 q n)) (u_*^{(2)}(\tau + 2 q n) - w_*^{(1)}(\tau + 2 q n - 2q)) U_{2n-1}^{(0)}, \\
\ddot{W}^{(2)}_{2n} = - 2\lambda^{(1)} \dot{W}^{(1)}_{2n} - 2\lambda^{(2)} \dot{W}^{(0)}_{2n} -
(\lambda^{(1)})^2 W^{(0)}_{2n} \\
\phantom{texttext}
+ V''(\varphi(\tau + 2 q n + 2 q)) (U^{(1)}_{2 n+1} - W_{2n}^{(0)}) +
V''(-\varphi(\tau + 2 q n)) (U^{(1)}_{2 n-1} - W_{2n}^{(0)}),
\end{array} \right.
\end{equation}
where corrections $u_*^{(2)}$ and $w_*^{(1)}$ are defined by expansion (\ref{expansion}).

To solve the linear inhomogeneous equations (\ref{eq:ADD-order-2}), the source terms have
to satisfy the Fredholm conditions because both operators $L$ and $L_0$
defined by (\ref{operator-1}) and (\ref{operator-2}) have one-dimensional kernels. Therefore,
we require the first equation of system (\ref{eq:ADD-order-2})
to be orthogonal to $\dot{\varphi}$ and the second
equation of system (\ref{eq:ADD-order-2})
to be orthogonal to $1$ on $[-\pi,\pi]$. Substituting (\ref{floquet-limit}) and
(\ref{floquet-order-1}) to the orthogonality conditions
and taking into account the symmetry between couplings of lattice sites on $\mathbb{Z}$,
we obtain the difference equations for $\{ c_{2n-1},a_{2n} \}_{n \in \Z}$:
\begin{equation}
\label{projection-equations}
\left\{ \begin{array}{l}
K \Lambda^2 c_{2n-1} = M_1 (c_{2n+1} + c_{2n-3} - 2 c_{2n-1}) + L_1 \Lambda (a_{2n} - a_{2n-2}), \\
\Lambda^2 a_{2n} = M_2 (a_{2n+2} + a_{2n-2} - 2 a_{2n}) + L_2 \Lambda (c_{2n+1} - c_{2n-1}),
\end{array} \right.
\end{equation}
where $\Lambda \equiv \lambda^{(1)}$, and $(K,M_1,M_2,L_1,L_2)$ are numerical coefficients to
be computed from the projections. In particular, we obtain
\begin{eqnarray*}
K & = & \int_{-\pi}^{\pi} \left( 2 \dot{v}(\tau) + \dot{\varphi}(\tau) \right) \dot{\varphi}(\tau) d \tau, \\
M_1 & = & \int_{-\pi}^{\pi} V''(-\varphi(\tau)) \dot{\varphi}(\tau) z_+(\tau + 2 q) d \tau =
\int_{-\pi}^{\pi} V''(\varphi(\tau)) \dot{\varphi}(\tau) z_-(\tau - 2 q) d \tau, \\
M_2 & = & \frac{1}{2\pi} \int_{-\pi}^{\pi} V''(\varphi(\tau + 2q)) y_-(\tau + 2q) d \tau =
\frac{1}{2\pi} \int_{-\pi}^{\pi} V''(-\varphi(\tau)) y_+(\tau) d \tau, \\
L_1 & = & -2\int_{-\pi}^{\pi} \dot{y}_-(\tau) \dot{\varphi}(\tau) d \tau =
2\int_{-\pi}^{\pi} \dot{y}_+(\tau) \dot{\varphi}(\tau) d \tau, \\
L_2 & = & \frac{1}{2\pi} \int_{-\pi}^{\pi} V''(\varphi(\tau + 2q)) v(\tau + 2q) d \tau =
-\frac{1}{2\pi} \int_{-\pi}^{\pi} V''(-\varphi(\tau)) v(\tau) d \tau.
\end{eqnarray*}

Note that the coefficients $M_1$ and $M_2$ need not to be computed at the diagonal terms
$c_{2n-1}$ and $a_{2n}$ thanks to the fact that the difference equations (\ref{projection-equations})
with $\Lambda = 0$ must have eigenvectors
with equal values of $\{ c_{2n-1} \}_{n \in \Z}$ and $\{ a_{2n} \}_{n \in \mathbb{Z}}$,
which correspond to the two symmetries of the linearized dimer system (\ref{eq:ADD-linear})
related to the symmetries (\ref{symmetry1}) and (\ref{symmetry2}).
This fact shows that the problem of limited smoothness of $V''$, which is $C$ but
not $C^1$ near zero, is not a serious obstacle in the derivation of the reduced system
(\ref{projection-equations}).

Difference equations (\ref{projection-equations}) give a necessary and
sufficient condition to solve
the linear inhomogeneous equations (\ref{eq:ADD-order-2}) at the $\mathcal{O}(\varepsilon^2)$ order
and to continue the perturbation expansions beyond this order. Before
justifying this formal perturbation theory, we shall explicitly compute
the coefficients $(K,M_1,M_2,L_1,L_2)$ of the difference equations (\ref{projection-equations}).

Note that the system of difference equations (\ref{projection-equations}) presents
a quadratic eigenvalue problem with respect to the spectral parameter $\Lambda$.
Such quadratic eigenvalue problem appear often in the context
of spectral stability of nonlinear waves \cite{ChPelQuad,Kollar}.

\subsection{Explicit computations of the coefficients}

We shall prove the following technical result.

\begin{lemma}
Coefficients $K$, $M_2$, $L_1$, and $L_2$ are independent of $q$ and are given by
$$
K = - \frac{4 \pi^2}{T'(E_0)}, \quad M_2 =
\frac{2}{\pi T'(E_0) (\dot{\varphi}(0))^2}, \quad
L_1 = 2\pi L_2 = \frac{2 (2\pi - T'(E_0) (\dot{\varphi}(0))^2)}{T'(E_0) \dot{\varphi}(0)}.
$$
Consequently, $K > 0$, whereas $M_2, L_1, L_2 < 0$. On the other hand,
coefficient $M_1$ depends on $q$ and is given by
$$
M_1 = -\frac{2}{\pi} (\dot{\varphi}(0))^2 + I(q),
$$
where
$$
I(q) = I(\pi - q) :=
- \int_{\pi-2q}^{\pi} \ddot{\varphi}(\tau) \ddot{\varphi}(\tau + 2q) d \tau, \quad q \in \left[0,\frac{\pi}{2}\right].
$$
\label{lemma-coefficients}
\end{lemma}

To prove Lemma \ref{lemma-coefficients}, we first
uniquely solve the linear inhomogeneous equations (\ref{eq-1}), (\ref{eq-2}),
and (\ref{eq-3}). For equation (\ref{eq-1}), we note that a general solution is
$$
v(\tau) = -\tau \dot{\varphi}(\tau) + b_1 \dot{\varphi}(\tau) + b_2 \partial_E \varphi_{E_0}(\tau),
\quad \tau \in [-\pi,\pi],
$$
where $(b_1,b_2)$ are arbitrary constants and $\partial_E \varphi_{E_0}$ is the derivative
of the $T(E)$-periodic solution $\varphi_E$ of the nonlinear oscillator equation (\ref{oscillator})
with the first integral (\ref{energy}) satisfying initial conditions
$\varphi_E(0) = 0$ and $\dot{\varphi}_E(0) = \sqrt{2E}$ at the value of energy $E = E_0$, for which
$T(E_0) = 2\pi$. We note the equation
\begin{equation}
\label{technical-equ}
\partial_E \varphi_{E_0}(\pm \pi) = \mp \frac{1}{2} T'(E_0) \dot{\varphi}(\pm \pi),
\end{equation}
that follows from the differentiation of equation $\varphi_E(\pm T(E)/2) = 0$ with respect
to $E$ at $E = E_0$.

To define $v$ uniquely, we require that $\langle \dot{\varphi}, v \rangle_{L^2_{\rm per}} = 0$.
Because $\dot{\varphi}$ is even in $\tau$, whereas $\tau \dot{\varphi}$ and $\partial_E \varphi_{E_0}$
are odd, we hence have $b_1 = 0$ and $v(0) = 0$. Hence $v$ is odd in $\tau$ and, in order
to satisfy the $2\pi$-periodicity, we shall only require $v(\pi) = 0$, which uniquely specifies
the value of $b_2$ by virtue of (\ref{technical-equ}),
$$
b_2 = \frac{\pi \dot{\varphi}(\pi)}{\partial_E \varphi_{E_0}(\pi)} = -\frac{2\pi}{T'(E_0)}.
$$
As a result, we obtain
\begin{equation}
\label{sol-1}
v(\tau) = -\tau \dot{\varphi}(\tau) - \frac{2\pi}{T'(E_0)} \partial_E \varphi_{E_0}(\tau), \quad
\tau \in [-\pi,\pi].
\end{equation}

For equation (\ref{eq-2}), we can use that $\varphi(\tau) \geq 0$ for $\tau \in [0,\pi]$
and $\varphi(\tau) \leq 0$ for $\tau \in [-\pi,0]$. We can also use
the symmetry $\dot{\varphi}(\pi) = -\dot{\varphi}(0)$.
Integrating equations for $y_{\pm}$ separately,
we obtain
\begin{eqnarray*}
y_+(\tau) & = & \left\{ \begin{array}{ll} 1 + a_+ \dot{\varphi} + b_+ \partial_E \varphi_{E_0}, & \tau \in [-\pi,0], \\
c_+ \dot{\varphi} + d_+ \partial_E \varphi_{E_0}, & \tau \in [0,\pi], \end{array} \right. \\
y_-(\tau) & = & \left\{ \begin{array}{ll} a_- \dot{\varphi} + b_- \partial_E \varphi_{E_0}, & \tau \in [-\pi,0], \\
1 + c_- \dot{\varphi} + d_- \partial_E \varphi_{E_0}, & \tau \in [0,\pi]. \end{array} \right.
\end{eqnarray*}
Continuity of $y_{\pm}$ and $\dot{y}_{\pm}$ across $\tau = 0$ defines uniquely
$d_{\pm} = b_{\pm}$ and $c_{\pm} = a_{\pm} \pm \frac{1}{\dot{\varphi}(0)}$.
With this definition, $\dot{y}_{\pm}(-\pi) = \dot{y}_{\pm}(\pi)$, whereas
condition $y_{\pm}(-\pi) = y_{\pm}(\pi)$ sets up uniquely
$$
b_{\pm} = \pm \frac{2}{T'(E_0) \dot{\varphi}(0)},
$$
whereas constants $a_{\pm}$ are not specified.

To define $y_{\pm}$ uniquely, we again require that
$\langle \dot{\varphi}, y_{\pm} \rangle_{L^2_{\rm per}} = 0$.
This yields the constraint on $a_{\pm}$,
$$
a_{\pm} = \mp \frac{1}{2 \dot{\varphi}(0)} \mp \frac{2 \langle \dot{\varphi}, \partial_E \varphi_{E_0} \rangle_{L^2_{\rm per}}}{T'(E_0) \dot{\varphi}(0) \langle \dot{\varphi}, \dot{\varphi} \rangle_{L^2_{\rm per}}}.
$$
As a result, we obtain
\begin{equation}
\label{sol-2-plus}
y_+(\tau) = a_+ \dot{\varphi}(\tau) + b_+ \partial_E \varphi_{E_0}(\tau) +
\left\{ \begin{array}{ll} 1, & \tau \in [-\pi,0], \\
\frac{\dot{\varphi}(\tau)}{\dot{\varphi}(0)},& \tau \in [0,\pi], \end{array} \right.
\end{equation}
and
\begin{equation}
\label{sol-2-minus}
y_-(\tau) = a_- \dot{\varphi}(\tau) + b_- \partial_E \varphi_{E_0}(\tau) +
\left\{ \begin{array}{ll} 0, & \tau \in [-\pi,0], \\
1 - \frac{\dot{\varphi}(\tau)}{\dot{\varphi}(0)}, & \tau \in [0,\pi], \end{array} \right.
\end{equation}
where $(a_{\pm},b_{\pm})$ are uniquely defined above.

For equation (\ref{eq-3}), we integrate separately on $[-\pi,0]$ and $[0,\pi]$ to obtain
$$
\dot{z}_+(\tau) = \left\{ \begin{array}{ll} c_+ - |\varphi(\tau)|^{\alpha}, & \tau \in [-\pi,0], \\
c_+, & \tau \in [0,\pi], \end{array} \right. \quad
\dot{z}_-(\tau) = \left\{ \begin{array}{ll} c_-, & \tau \in [-\pi,0], \\
c_- + |\varphi(\tau)|^{\alpha}, & \tau \in [0,\pi], \end{array} \right.
$$
where $(c_+,c_-)$ are constants of integration and continuity of $\dot{z}_{\pm}$ across $\tau = 0$
have been used. To define $z_{\pm}$ uniquely, we require that
$\langle 1, z_{\pm} \rangle_{L^2_{\rm per}} = 0$.
Integrating the equations above under this condition, we obtain:
$$
z_+(\tau) = \left\{ \begin{array}{ll} c_+ \tau + d_+ - \dot{\varphi}(\tau), & \tau \in [-\pi,0], \\
c_+ \tau - d_+, & \tau \in [0,\pi], \end{array} \right. \quad
z_-(\tau) = \left\{ \begin{array}{ll} c_- \tau + d_-, & \tau \in [-\pi,0], \\
c_- \tau - d_- - \dot{\varphi}(\tau), & \tau \in [0,\pi], \end{array} \right.
$$
where $(d_+,d_-)$ are constants of integration.
Continuity of $z_{\pm}$ across $\tau = 0$ uniquely sets coefficients
$d_{\pm} = \pm \frac{1}{2} \dot{\varphi}(0)$. Periodicity of $\dot{z}_{\pm}(-\pi)
= \dot{z}_{\pm}(\pi)$ is satisfied. Periodicity of $z_{\pm}(-\pi) = z_{\pm}(\pi)$
uniquely defines coefficients $c_{\pm} = \pm \frac{1}{\pi} \dot{\varphi}(0)$. As a result, we obtain
\begin{equation}
\label{sol-3-plus}
z_+(\tau) = \frac{1}{2\pi} \left\{ \begin{array}{ll}
\dot{\varphi}(0) (2 \tau + \pi) - 2 \pi \dot{\varphi}(\tau), & \tau \in [-\pi,0], \\
\dot{\varphi}(0) (2 \tau - \pi), & \tau \in [0,\pi], \end{array} \right.
\end{equation}
and
\begin{equation}
\label{sol-3-minus}
z_-(\tau) = \frac{1}{2\pi} \left\{ \begin{array}{ll} -\dot{\varphi}(0) (2 \tau + \pi),
& \tau \in [-\pi,0], \\
-\dot{\varphi}(0) (2 \tau - \pi) - 2 \pi \dot{\varphi}(\tau), & \tau \in [0,\pi]. \end{array} \right.
\end{equation}

We can now compute the coefficients $(K,M_1,M_2,L_1,L_2)$ of the difference equations
(\ref{projection-equations}). For coefficients $K$, we integrate by parts, use
equations (\ref{oscillator}), (\ref{energy}), (\ref{sol-1}), and obtain
\begin{eqnarray*}
K & = & \int_{-\pi}^{\pi} \dot{\varphi} (\dot{\varphi} + 2 \dot{v}) d \tau =
\int_{-\pi}^{\pi} (\dot{\varphi}^2 - 2 v \ddot{\varphi}) d \tau \\
& = & \left[ \tau \dot{\varphi}^2 + \frac{2\pi}{T'(E_0)}
\partial_E \varphi_{E_0} \dot{\varphi} \right] \biggr|_{\tau = -\pi}^{\tau = \pi}
+ \frac{2 \pi}{T'(E_0)} \int_{-\pi}^{\pi} \left(
\partial_E \varphi_{E_0} \ddot{\varphi} - \partial_E \dot{\varphi}_{E_0} \dot{\varphi} \right) d \tau \\
& = & - \frac{4 \pi}{T'(E_0)} \int_{0}^{\pi}
\partial_E \left( \frac{1}{2} \dot{\varphi}^2 + \frac{1}{1 + \alpha} \varphi^{1 + \alpha} \right)_{E_0} d \tau
= - \frac{4 \pi^2}{T'(E_0)}.
\end{eqnarray*}
Because $T'(E_0) < 0$, we find that $K > 0$.

For $M_1$, we use equation (\ref{eq-3}), solution (\ref{sol-3-minus}), and obtain
\begin{eqnarray*}
M_1 & = & \int_{-\pi}^{\pi} V''(-\varphi(\tau)) \dot{\varphi}(\tau) z_+(\tau + 2 q) d \tau =
\int_{-\pi}^{\pi} \ddot{z}_-(\tau) z_+(\tau + 2 q) d \tau \\
& = & - \int_{-\pi}^{\pi} \dot{z}_-(\tau) \dot{z}_+(\tau + 2 q) d \tau =
\int_0^{\pi} \ddot{\varphi}(\tau) \dot{z}_+(\tau + 2 q) d \tau,
\end{eqnarray*}
hence, the sign of $M_1$ depends on $q$. Using solution (\ref{sol-3-minus}),
for $q \in \left[0,\frac{\pi}{2}\right]$, we obtain
\begin{eqnarray*}
M_1 & = & \frac{1}{\pi} \dot{\varphi}(0) \int_{0}^{\pi}  \ddot{\varphi}(\tau) d \tau
- \int_{\pi-2q}^{\pi} \ddot{\varphi}(\tau) \ddot{\varphi}(\tau + 2q) d \tau \\
& = & -\frac{2}{\pi} (\dot{\varphi}(0))^2 + I(q), \quad I(q) := - \int_{\pi-2q}^{\pi} \ddot{\varphi}(\tau) \ddot{\varphi}(\tau + 2q) d \tau.
\end{eqnarray*}
On the other hand, for $q \in \left[\frac{\pi}{2},\pi \right]$, we obtain
\begin{eqnarray*}
M_1 = -\frac{2}{\pi} (\dot{\varphi}(0))^2 + \tilde{I}(q), \quad
\tilde{I}(q) := - \int_{0}^{2 \pi - 2q}
\ddot{\varphi}(\tau) \ddot{\varphi}(\tau + 2q) d \tau,
\end{eqnarray*}
so that
\begin{eqnarray*}
\tilde{I}(\pi - q) = - \int_{0}^{2q}
\ddot{\varphi}(\tau) \ddot{\varphi}(\tau - 2q) d \tau
= - \int_{-2q}^{0}
\ddot{\varphi}(\tau) \ddot{\varphi}(\tau + 2q) d \tau = I(q),
\end{eqnarray*}
because the mean value of a periodic function does not depend on the limits of integration.

For $M_2$, we use equation (\ref{eq-2}) and obtain
\begin{eqnarray*}
M_2 & = &  \frac{1}{2\pi} \int_{-\pi}^{\pi} V''(-\varphi) y_+ d \tau =
\frac{\alpha}{2\pi} \int_{0}^{\pi} \varphi^{\alpha-1} y_+ d \tau \\
& = & -\frac{1}{2\pi} \int_0^{\pi} \ddot{y}_+ d \tau =
\frac{1}{\pi} b_+ \partial_E \dot{\phi}_{E_0}(0) =
\frac{2}{\pi T'(E_0) (\dot{\varphi}(0))^2},
\end{eqnarray*}
hence, $M_2 < 0$.

For $L_1$, we use equations (\ref{oscillator}), (\ref{energy}), (\ref{sol-2-minus}), and obtain
\begin{eqnarray*}
L_1 & = & -2\int_{-\pi}^{\pi} \dot{y}_- \dot{\varphi} d \tau  = - 2 b_-
\int_{-\pi}^{\pi} \partial_E \dot{\varphi}_{E_0} \dot{\varphi} d \tau\\
& = & \frac{4}{T'(E_0) \dot{\varphi}(0)} \left[
\int_{0}^{\pi} \left( \partial_E \dot{\varphi}_{E_0} \dot{\varphi} - \partial_E \varphi_{E_0} \ddot{\varphi}
\right) d \tau +
\dot{\varphi} \partial_E \varphi_{E_0} \biggr|_{\tau = 0}^{\tau = \pi} \right] \\
& = & \frac{2 (2\pi - T'(E_0) (\dot{\varphi}(0))^2)}{T'(E_0) \dot{\varphi}(0)}.
\end{eqnarray*}
Because $\dot{\varphi}(0) > 0$ and $T'(E_0) < 0$, we find that $L_1 < 0$.

For $L_2$, we use equations (\ref{oscillator}), (\ref{energy}), (\ref{sol-1}), and obtain
\begin{eqnarray*}
L_2 & = & -\frac{1}{2\pi} \int_{-\pi}^{\pi} V''(-\varphi) v d \tau =
-\frac{\alpha}{2\pi} \int_0^{\pi} \varphi^{\alpha - 1} v d \tau \\
& = & \frac{1}{T'(E_0)} \int_0^{\pi} \partial_E \left( \varphi_{E_0} \right)^{\alpha} d\tau -
\frac{1}{2\pi} \int_0^{\pi} \varphi^{\alpha} d\tau \\
& = & \left[ \frac{1}{2\pi} \dot{\varphi}
- \frac{1}{T'(E_0)} \partial_E \dot{\varphi}_{E_0}  \right] \biggr|_{\tau = 0}^{\tau = \pi} \\
& = & \frac{2\pi - T'(E_0) (\dot{\varphi}(0))^2}{\pi T'(E_0) \dot{\varphi}(0)} = \frac{1}{2 \pi} L_1,
\end{eqnarray*}
hence, $L_2 < 0$.

The proof of Lemma \ref{lemma-coefficients} is complete.

\subsection{Eigenvalues of the difference equations}

Because the coefficients $(K,M_1,M_2,L_1,L_2)$ of the difference equations
(\ref{projection-equations}) are independent of $n$, we can solve these equations by
the discrete Fourier transform. Substituting
\begin{equation}
\label{dFT}
c_{2n-1} = C e^{i \theta (2n-1)}, \quad a_{2n} = A e^{i 2 \theta n},
\end{equation}
where $\theta \in [0,\pi]$ is the Fourier spectral parameter, we obtain
the system of linear homogeneous equations,
\begin{equation}
\label{linear-equations}
\left\{ \begin{array}{l}
K \Lambda^2 C = 2 M_1 (\cos(2\theta) - 1) C + 2 i L_1 \Lambda \sin(\theta) A, \\
\Lambda^2 A = 2 M_2 (\cos(2\theta) - 1) A + 2 i L_2 \Lambda \sin(\theta) C.
\end{array} \right.
\end{equation}

A nonzero solution of system (\ref{linear-equations}) exists if and only if $\Lambda$
is a root of the characteristic polynomial,
\begin{equation}
\label{characteristic-eq}
D(\Lambda;\theta) = K \Lambda^4 + 4 \Lambda^2 (M_1 + K M_2 + L_1 L_2) \sin^2(\theta) +
16 M_1 M_2 \sin^4(\theta) = 0.
\end{equation}
Since this equation is bi-quadratic, it has two pairs of roots for each $\theta \in [0,\pi]$.
For $\theta = 0$, both pairs are zero, which recovers the characteristic exponent
$\lambda = 0$ of algebraic multiplicity of (at least) 4 in the linear eigenvalue problem
(\ref{eq:ADD-eigenvalue}). For a fixed $\theta \in (0,\pi)$, the two pairs
of roots are generally nonzero, say $\Lambda_1^2$ and $\Lambda_2^2$.
The following result specifies their location.

\begin{lemma}
There exists a $q_0 \in \left(0,\frac{\pi}{2}\right)$ such that
$\Lambda_1^2 \leq \Lambda_2^2 < 0$ for $q \in [0,q_0) \cup (\pi - q_0,\pi]$
and $\Lambda_1^2 < 0 < \Lambda_2^2$ for $q \in (q_0,\pi - q_0)$.
\label{lemma-roots}
\end{lemma}

To classify the nonzero roots of the characteristic polynomial
\eqref{characteristic-eq}, we define
\begin{equation}
\label{parameters}
\Gamma := M_1 + K M_2 + L_1 L_2, \quad \Delta := 4 K M_1 M_2.
\end{equation}
The two pairs of roots are determined in Table I.

\begin{center}
\begin{tabular}{|l|l|}
  \hline
Coefficients &  Roots \\
  \hline
$\Delta < 0$ & $\Lambda_1^2 < 0 < \Lambda_2^2$ \\
 \hline
$0 < \Delta \leq \Gamma^2$, $\Gamma > 0$ & $\Lambda_1^2 \leq \Lambda_2^2 < 0$ \\
 \hline
$0 < \Delta \leq \Gamma^2$, $\Gamma < 0$ & $\Lambda_2^2 \geq \Lambda_1^2 > 0$ \\
 \hline
$\Delta > \Gamma^2$ & ${\rm Im}(\Lambda_1^2) > 0, \; {\rm Im}(\Lambda_2^2) < 0$ \\
\hline
\end{tabular}
\end{center}
{\bf Table I:}  Squared roots of the characteristic equation (\ref{characteristic-eq}).

\vspace{0.25cm}

Using the explicit computations of the coefficients $(K,M_1,M_2,L_1,L_2)$, we obtain
$$
\Gamma = -\frac{8}{T'(E_0)} + I(q), \quad \Delta = \frac{64}{(T'(E_0))^2} \left(
1 - \frac{\pi I(q)}{2 (\dot{\varphi}(0))^2} \right).
$$
Because $I(q)$ is symmetric about $q = \frac{\pi}{2}$, we can restrict our consideration
to the values $q \in \left[0,\frac{\pi}{2}\right]$ and use the explicit definition
from Lemma \ref{lemma-coefficients}:
$$
I(q) = -\int_{\pi-2q}^{\pi} \ddot{\varphi}(\tau) \ddot{\varphi}(\tau + 2q) d \tau,
\quad q \in \left[0,\frac{\pi}{2}\right].
$$
We claim that $I(q)$ is a positive, monotonically increasing function in $\left[0,\frac{\pi}{2}\right]$
starting with $I(0) = 0$.

Because $\ddot{\varphi}(\tau) = -|\varphi(\tau)|^{\alpha - 1} \varphi(\tau)$,
we realize that $\ddot{\varphi}(\tau) \leq 0$ for $\tau \in [0,\pi]$,
whereas $\ddot{\varphi}(\tau + 2q) \geq 0$ for $\tau \in [\pi-2q,\pi]$.
Hence, $I(q) \geq 0$ for any $2q \in [0,\pi]$. Moreover, $I$ is a continuously
differentiable function of $q$, because the first derivative,
\begin{eqnarray*}
I'(q) & = & - 2 \int_{\pi-2q}^{\pi} \ddot{\varphi}(\tau) \dddot{\varphi}(\tau + 2q) d \tau =
2 \int_{\pi-2q}^{\pi} \dddot{\varphi}(\tau) \ddot{\varphi}(\tau + 2q) d \tau \\
& = & -2 \alpha \int_{\pi-2q}^{\pi} |\varphi(\tau)|^{\alpha-1}
\dot{\varphi}(\tau) \ddot{\varphi}(\tau + 2q) d \tau,
\end{eqnarray*}
is continuous for all $2q \in [0,\pi]$.
Because $\dot{\varphi}(\tau)$ and $\ddot{\varphi}(\tau)$
are odd and even with respect to $\tau = \frac{\pi}{2}$, respectively,
and $\dot{\varphi}(\tau) \geq 0$ for $\tau \in \left[0,\frac{\pi}{2}\right]$,
we have $I'(q) \geq 0$ for any $2q \in [0,\pi]$. Therefore,
$I(q)$ is monotonically increasing from $I(0) = 0$
to
$$
I\left(\frac{\pi}{2}\right) = - \int_{0}^{\pi} \ddot{\varphi}(\tau) \ddot{\varphi}(\tau + \pi) d \tau
= \int_{0}^{\pi} (\ddot{\varphi}(\tau))^2 d \tau > 0.
$$
Hence, for all $q \in \left[0,\frac{\pi}{2}\right]$, we have $\Gamma > 0$ and
$$
\Gamma^2 - \Delta = I(q) \left( I(q) - \frac{16}{T'(E_0)} + \frac{32 \pi}{(T'(E_0) \dot{\varphi}(0))^2} \right) \geq 0,
$$
where $\Delta = \Gamma^2$ if and only if $q = 0$. Therefore,
only the first two lines of Table I can occur.

For $q = 0$, $I(0) = 0$, hence $M_1 < 0$, $\Delta > 0$ and $\Delta = \Gamma^2$.
The second line of Table I gives $\Lambda_1^2  = \Lambda_2^2 < 0$.
All characteristic exponents are purely imaginary and degenerate, thanks
to the explicit computations:
\begin{equation}
\label{limiting-roots}
\Lambda_1^2 = \Lambda_2^2 = - \frac{4}{\pi^2} \sin^2(\theta).
\end{equation}

The proof of Lemma \ref{lemma-roots} is achieved if there is $q_0 \in \left(0,\frac{\pi}{2}\right)$ such that
the first line of Table I yields $\Lambda_1^2  < 0 < \Lambda_2^2$ for $q \in \left(q_0,\frac{\pi}{2}\right]$
and the second line of Table II yields $\Lambda_1^2 < \Lambda_2^2 < 0$  for $q \in (0,q_0)$.
Because $I$ is a monotonically increasing function of $q$ and $\Delta > 0$ for $q = 0$,
the existence of $q_0 \in \left(0,\frac{\pi}{2}\right)$
follows by continuity if $\Delta < 0$ for $q = \frac{\pi}{2}$.
Since $K > 0$ and $M_2 < 0$, we need to prove that $M_1 > 0$ for $q = \frac{\pi}{2}$
or equivalently,
$$
I\left(\frac{\pi}{2}\right) > \frac{2}{\pi} (\dot{\varphi}(0))^2.
$$

Because $\dot{\varphi}$ is a $2\pi$-periodic function with zero mean,
Poincar\'e inequality yields
$$
I\left(\frac{\pi}{2}\right) = \frac{1}{2} \int_{-\pi}^{\pi} (\ddot{\varphi}(\tau))^2 d \tau \geq
\frac{1}{2} \int_{-\pi}^{\pi} (\dot{\varphi}(\tau))^2 d \tau.
$$
On the other hand, using equations (\ref{oscillator}), (\ref{energy}), and integration
by parts, we obtain
$$
\frac{1}{2} \int_{-\pi}^{\pi} (\dot{\varphi}(\tau))^2 d \tau =
-\frac{1}{2} \int_{-\pi}^{\pi} \varphi(\tau) \ddot{\varphi}(\tau) d \tau =
\frac{1}{2} \int_{-\pi}^{\pi} |\varphi(\tau)|^{\alpha + 1} d \tau = \frac{2\pi (\alpha+1)}{(\alpha + 3)} E,
$$
where the last equality is obtained by integrating the first invariant (\ref{energy}) on $[-\pi,\pi]$.
Therefore, we obtain
$$
I\left(\frac{\pi}{2}\right) \geq
\frac{2\pi (\alpha+1)}{(\alpha + 3)} E = \frac{\pi (\alpha+1)}{(\alpha + 3)} (\dot{\varphi}(0))^2 > \frac{2}{\pi} (\dot{\varphi}(0))^2,
$$
where the last inequality is obtained for $\alpha = \frac{3}{2}$ based on the fact that
$\frac{5 \pi^2}{18} \approx 2.74 > 1$. Therefore, $M_1 > 0$ and hence, $\Delta < 0$
for $q = \frac{\pi}{2}$. The proof of Lemma \ref{lemma-roots} is complete.

\begin{figure}
\begin{center}
\includegraphics[width=7cm,height=5cm]{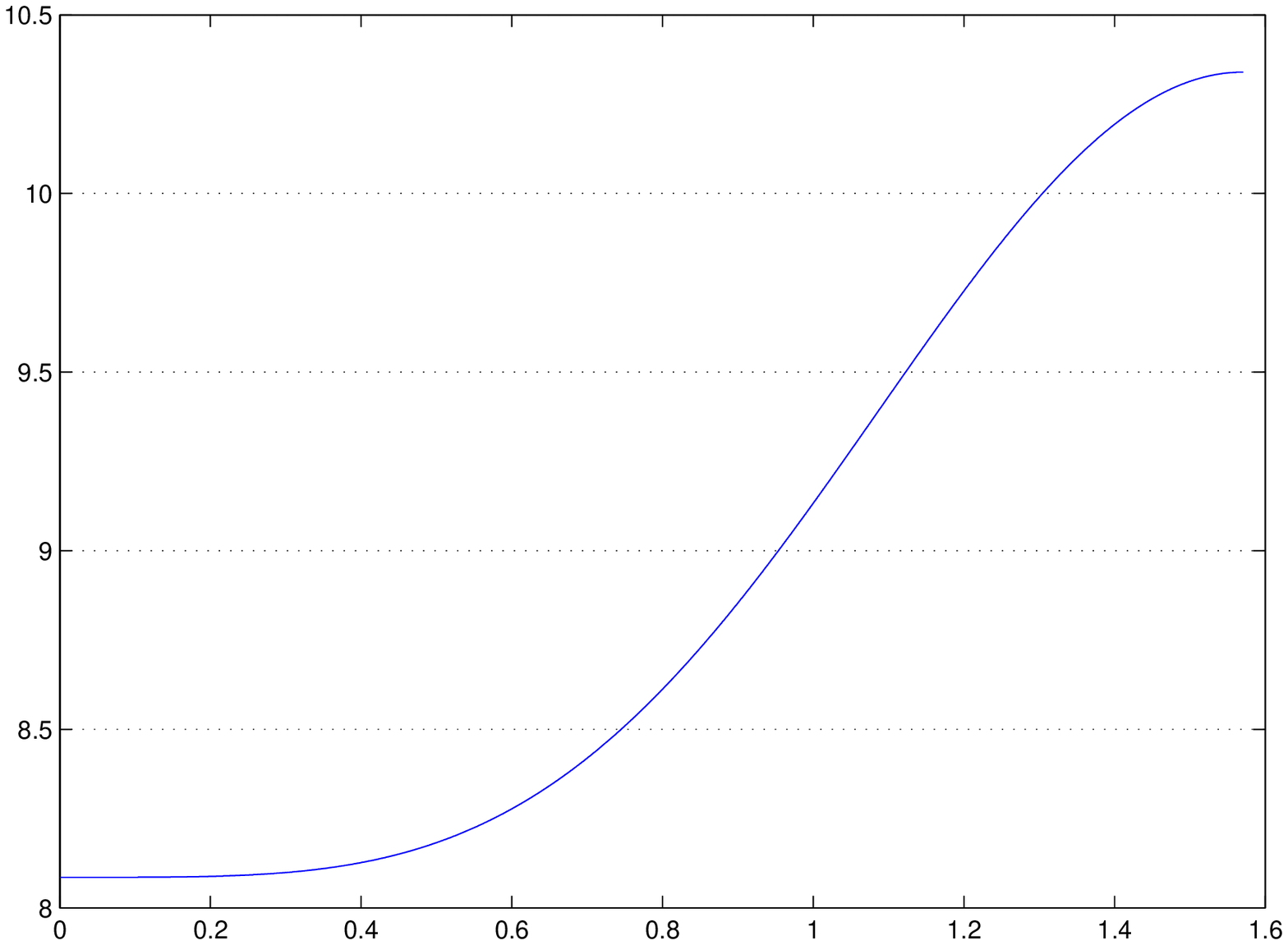} \hspace{0.1cm}
\includegraphics[width=7cm,height=5cm]{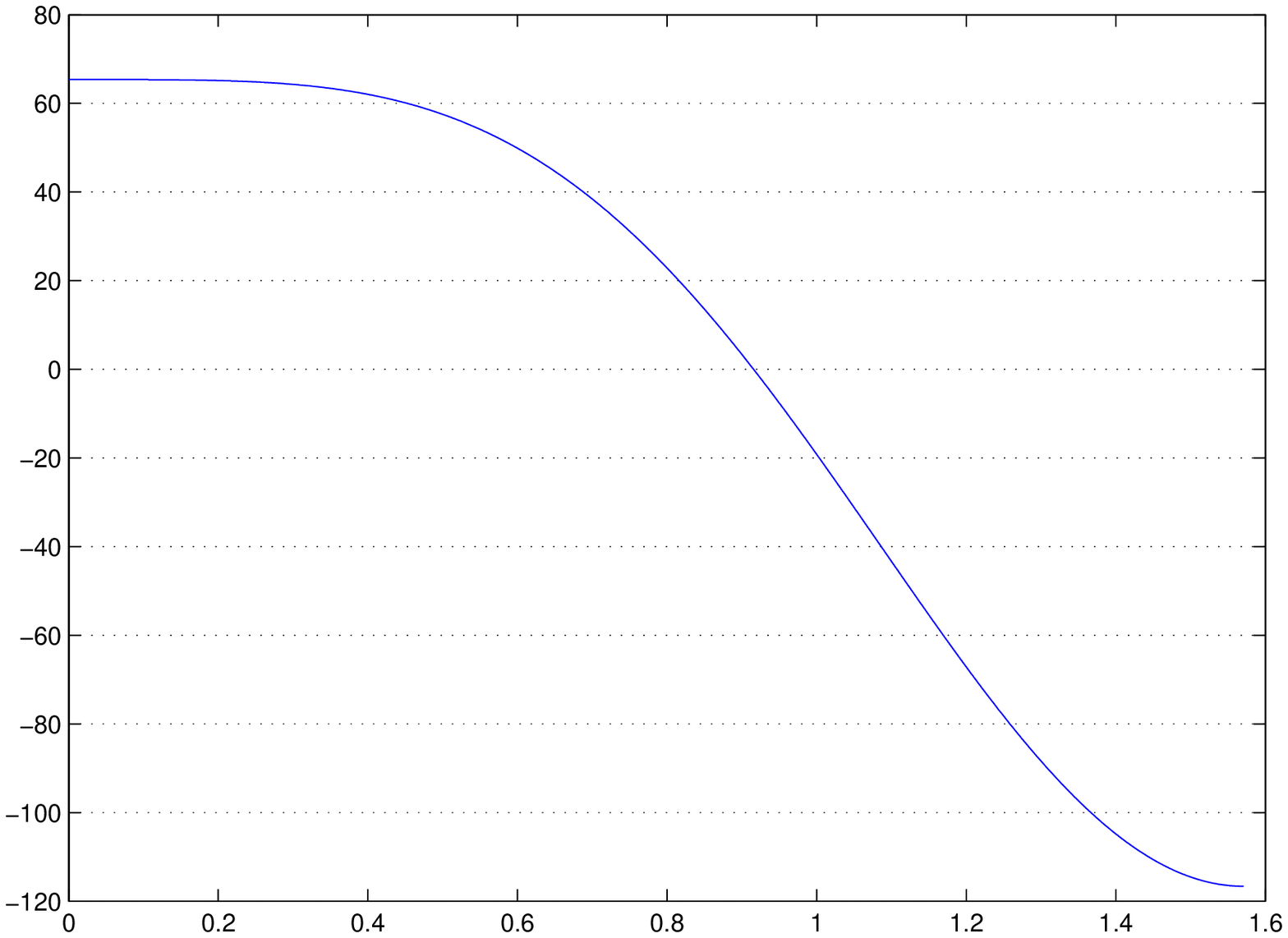}
\end{center}
\caption{Coefficients $\Gamma$ (left) and $\Delta$ (right) versus $q$.}
\label{fig:coeff}
\end{figure}

Numerical approximations of coefficients $\Gamma$ and $\Delta$ versus $q$ is shown on
Figure \ref{fig:coeff}. We can see from the figure that the sign change of $\Delta$
occurs at $q_0 \approx 0.915$.

\subsection{Krein signature of eigenvalues}

Because the eigenvalue problem (\ref{linear-equations}) is symmetric
with respect to reflection of $\theta$ about $\frac{\pi}{2}$, that is,
$\sin(\theta) = \sin(\pi - \theta)$, some roots $\Lambda \in \C$ of
the characteristic polynomial (\ref{characteristic-eq}) produce multiple
eigenvalues $\lambda$ in the linear eigenvalue problem (\ref{eq:ADD-eigenvalue})
at the $\mathcal{O}(\epsilon)$ order of the asymptotic expansion (\ref{expansions-eigenvalue}).
To control splitting and persistence of eigenvalues $\lambda \in i \R_+$ with respect to perturbations, we shall look
at the Krein signature of the $2$-form $\sigma$ defined by (\ref{Krein}).
The following result allows us to compute $\sigma$ asymptotically as $\epsilon \to 0$.

\begin{lemma}
For every $q \in (0,q_0)$, the $2$-form $\sigma$ for every eigenvector
of the linear eigenvalue problem (\ref{eq:ADD-eigenvalue}) generated by the perturbation expansion
(\ref{expansions-eigenvectors})
associated with the root $\Lambda \in i \R_+$ of the characteristic equation (\ref{characteristic-eq})
is nonzero.
\label{lemma-Krein}
\end{lemma}

Using the representation (\ref{floquet-eigenvectors}) for $\lambda = i \omega$ with $\omega \in \R_+$,
we rewrite $\sigma$ in the form:
$$
\sigma = 2 \omega \sum_{n \in \Z} \left[ |U_{2n-1}|^2 + |W_{2n}|^2 \right] +
i \sum_{n \in \mathbb{Z}} \left[ U_{2n-1} \dot{\bar{U}}_{2n-1} - \bar{U}_{2n-1} \dot{U}_{2n-1}
+ W_{2n} \dot{\bar{W}}_{2n} - \bar{W}_{2n} \dot{W}_{2n} \right].
$$
Now using perturbation expansion $\omega = \epsilon \Omega + \mathcal{O}(\epsilon^2)$,
where $\Lambda = i \Omega \in i \R_+$ is a root of the characteristic equation (\ref{characteristic-eq}),
and the perturbation expansions (\ref{expansions-eigenvectors}) for the eigenvector, we compute
$$
\sigma = \epsilon \sum_{n \in \Z} \sigma_n^{(1)} + \mathcal{O}(\epsilon^2),
$$
where
\begin{eqnarray*}
\sigma_n^{(1)} & = & 2 \Omega \left[ |c_{2n-1}|^2 \dot{\varphi}^2(\tau + 2 q n) + |a_{2n}|^2 \right] +
i (c_{2n-1} \dot{\bar{U}}^{(1)}_{2n-1} - \bar{c}_{2n-1} \dot{U}^{(1)}_{2n-1}) \dot{\varphi}(\tau + 2 q n)\\
& \phantom{t} &
- i (c_{2n-1} \bar{U}^{(1)}_{2n-1} - \bar{c}_{2n-1} U^{(1)}_{2n-1}) \ddot{\varphi}(\tau + 2 q n)
+ i (a_{2n} \dot{\bar{W}}^{(1)}_{2n} - \bar{a}_{2n} \dot{W}^{(1)}_{2n}).
\end{eqnarray*}
Using representation (\ref{floquet-order-1}), this becomes
$$
\sigma_n^{(1)} = 2 \Omega (|c_{2n-1}|^2 E_0 + |a_{2n}|^2) +
i (c_{2n-1} \bar{a}_{2n} - \bar{c}_{2n-1} a_{2n}) E_- +
i (c_{2n-1} \bar{a}_{2n-2} - \bar{c}_{2n-1} a_{2n-2}) E_+,
$$
where $E_0$ and $E_{\pm}$ are numerical coefficients given by
\begin{eqnarray*}
E_0 & = & \dot{\varphi}^2 + \dot{\varphi} \dot{v} - \ddot{\varphi} v, \\
E_{\pm} & = & \dot{\varphi} \dot{y}_{\pm} - \ddot{\varphi} y_{\pm} - \dot{z}_{\pm}.
\end{eqnarray*}
Using explicit computations of functions $v$, $y_{\pm}$, and $z_{\pm}$ in Lemma \ref{lemma-coefficients},
we obtain
\begin{eqnarray*}
E_0 = -\frac{2\pi}{T'(E_0)}, \quad E_{\pm} = \pm \frac{2\pi - T'(E_0) (\dot{\varphi}(0))^2}{\pi T'(E_0) \dot{\varphi}(0)},
\end{eqnarray*}
and hence we have
$$
\sigma_n^{(1)} = 2 \Omega \left(\frac{K}{2\pi} |c_{2n-1}|^2 + |a_{2n}|^2 \right) -
i L_2 (c_{2n-1} \bar{a}_{2n} - \bar{c}_{2n-1} a_{2n} - c_{2n-1} \bar{a}_{2n-2} + \bar{c}_{2n-1} a_{2n-2}).
$$

Substituting the eigenvector of the reduced eigenvalue problem (\ref{projection-equations}) in
the discrete Fourier transform form (\ref{dFT}), we obtain
\begin{eqnarray*}
\sigma_n^{(1)} & = & 2 \Omega \left(\frac{K}{2\pi} C^2 + A^2 \right) -
4 L_2 \sin(\theta) C A \\
& = & \frac{1}{\pi \Omega} \left( \Omega^2 K C^2 + 8 \pi M_2 \sin^2(\theta) A^2 \right),
\end{eqnarray*}
where the second equation of system (\ref{linear-equations}) has been used.
Using now the first equation of system (\ref{linear-equations}), we obtain
\begin{equation}
\label{sigma-n}
\sigma_n^{(1)} = \frac{C^2}{\pi L_1 L_2 \Omega^3} \left[ K L_1 L_2 \Omega^4 +
M_2 (K \Omega^2 - 4 M_1 \sin^2(\theta))^2 \right].
\end{equation}
Note that $\sigma_n^{(1)}$ is independent of $n$, hence periodic boundary conditions
are used to obtain a finite expression for the $2$-form $\sigma$.

We consider $q \in (0,q_0)$ and $\theta \in (0,\pi)$, so that $\Omega \neq 0$ and $C \neq 0$.
Then, $\sigma_n^{(1)} = 0$ if and only if
$$
K L_1 L_2 \Omega^4 + M_2 (K \Omega^2 - 4 M_1 \sin^2(\theta))^2 = 0.
$$
Using the explicit coefficients in Lemma \ref{lemma-coefficients}, we factorize
the left hand side as follows:
\begin{eqnarray}
\nonumber
K L_1 L_2 \Omega^4 + M_2 (K \Omega^2 - 4 M_1 \sin^2(\theta))^2 = \left( \Omega^2 + T'(E_0) M_1 M_2
\sin^2(\theta) \right) \\
\times \left( \frac{32 \pi^2}{(T'(E_0))^2} \left( 1 - \frac{T'(E_0) (\dot{\varphi}(0))^2}{4 \pi}
\right) \Omega^2 + \frac{16}{T'(E_0)} M_1 \sin^2(\theta) \right). \label{factorization}
\end{eqnarray}
For every $q \in (0,q_0)$, $M_1 < 0$, so that the second bracket is strictly positive (recall that
$T'(E_0) < 0$). Now the first bracket vanishes at
$$
\Omega^2 = \frac{-2 M_1}{\pi (\dot{\varphi}(0))^2} \sin^2(\theta).
$$
Substituting this constraint to the characteristic equation (\ref{characteristic-eq})
yields after straightforward computations:
$$
D(i \Omega; \theta) = \frac{8 M_1 \sin^4(\theta)}{\pi \dot{\varphi}^2(0)}
\left(1 - \frac{2\pi}{T'(E_0) \dot{\varphi}^2(0)} \right) I(q),
$$
which is nonzero for all $q \in (0,q_0)$ and $\theta \in (0,\pi)$.
Therefore, $\sigma_n^{(1)}$ does not vanish if
$q \in (0,q_0)$ and $\theta \in (0,\pi)$. By continuity of
the perturbation expansions in $\epsilon$, $\sigma$ does not vanish too.
The proof of Lemma \ref{lemma-Krein} is complete.

\begin{remark}
For every $q \in (0,q_0)$, all roots $\Lambda \in i \R_+$ of the characteristic
equation (\ref{characteristic-eq})
are divided into two equal sets, one has $\sigma_n^{(1)} > 0$
and the other one has $\sigma_n^{(1)} < 0$. This follows from the
factorization
$$
D(i \Omega;\theta) = -\frac{4 \pi^2}{T'(E_0)} \left( \Omega^2 - \frac{4}{\pi^2} \sin^2(\theta) \right)^2
- 4 I(q) \left(\Omega^2 - \frac{8}{\pi T'(E_0) (\dot{\varphi}(0))^2} \sin^2(\theta) \right) \sin^2(\theta).
$$
As $q \to 0$, $I(q) \to 0$ and perturbation theory for double roots (\ref{limiting-roots}) for $q = 0$ yields
$$
\Omega^2 = \frac{4}{\pi^2} \sin^2(\theta) \pm \frac{2}{\pi^2} \sin^2(\theta) \sqrt{|T'(E_0)|
I(q)\left(1 - \frac{2 \pi}{T'(E_0) (\dot{\varphi}(0))^2} \right)} + \mathcal{O}(I(q)).
$$
Using the factorization formula (\ref{factorization}),
the sign of $\sigma_n^{(1)}$ is determined by the expression
$$
 \Omega^2 + T'(E_0) M_1 M_2 \sin^2(\theta) = \pm \frac{2}{\pi^2} \sin^2(\theta) \sqrt{|T'(E_0)|
I(q)\left(1 - \frac{2 \pi}{T'(E_0) (\dot{\varphi}(0))^2} \right)} + \mathcal{O}(I(q)),
$$
which justifies the claim for small positive $q$. By Lemma
\ref{lemma-Krein}, the Krein signature of $\sigma_n^{(1)}$ does not vanish for all $q \in (0,q_0)$
and $\theta \in (0,\pi)$, therefore
the splitting of all roots $\Lambda \in i \R_+$ into two equal sets persists for all values of
$q \in (0,q_0)$.\label{remark-Krein}
\end{remark}

\subsection{Proof of Theorem \ref{theorem-2}}

To conclude the proof of Theorem \ref{theorem-2}, we develop rigorous perturbation
theory in the case when $q = \frac{\pi m}{N}$ for some positive integers $m$ and $N$
such that $1 \leq m \leq N$. In this case, the linear eigenvalue problem (\ref{eq:ADD-eigenvalue})
can be closed at $2mN$ second-order differential equations subject to $2mN$-periodic
boundary conditions (\ref{boundary-conditions}) and we are looking for $4 m N$ eigenvalues $\lambda$,
which are characteristic values of a $4mN \times 4mN$ Floquet matrix.

At $\varepsilon = 0$, we have $2 mN$ double Jordan blocks for
$\lambda = 0$. The $2 mN$ eigenvectors are given by (\ref{floquet-limit}).
The $2mN$-periodic boundary conditions
are incorporated in the discrete Fourier transform (\ref{dFT}) if
$$
\theta = \frac{\pi k}{m N} \equiv \theta_k(m,N), \quad k = 0,1,\ldots, mN - 1.
$$
Because the characteristic equation (\ref{characteristic-eq}) for each $\theta_k(m,N)$ returns $4$ roots,
we count $4 m N$ roots of the characteristic equation
(\ref{characteristic-eq}), as many as there are eigenvalues
$\lambda$ in the linear eigenvalue problem (\ref{eq:ADD-eigenvalue}).
As long as the roots are non-degenerate (if $\Delta \neq \Gamma^2$) and
different from zero (if $\Delta \neq 0$), the first-order perturbation theory
predicts splitting of $\lambda = 0$ into symmetric pairs of non-zero eigenvalues. The
zero eigenvalue of multiplicity $4$ persists and corresponds to the value $\theta_0(m,N) = 0$.
It is associated with the symmetries of the dimer equations (\ref{symmetry1}) and (\ref{symmetry2}).

The non-zero eigenvalues are located hierarchically with respect to the values of
$\sin^2(\theta)$ for $\theta = \theta_k(m,N)$ with $1 \leq k \leq mN - 1$.
Because $\sin(\theta) = \sin(\pi - \theta)$, every non-zero eigenvalue
corresponding to $\theta_k(m,N) \neq \frac{\pi}{2}$ is double. Because
all eigenvalues $\lambda \in i \R_+$ have a definite Krein signature by Lemma \ref{lemma-Krein}
and the sign of $\sigma_n^{(1)}$ in (\ref{sigma-n}) is same for both eigenvalues with
$\sin(\theta) = \sin(\pi - \theta)$, the double eigenvalues $\lambda \in i \R$ are structurally stable with
respect to parameter continuations \cite{ChPel} in the sense that they
split along the imaginary axis beyond the leading-order perturbation theory.

\begin{remark}
The argument based on the Krein signature does not cover the case of double
real eigenvalues $\Lambda \in \R_+$, which may split off the real axis to the complex domain.
However, both real and complex eigenvalues contribute to the count of unstable
eigenvalues with the account of their multiplicities.
\end{remark}

It remains to address the issue that the first-order perturbation theory
uses computations of $V'''$, which is not a continuous function of its argument.
To deal with this issue, we use a renormalization technique. We note that
if $(u_*,w_*)$ is a solution of the differential advance-delay equations (\ref{eq:advance-delay})
given by Theorem \ref{theorem-1}, then
\begin{eqnarray}
\nonumber
\dddot{u}_*(\tau) & = & V''(\varepsilon w_*(\tau) - u_*(\tau))
(\varepsilon \dot{w}_*(\tau) - \dot{u}_*(\tau)) \\
\label{eq:derivative}
& \phantom{t} & \phantom{texttext} -
V''(u_*(\tau) - \varepsilon w_*(\tau - 2 q)) (\dot{u}_*(\tau) -
\varepsilon \dot{w}_*(\tau - 2 q)),
\end{eqnarray}
where the right-hand side is a continuous function of $\tau$.

Using (\ref{eq:derivative}), we substitute
$$
U_{2n-1} = c_{2n-1} \dot{u}_*(\tau + 2 q n) + \mathcal{U}_{2n-1}, \quad
W_{2n} = \mathcal{W}_{2n},
$$
for an arbitrary choice of $\{ c_{2n-1} \}_{n \in \Z}$, into
the linear eigenvalue problem (\ref{eq:ADD-eigenvalue})
and obtain:
\begin{equation}
\label{eq:ADD-eigenvalue-renormalized}
\left\{ \begin{array}{l}
\ddot{\mathcal{U}}_{2n-1} + 2\lambda \dot{\mathcal{U}}_{2n-1} + \lambda^2 \mathcal{U}_{2n-1}
= V''(\varepsilon w_*(\tau + 2qn) - u_*(\tau + 2qn)) (\varepsilon \mathcal{W}_{2n} - \mathcal{U}_{2n-1})
 \\ \phantom{texttexttexttexttext}
- V''(u_*(\tau + 2qn) - \varepsilon w_*(\tau + 2qn - 2q)) (\mathcal{U}_{2n-1} - \varepsilon \mathcal{W}_{2 n - 2}),
 \\ \phantom{texttexttexttexttext}
- (2 \lambda \ddot{u}_*(\tau + 2 qn) + \lambda^2 \dot{u}_*(\tau + 2 qn)) c_{2n-1}  \\
\phantom{texttexttexttexttext}
- \varepsilon V''(\varepsilon w_*(\tau + 2qn)-u_*(\tau + 2q n)) \dot{w}_*(\tau + 2qn) c_{2n-1} \\
\phantom{texttexttexttexttext}
- \varepsilon V''(u_*(\tau + 2qn) - \varepsilon w_*(\tau + 2qn - 2q)) \dot{w}_*(\tau + 2qn - 2q) c_{2n-1}, \\
\ddot{\mathcal{W}}_{2n} + 2\lambda \dot{\mathcal{W}}_{2n} + \lambda^2 \mathcal{W}_{2n} =
\varepsilon V''(u_*(\tau + 2qn + 2 q) - \varepsilon w_*(\tau + 2qn)) (\mathcal{U}_{2 n+1}- \varepsilon \mathcal{W}_{2n}) \\
\phantom{texttexttexttexttext}
- \varepsilon V''(\varepsilon w_*(\tau + 2qn) - u_*(\tau + 2qn))
(\varepsilon \mathcal{W}_{2n} - \mathcal{U}_{2 n-1}) \\
\phantom{texttexttexttexttext}
+ \varepsilon V''(u_*(\tau + 2qn + 2 q) - \varepsilon w_*(\tau + 2qn)) \dot{u}_*(\tau + 2qn + 2 q) c_{2n-1} \\
\phantom{texttexttexttexttext}
+ \varepsilon V''(\varepsilon w_*(\tau + 2qn) - u_*(\tau + 2qn)) \dot{u}_*(\tau + 2qn) c_{2n-1}.
\end{array} \right.
\end{equation}

When we repeat decompositions of the first-order perturbation theory, we write
\begin{eqnarray*}
\lambda & = & \varepsilon \lambda^{(1)} + \varepsilon^2 \lambda^{(2)} + {\rm o}(\varepsilon^2), \\
\mathcal{U}_{2n-1} & = & \varepsilon \mathcal{U}_{2n-1}^{(1)} + \varepsilon^2 \mathcal{U}_{2n-1}^{(2)}
+ {\rm o}(\varepsilon^2), \\
\mathcal{W}_{2n} & = & a_{2n} + \varepsilon \mathcal{W}_{2n}^{(1)}
+ \varepsilon^2 \mathcal{W}_{2n}^{(2)} + {\rm o}(\varepsilon^2),
\end{eqnarray*}
for an arbitrary choice of $\{ a_{2n} \}_{n \in \Z}$. Substituting this decomposition
to system (\ref{eq:ADD-eigenvalue-renormalized}), we obtain equations
at the $\mathcal{O}(\varepsilon)$ and $\mathcal{O}(\varepsilon^2)$ orders, which do not
require computations of $V'''$. Hence, the system of difference equations
(\ref{projection-equations}) is justified and the splitting of the
eigenvalues $\lambda$ at the first order of the perturbation theory
obeys roots of the characteristic equation (\ref{characteristic-eq}).
Persistence of roots beyond the ${\rm o}(\varepsilon^2)$ order
holds by the standard perturbation theory for isolated eigenvalues of
the Floquet matrix. The proof of Theorem \ref{theorem-2} is complete.

\section{Numerical Results}
\label{sec:num}

We obtain numerical approximations of the periodic travelling waves
(\ref{traveling}) in the case $q = \frac{\pi}{N}$, where $N$ is an integer,
when the dimer system \eqref{eq:ADD} can be closed as the following system
of $2N$ differential equations:
\begin{equation}
 \label{eq:ODE}
 \left\{ \begin{array}{l} \ddot{u}_{2n-1}(t) = (\varepsilon w_{2n}(t) - u_{2n-1}(t))_+^{\alpha}-(u_{2n-1}(t)-\varepsilon w_{2n-2}(t))_+^{\alpha},\\
\ddot{w}_{2n}(t) = \varepsilon(u_{2n-1}(t)-\varepsilon w_{2n}(t))_+^{\alpha}-\varepsilon(\varepsilon w_{2n}(t)-u_{2n+1}(t))_+^{\alpha},
\end{array} \right. \quad  1 \leq n \leq N,
\end{equation}
subject to the periodic boundary conditions
\begin{equation}
u_{-1} = u_{2 N-1}, \quad u_{2 N+1} = u_1, \quad w_0 = w_{2 N}, \quad w_{2 N+2} = w_2.
\end{equation}
The periodic travelling waves (\ref{traveling}) corresponds to $2\pi$-periodic
solutions of system (\ref{eq:ODE}) satisfying the reduction
\begin{equation}
\label{traveling-wave-reduction}
u_{2n+1}(t) = u_{2n-1}\left(t + \frac{2\pi}{N}\right), \quad
w_{2n+2}(t) = w_{2n}\left(t + \frac{2\pi}{N}\right),
\quad t \in \R, \quad 1 \leq n \leq N.
\end{equation}
For convenience and uniqueness, we look for an odd function $u_1(t) = -u_1(-t)$
with
\begin{equation}
\label{convention}
u_1(0) = 0 \quad \mbox{\rm and} \quad \dot{u}_1(0) > 0.
\end{equation}
By Theorem \ref{theorem-1}, the travelling wave solutions satisfying
(\ref{traveling-wave-reduction}) and (\ref{convention})
exist uniquely at least for small values of $\varepsilon$. We can continue this branch
of solutions with respect to parameter $\varepsilon$
in the interval $[0,1]$ starting from the limiting solutions
obtained at $\varepsilon = 0$.

\subsection{Existence of travelling periodic wave solutions}

In order to obtain $2\pi$-periodic traveling wave solutions to the nonlinear system
\eqref{eq:ODE}, we use the shooting method. Our shooting parameters are given
by the initial conditions
$$
\{ (u_{2n-1}(0),\dot{u}_{2n-1}(0),w_{2n}(0),\dot{w}_{2n}(0) \}_{1 \leq n \leq N}.
$$
Since $u_1(0) = 0$, this gives a set of $2N-1$ shooting parameters. However,
for solutions satisfying the travelling wave reduction (\ref{traveling-wave-reduction}),
we can use symmetries of the nonlinear system of differential equations
(\ref{eq:ODE}) to reduce the number of shooting parameters to $N$ parameters.

For two particles ($N = 1$ or $q = \pi$), the existence and stability
problems are trivial. The exact solution (\ref{exact-2}) is uniquely continued
for all $\varepsilon \in [0,1]$ and matches the
exact solution of the granular chain of two identical particles at $\varepsilon = 1$
considered in \cite{James2}. This solution is spectrally stable with respect to
$2$-periodic perturbations for all $\varepsilon \in [0,1]$ because the characteristic
value $\lambda = 0$ has algebraic multiplicity four, which coincides with the total number
of admissible characteristic values $\lambda$.

For four particles ($N=2$ or $q = \frac{\pi}{2}$),
the nonlinear system (\ref{eq:ODE}) is written explicitly as
\begin{equation}
 \label{eq:four}
 \left\{ \begin{array}{l}
\ddot{u}_1(t) = (\varepsilon w_4(t)-u_1(t))_+^{\alpha}-(u_1(t)-\varepsilon w_2(t))_+^{\alpha},\\
\ddot{w}_2(t) = \varepsilon[(u_1(t)-\varepsilon w_2(t))_+^{\alpha}-(\varepsilon w_2(t)-u_3(t))_+^{\alpha},\\
\ddot{u}_3(t) = (\varepsilon w_2(t)-u_3(t))_+^{\alpha}-(u_3(t)-\varepsilon w_4(t))_+^{\alpha},\\
\ddot{w}_4(t) = \varepsilon[(u_3(t)-\varepsilon w_4(t))_+^{\alpha}-(\varepsilon w_4(t)-u_1(t))_+^{\alpha}.
\end{array} \right.
\end{equation}
We are looking for $2\pi$-periodic functions satisfying the travelling wave reduction:
\begin{equation}
\label{travelling-reductions}
u_3(t) = u_1(t + \pi), \quad w_4(t) = w_2(t + \pi).
\end{equation}
We note that the system (\ref{eq:four}) is invariant with respect to the following transformation:
\begin{equation}
\label{reversibility-reduction}
u_1(-t) = -u_1(t), \quad w_2(-t) = -w_4(t), \quad u_3(-t) = -u_3(t), \quad w_4(-t) = -w_2(t).
\end{equation}
A $2\pi$-periodic solution of this system satisfying (\ref{reversibility-reduction})
must also satisfy $u_1(\pi) = u_3(\pi) = 0$ and $w_2(\pi) = -w_4(\pi)$. Then,
the constraints of the travelling wave reduction (\ref{travelling-reductions})
yields the additional condition $w_4(\pi) = w_2(0)$.

To approximate a solution of the initial-value problem for the nonlinear
system (\ref{eq:four}) satisfying (\ref{reversibility-reduction}),
we only need four shooting parameters $(a_1,a_2,a_3,a_4)$ in the initial condition:
\begin{eqnarray*}
& u_1(0) = 0, \quad \dot{u}_1(0) = a_1, \quad w_2(0) = a_2, \quad \dot{w}_2(0) = a_3, \\
& u_3(0) = 0, \quad \dot{u}_3(0) = a_4, \quad w_4(0) = -a_2, \quad \dot{w}_4(0) = a_3.
\end{eqnarray*}
The solution of the initial-value problem corresponds to a $2\pi$-periodic
travelling wave solution only if the following four conditions are satisfied:
\begin{equation}
\label{shooting-four-conditions}
u_1(\pi) = 0, \quad w_2(\pi) + w_4(\pi) = 0, \quad w_2(0)-w_4(\pi) = 0, \quad u_3(\pi) = 0.
\end{equation}
These four conditions fully specify the shooting method for the four parameters
$(a_1,a_2,a_3,a_4)$. Additionally, the solution of the initial-value problem
must satisfy two more conditions:
\begin{equation}
\label{shooting-four-conditions-extra}
\dot{w}_2(\pi) - \dot{w}_4(\pi) = 0, \quad \dot{w}_2(0) - \dot{w}_4(\pi) = 0,
\end{equation}
but these additional conditions are redundant for the shooting method. We
have been checked conditions (\ref{shooting-four-conditions-extra})
apostoreori, after the shooting method has converged to a solution.

We are now able to run the shooting method based on conditions (\ref{shooting-four-conditions}).
The error of this numerical
method is composed from the error of an ODE solver and the error in finding zeros for
the functions above. We use the built-in MATLAB function \verb"ode113"
on the interval $[0,\pi]$ as an ODE solver
and then use the transformation (\ref{reversibility-reduction})
to extend the solutions to the interval $[-\pi,\pi]$ or $[0,2\pi]$.

\begin{figure}[t]
\begin{center}
\includegraphics[width=7cm,height=5cm]{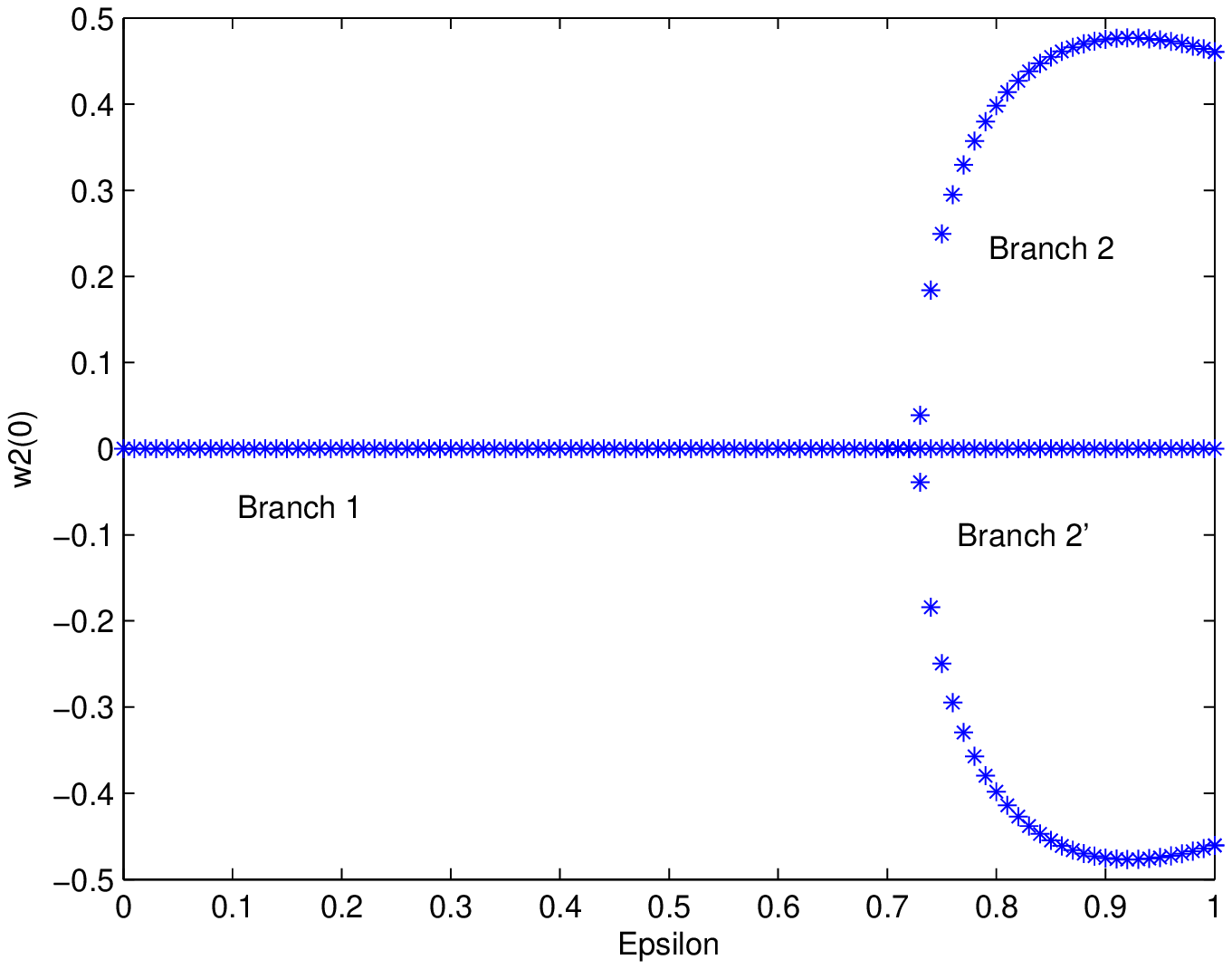} \hspace{0.1cm}
\includegraphics[width=7cm,height=5cm]{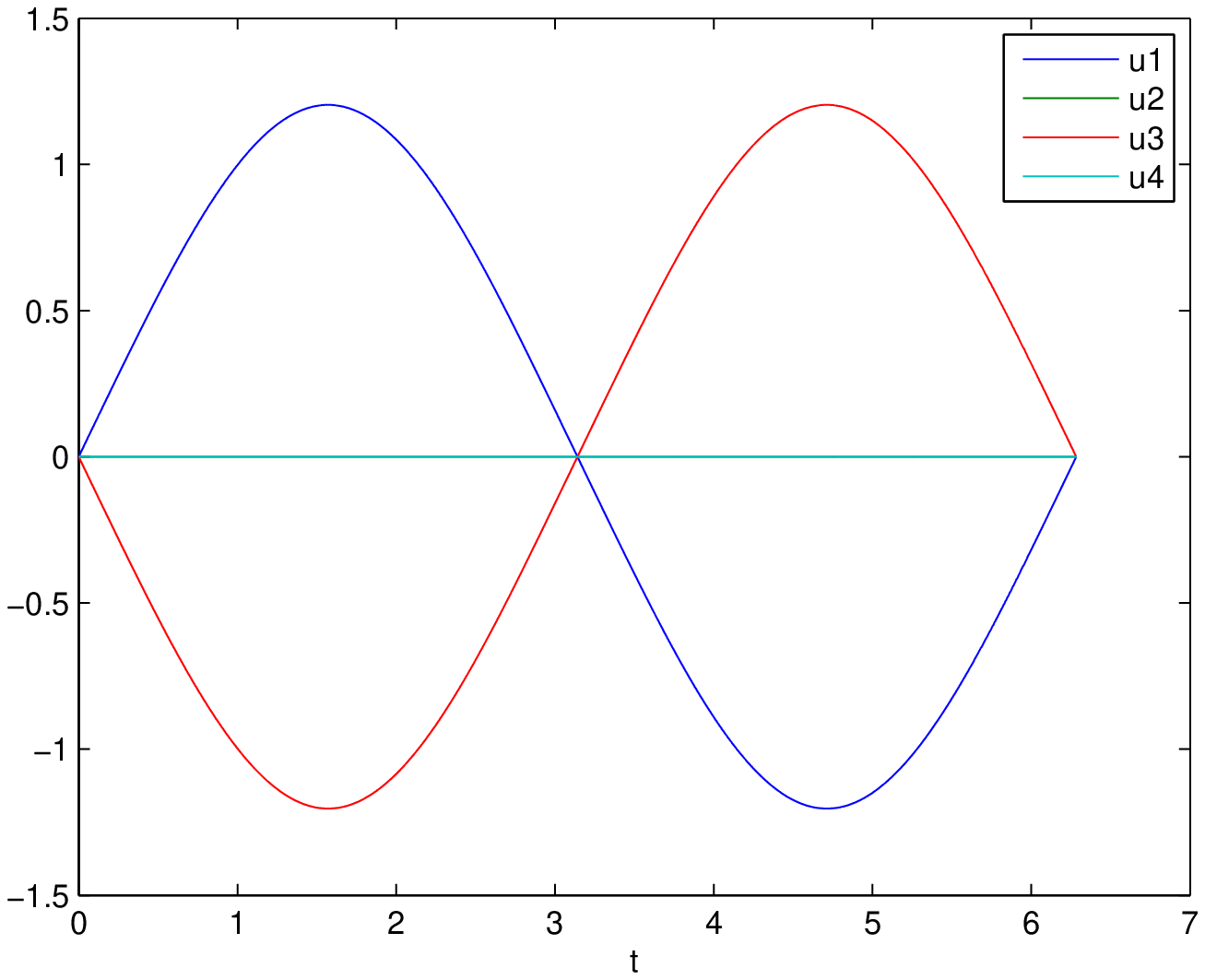} \\
\includegraphics[width=7cm,height=5cm]{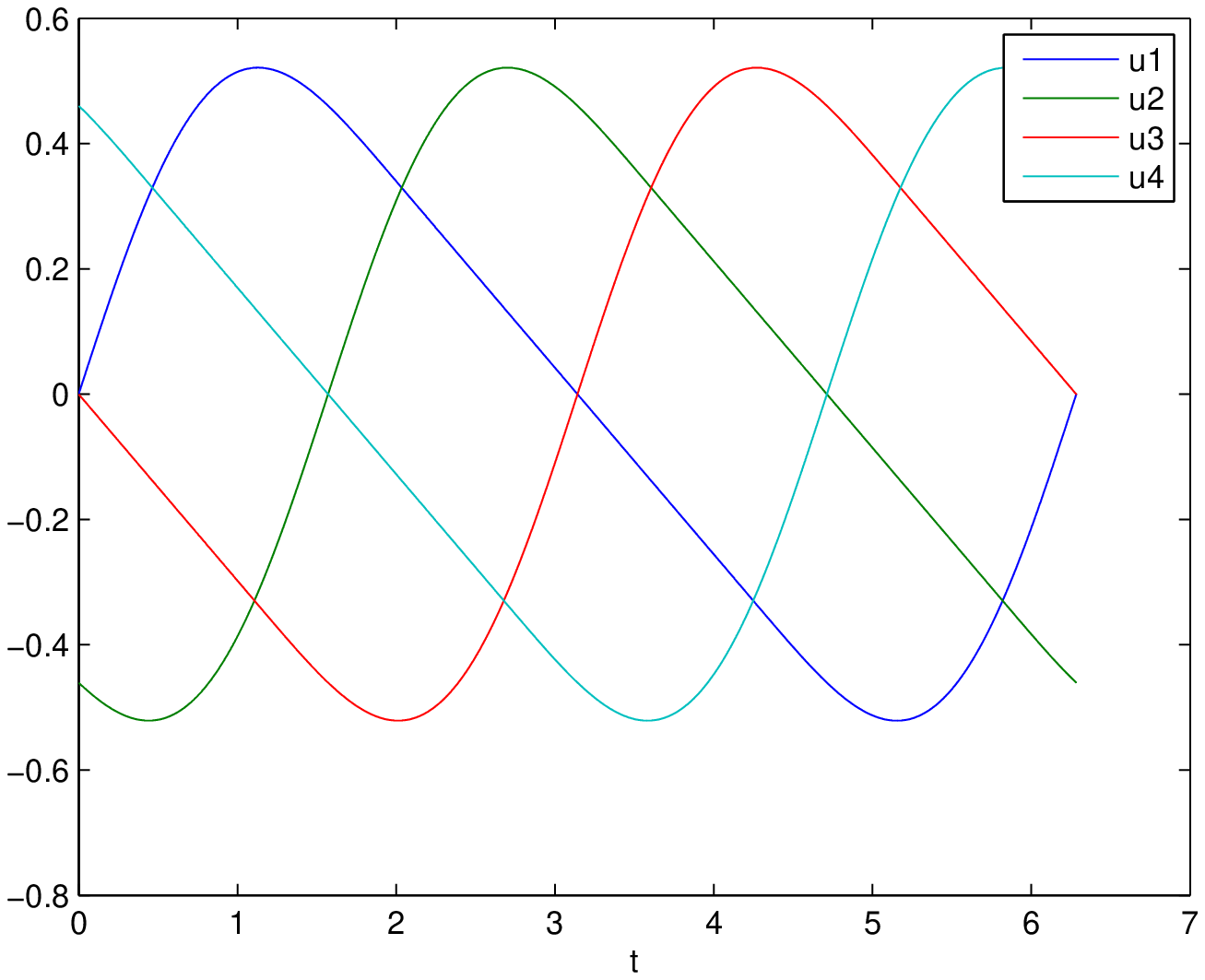} \hspace{0.1cm}
\includegraphics[width=7cm,height=5cm]{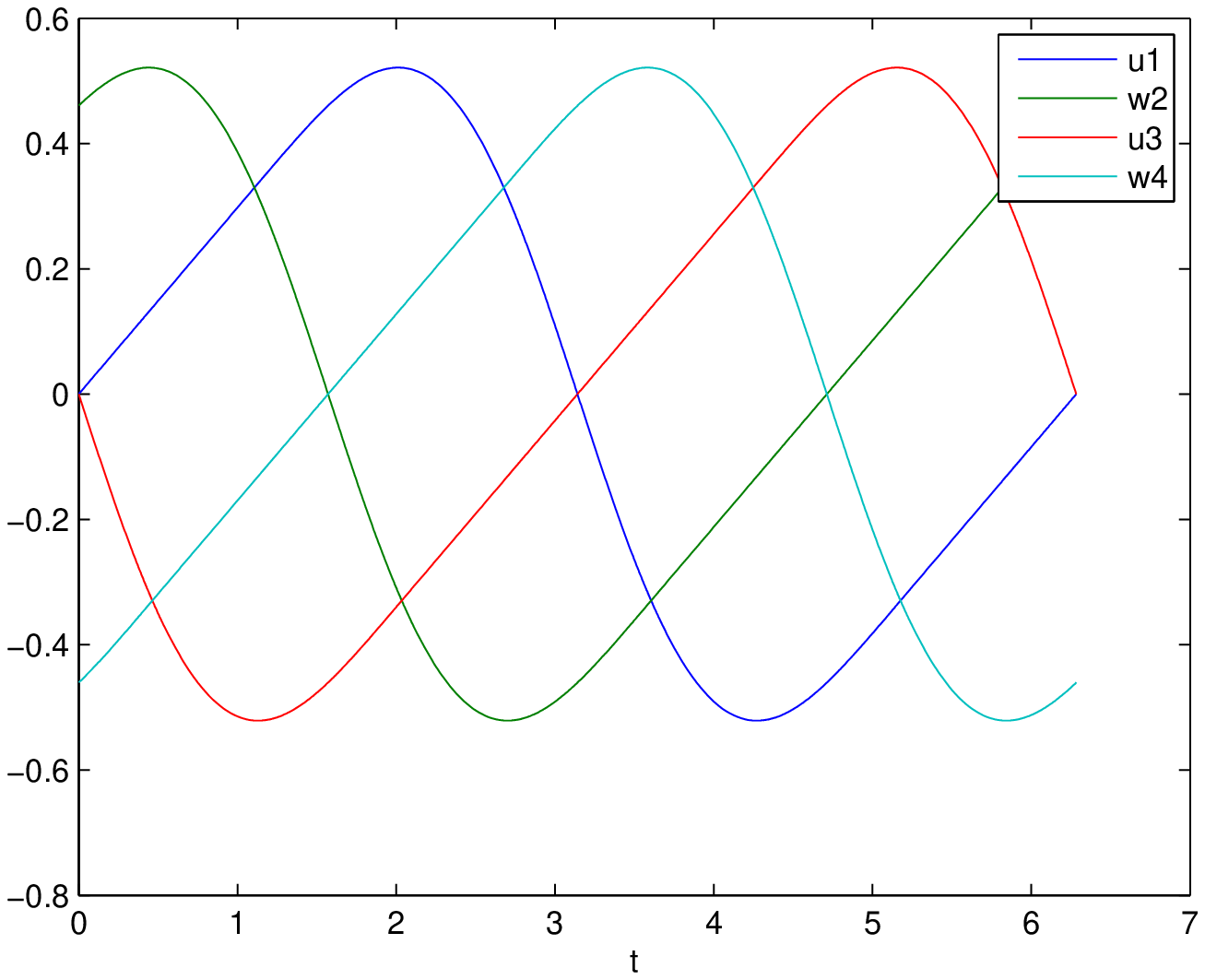}
\end{center}
\caption{Travelling wave solutions for $N = 2$:
the solution of the dimer chain continued from $\varepsilon=0$
to $\varepsilon=1$ (top right) and two solutions of the monomer chain
at $\varepsilon=1$ (bottom left and right).
The top left panel shows the value of $w_2(0)$ for all three solutions branches
versus $\varepsilon$. }
\label{fig:4soln}
\end{figure}

Figure \ref{fig:4soln} (top left) shows three solution branches obtained by the shooting method
by plotting $w_2(0)$ versus $\varepsilon$.
The first solution branch (labeled as branch 1) exists for all $\varepsilon \in [0,1]$ and is shown on
the top right panel for $\varepsilon = 1$. This branch
coincides with the exact solution (\ref{exact-1}). The error in the supremum norm between the numerical
and exact solutions $\| u_1 - \varphi \|_{L^{\infty}}$ can be found in Table II.

\begin{center}
\label{error}
  \begin{tabular}{ |l | c | r | }
    \hline
     \verb"AbsTol" of Shooting Method & \verb"AbsTol" of ODE solver & $L^\infty$ error \\ \hline
    $\mathcal{O}(10^{-12})$ & $\mathcal{O}(10^{-15})$ & $4.5 \times 10^{-14}$  \\
    $\,$ & $\mathcal{O}(10^{-10})$ &  $3.0 \times10^{-11}$  \\ \hline
    $\mathcal{O}(10^{-8})$ & $\mathcal{O}(10^{-15})$ & $4.5 \times 10^{-14}$  \\
    $\,$ & $\mathcal{O}(10^{-10})$ &  $3.0 \times 10^{-11}$  \\
    \hline
  \end{tabular}
\end{center}
{\bf Table II:} Error between numerical and exact solutions for branch $1$.

\vspace{0.5cm}

We can see from the top left panel of Figure \ref{fig:4soln} that a pitchfork bifurcation
occurs at $\varepsilon = \varepsilon_0 \approx 0.72$ and results in the appearance
of two symmetrically reflected branches (labeled as branches $2$ and $2'$).
These branches with $w_2(0) \neq 0$ extend to $\varepsilon = 1$ (bottom panels)
to recover two travelling wave solutions of the monomer chain (\ref{eq:Mono}).
The solution of branch $2$ satisfies the travelling wave reduction
$U_{n+1}(t) = U_n\left(t + \frac{\pi}{2}\right)$ and
was previously approximated numerically by James \cite{James2}.
The other solution of branch $2'$ satisfies the travelling wave reduction
$U_{n+1}(t) = U_n\left(t - \frac{\pi}{2}\right)$
and was previously obtained numerically by Starosvetsky and Vakakis \cite{Star1}.

For $N = 2$ ($q = \frac{\pi}{2}$), the solution of branch $2'$ given by
$\{ \tilde{u}_{2n-1},\tilde{w}_{2n}\}_{n \in \{1,2\}}$ is obtained from
the solution of branch $2$ given by $\{ u_{2n-1},w_{2n}\}_{n \in \{1,2\}}$,
by means of the symmetry
\begin{equation}
\label{four-particles-symmetry}
\tilde{u}_1(t) = -u_3(t), \quad \tilde{w}_2(t) = - w_2(t), \quad
\tilde{u}_3(t) = -u_1(t), \quad \tilde{w}_4(t) = - w_4(t),
\end{equation}
which holds for any $\varepsilon > 0$. (Of course, both solutions $2$ and $2'$ exist
only for $\varepsilon \in (\varepsilon_0,1]$ because of the pitchfork bifurcation at $\varepsilon
= \varepsilon_0 \approx 0.72$.) The solution of branch $1$ is the invariant reduction
$\tilde{u}_{2n-1} = u_{2n-1}$, $\tilde{w}_{2n} = w_{2n}$
with respect to the symmetry (\ref{four-particles-symmetry})
so that it satisfies $w_2(t) = w_4(t) = 0$ for all $t$.

For six particles ($N=3$ or $q = \frac{\pi}{3}$),
the nonlinear system (\ref{eq:ODE}) is written explicitly as
\begin{equation}
 \label{eq:six}
 \left\{ \begin{array}{l}
\ddot{u}_1(t) = (\varepsilon w_6(t)-u_1(t))_+^{\alpha}-(u_1(t)-\varepsilon w_2(t))_+^{\alpha},\\
\ddot{w}_2(t) = \varepsilon[(u_1(t)-\varepsilon w_2(t))_+^{\alpha}-(\varepsilon w_2(t)-u_3(t))_+^{\alpha},\\
\ddot{u}_3(t) = (\varepsilon w_2(t)-u_3(t))_+^{\alpha}-(u_3(t)-\varepsilon w_4(t))_+^{\alpha},\\
\ddot{w}_4(t) = \varepsilon[(u_3(t)-\varepsilon w_4(t))_+^{\alpha}-(\varepsilon w_4(t)-u_5(t))_+^{\alpha},\\
\ddot{u}_5(t) = (\varepsilon w_4(t)-u_5(t))_+^{\alpha}-(u_5(t)-\varepsilon w_6(t))_+^{\alpha},\\
\ddot{w}_6(t) = \varepsilon[(u_5(t)-\varepsilon w_6(t))_+^{\alpha}-(\varepsilon w_6(t)-u_1(t))_+^{\alpha}.
\end{array} \right.
\end{equation}
We are looking for $2\pi$-periodic functions satisfying the travelling wave reduction:
\begin{equation}
\label{travelling-reductions-six}
u_5(t) = u_3\left(t + \frac{2\pi}{3}\right) = u_1\left(t + \frac{4\pi}{3}\right), \quad
w_6(t) = w_4\left(t + \frac{2\pi}{3}\right) = w_2\left(t + \frac{4\pi}{3}\right).
\end{equation}
We note that the system (\ref{eq:six}) is invariant with respect to the following transformation:
\begin{equation}
\label{reversibility-reduction-six}
u_1(-t) = -u_1(t), \quad w_2(-t) = -w_6(t), \quad u_3(-t) = -u_5(t), \quad w_4(-t) = -w_4(t).
\end{equation}
A $2\pi$-periodic solution of this system satisfying (\ref{reversibility-reduction-six})
must also satisfy $u_1(\pi) = w_4(\pi) = 0$, $w_2(\pi) = -w_6(\pi)$, and $u_3(\pi) = -u_5(\pi)$. Then,
the constraints of the travelling wave reduction (\ref{travelling-reductions-six}) yield
the conditions $u_3(\pi) = u_1\left(\frac{\pi}{3}\right)$ and $w_4(\pi) = w_2\left(\frac{\pi}{3}\right)$.

To approximate a solution of the initial-value problem for the nonlinear
system (\ref{eq:six}) satisfying (\ref{reversibility-reduction-six}),
we only need six shooting parameters $(a_1,a_2,a_3,a_4,a_5,a_6)$ in the initial condition:
\begin{eqnarray*}
& u_1(0) = 0, \quad \dot{u}_1(0) = a_1, \quad w_2(0) = a_2, \quad \dot{w}_2(0) = a_3, \\
& u_3(0) = a_4, \quad \dot{u}_3(0) = a_5, \quad w_4(0) = 0, \quad \dot{w}_4(0) = a_6, \\
& u_5(0) = -a_4, \quad \dot{u}_5(0) = a_5, \quad w_6(0) = -a_2, \quad \dot{w}_6(0) = a_3.
\end{eqnarray*}
This solution corresponds to a $2\pi$-periodic travelling wave solution only
if it satisfies the following six conditions:
\begin{eqnarray*}
& u_1(\pi) = 0, \quad w_2(\pi) + w_6(\pi) = 0, \quad u_3(\pi) + u_5(\pi) = 0, \\
& \quad u_1\left(\frac{\pi}{3}\right)-u_3(\pi) = 0 \quad w_2\left(\frac{\pi}{3}\right)-w_4(\pi) = 0,
\quad w_4(\pi) = 0.
\end{eqnarray*}
The six conditions determines the shooting method for the six parameters $(a_1,a_2,a_3,a_4,a_5,a_6)$.
Additional conditions,
$$
\dot{w}_2(\pi) - \dot{w}_6(\pi) = 0, \quad \dot{u}_3(\pi) - \dot{u}_5(\pi) = 0, \quad
\dot{u}_1\left(\frac{\pi}{3}\right) - \dot{u}_3(\pi) = 0, \quad
\dot{w}_2\left(\frac{\pi}{3}\right) - \dot{w}_4(\pi) = 0,
$$
are to be checked aposteriori, after the shooting method has converged to a solution.

\begin{figure}[t]
\begin{center}
\includegraphics[width=7cm,height=5cm]{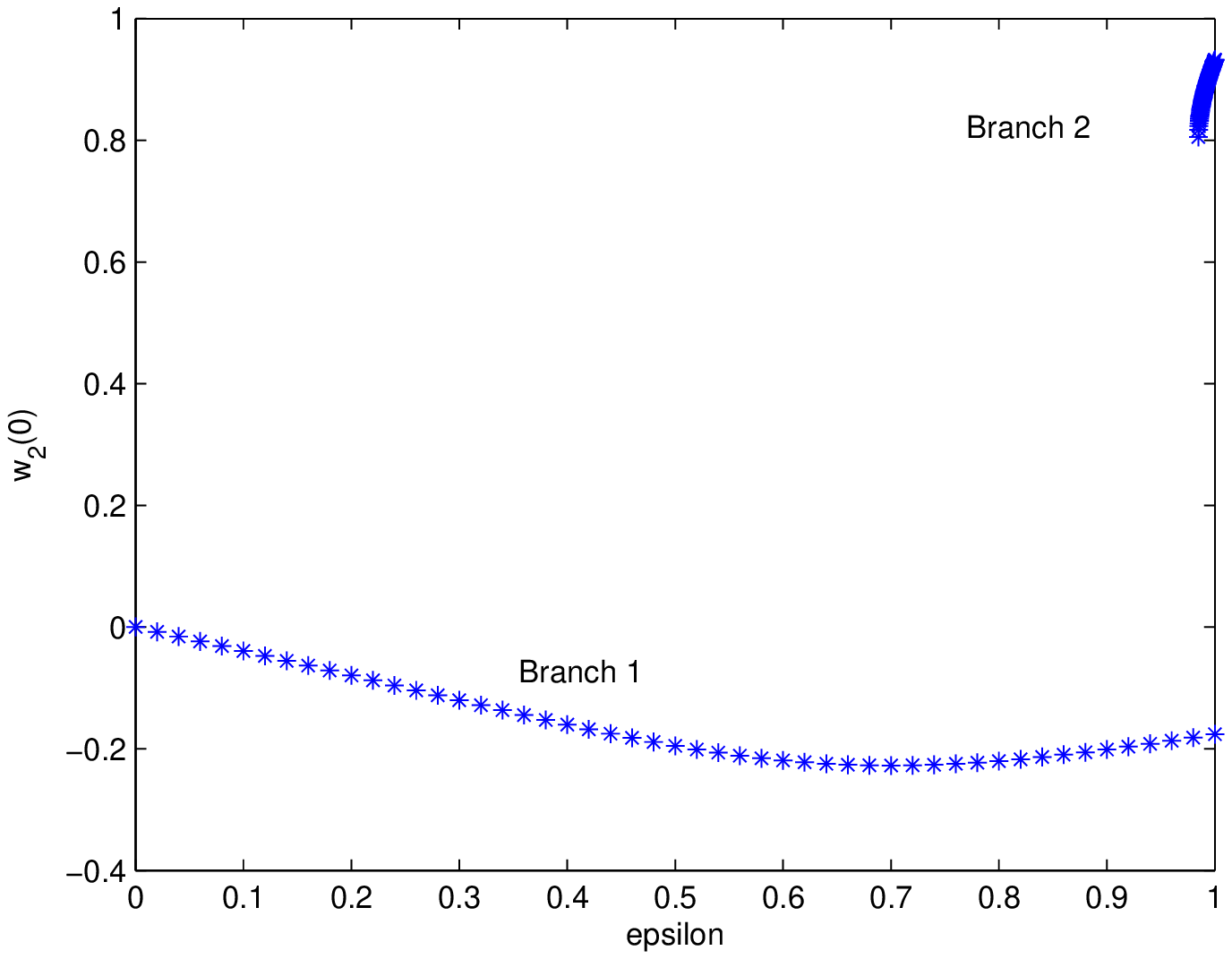} \hspace{0.1cm}
\includegraphics[width=7cm,height=5cm]{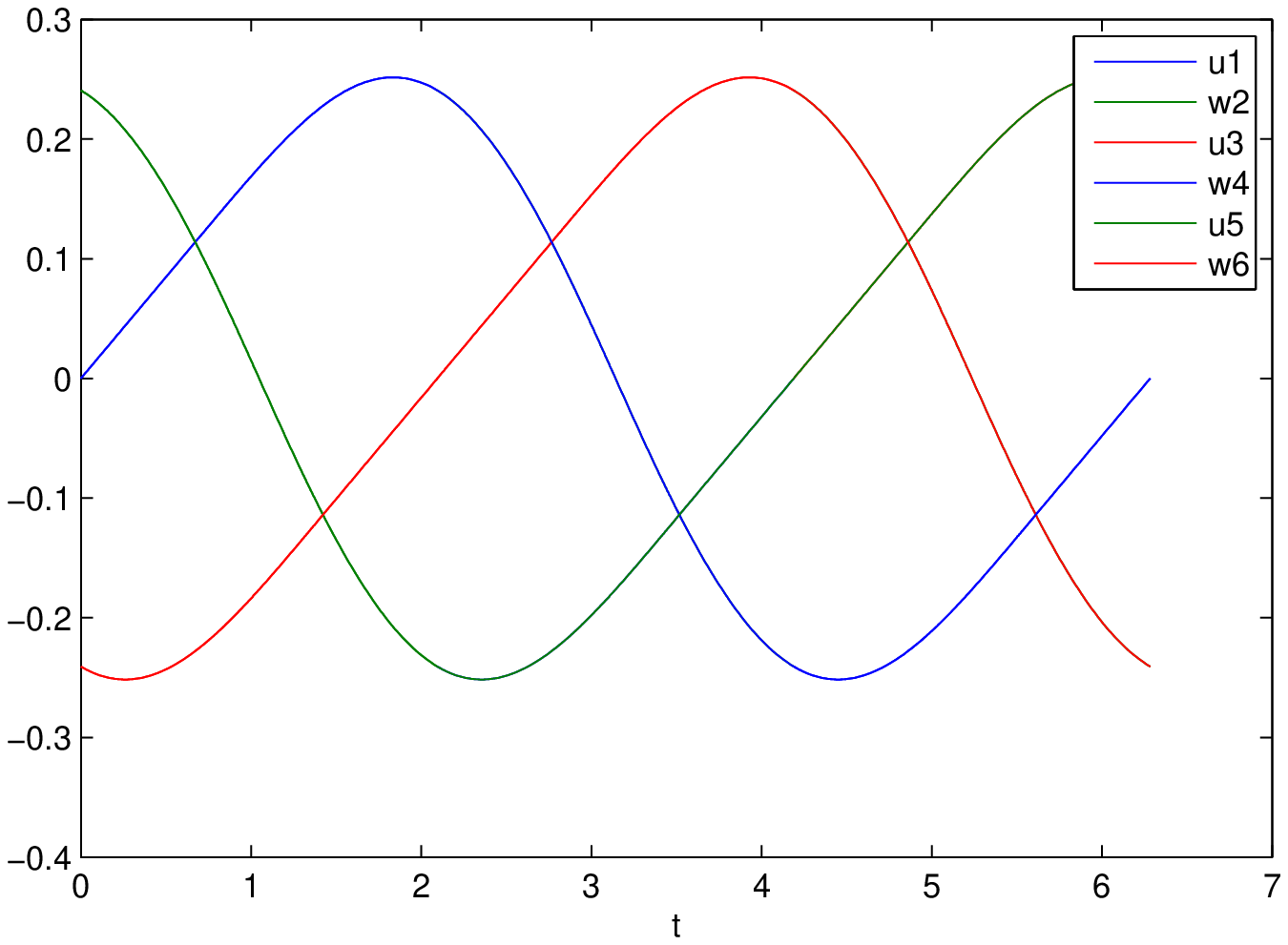} \\
\includegraphics[width=7cm,height=5cm]{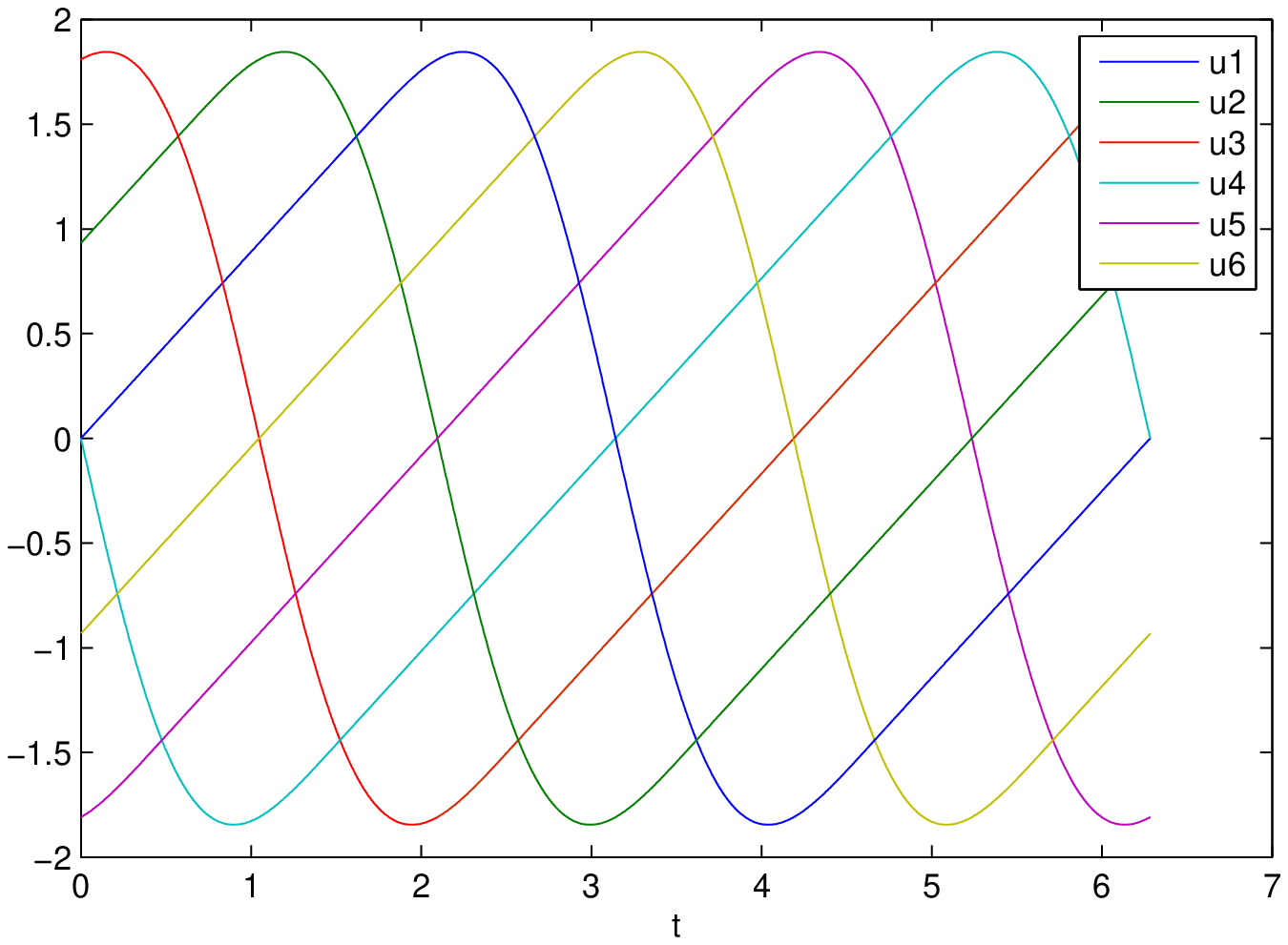} \hspace{0.1cm}
\includegraphics[width=7cm,height=5cm]{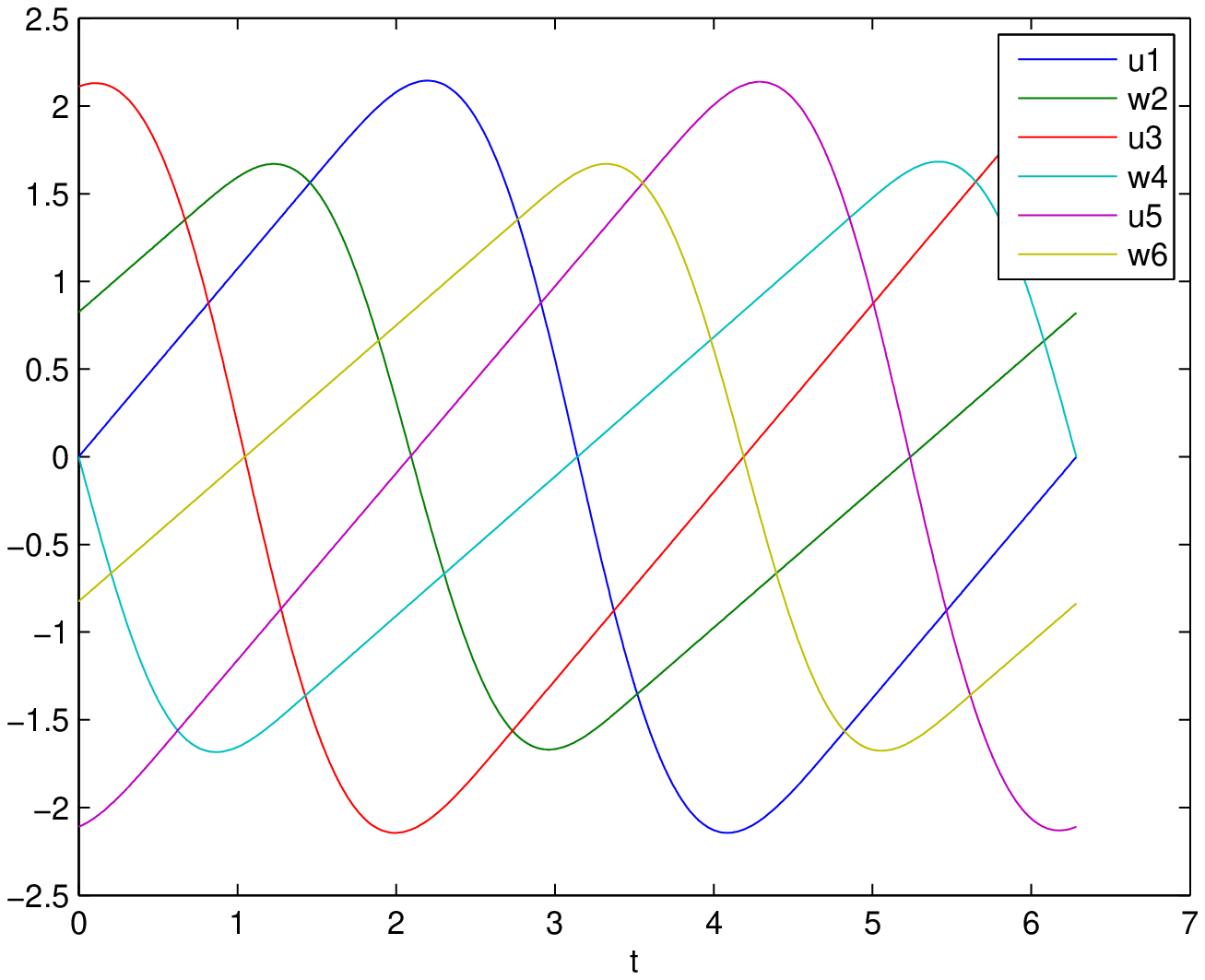}
\end{center}
\caption{Travelling wave solutions for $N = 3$:
the solution of branch 1 is continued from $\varepsilon=0$
to $\varepsilon=1$ (top right) and the solution of branch 2 is continued
from $\varepsilon=1$ (bottom left) to $\varepsilon = 0.985$ (bottom right).
The top left panel shows the value of $w_2(0)$ for solution branches 1 and 2
versus $\varepsilon$. }
\label{fig:6soln}
\end{figure}

Figure \ref{fig:6soln} (top left) shows two solution branches obtained
by the shooting method. Again, $w_2(0)$ is plotted versus $\varepsilon$.
Branch 1 is continued from $\varepsilon = 0$ to $\varepsilon = 1$ (top right)
without any pitchfork bifurcation in $\varepsilon \in (0,1)$.
Branch 2 is continued from $\varepsilon = 1$ (bottom left)
starting with a numerical solution of the monomer chain (\ref{eq:Mono})
satisfying the reduction $U_{n+1}(t) = U_n\left(t + \frac{\pi}{3}\right)$
to $\varepsilon = 0.985$ (bottom right), where the branch disappears
from the radars of our shooting method. We have not been able so far to detect
numerically any other branch of travelling wave solutions near branch 2 for
$\varepsilon = 0.985$, hence the nature of this bifurcation will remain
opened for further studies.

\begin{figure}[t]
\begin{center}
\includegraphics[width=7cm,height=5cm]{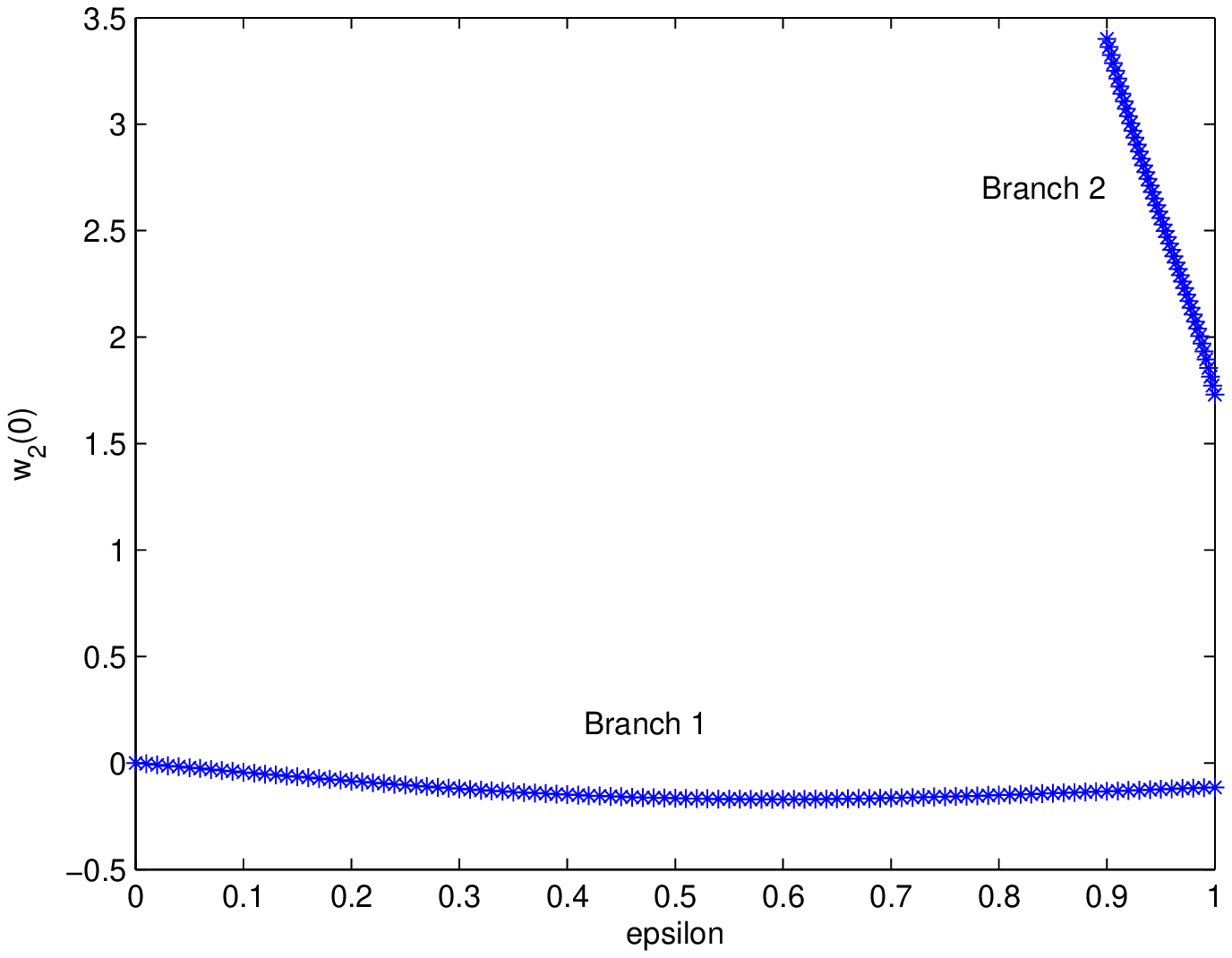} \hspace{0.1cm}
\includegraphics[width=7cm,height=5cm]{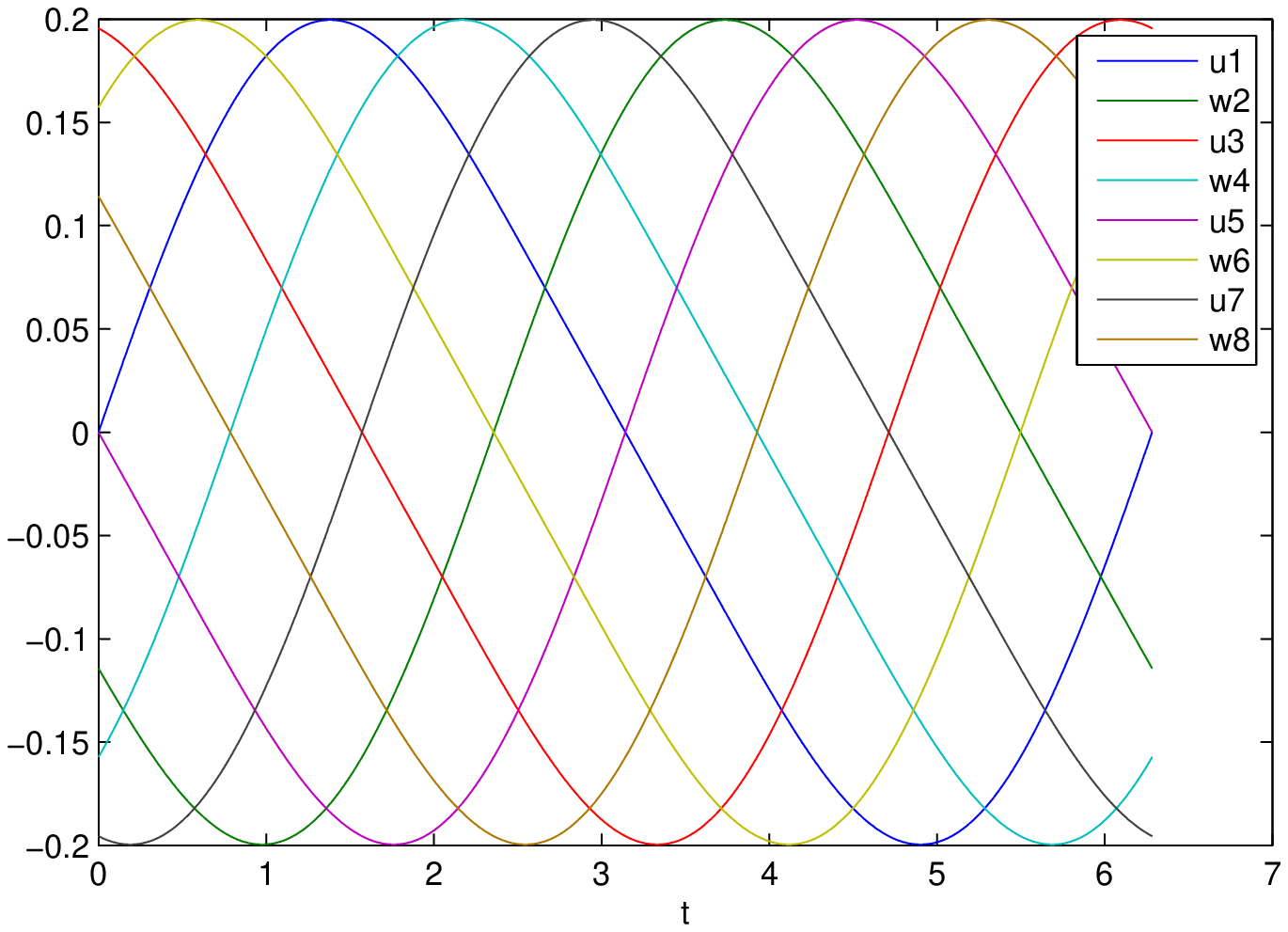} \\
\includegraphics[width=7cm,height=5cm]{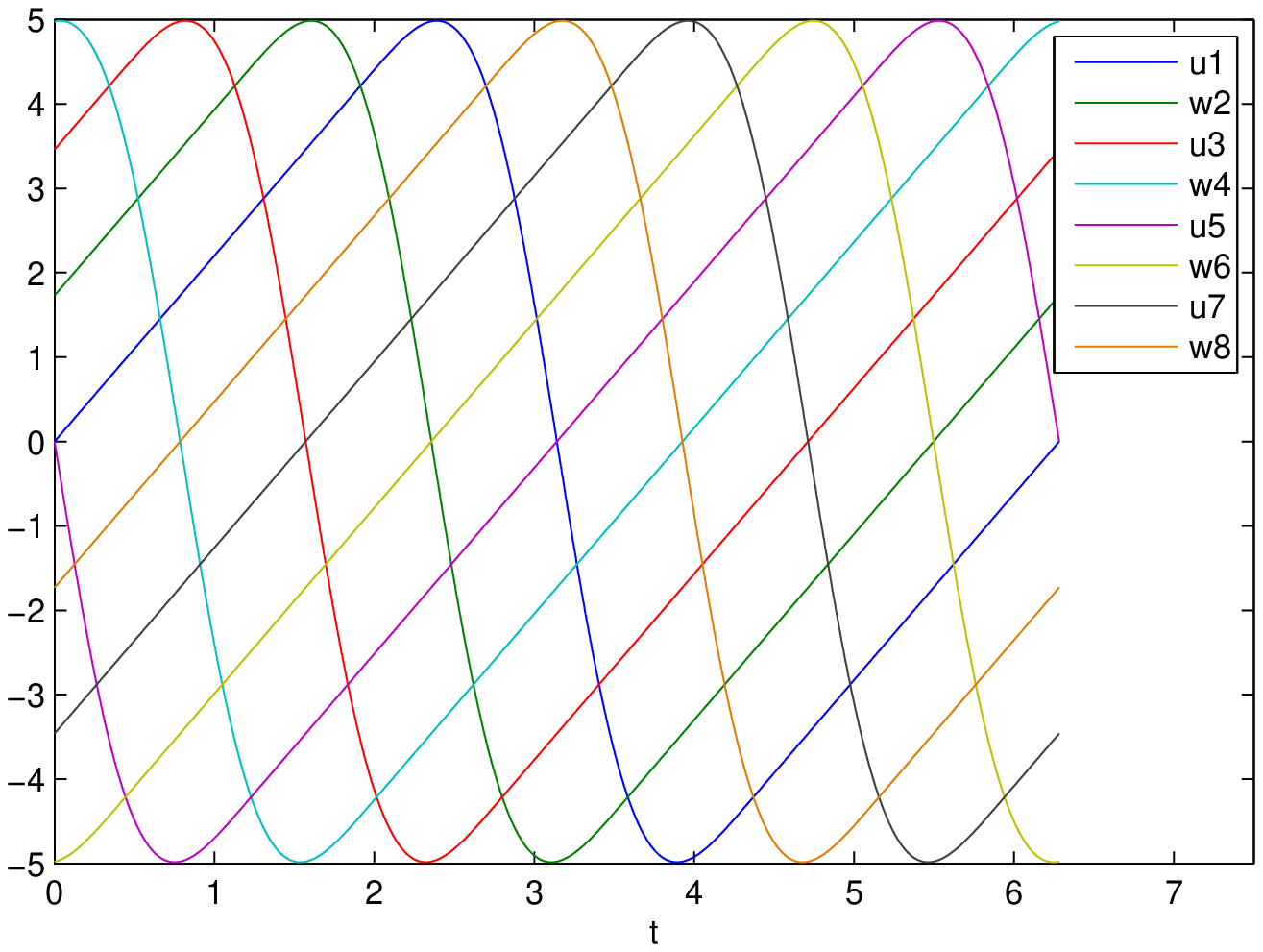} \hspace{0.1cm}
\includegraphics[width=7cm,height=5cm]{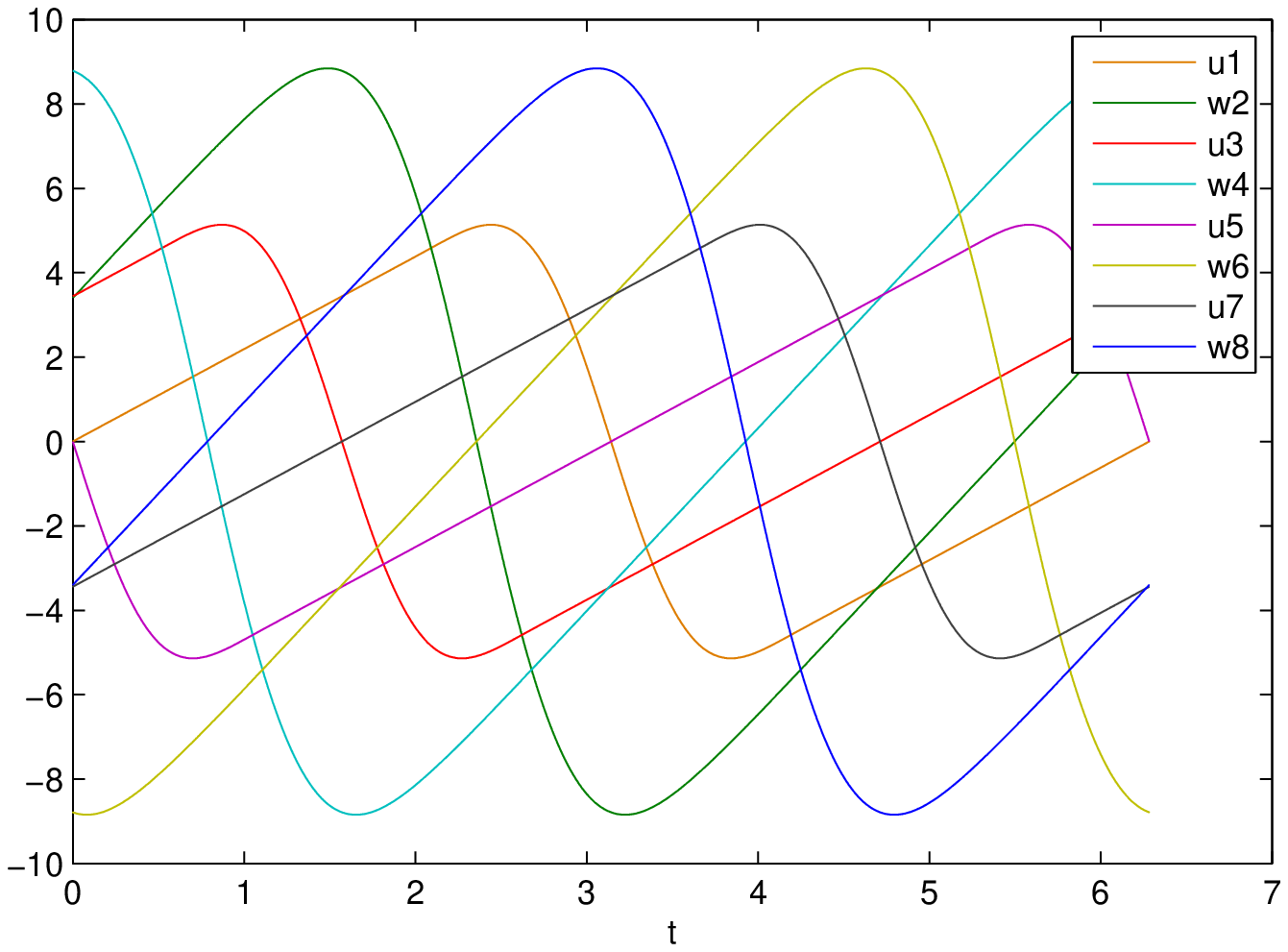}
\end{center}
\caption{Travelling wave solutions for $N=4$:
the solution of branch 1 continued from $\varepsilon=0$
to $\varepsilon = 1$ (top right) and the solution of
branch 1 continued from $\varepsilon=1$ (bottom left) to $\varepsilon = 0.9$ (bottom right).
The top left panel shows the value of $w_2(0)$ for solution branches 1 and 2
versus $\varepsilon$.}
\label{fig:8soln}
\end{figure}

We use the same technique for $N=4$ and show similar results on Figure \ref{fig:8soln}.
Branch 1 is uniquely continued from $\varepsilon = 0$ to $\varepsilon = 1$ (top right),
whereas branch 2 is continued from $\varepsilon = 1$ (bottom left)
starting with a numerical solution of the monomer chain (\ref{eq:Mono})
satisfying the reduction $U_{n+1}(t) = U_n\left(t + \frac{\pi}{4}\right)$
to $\varepsilon = 0.9$ (bottom right), where the branch terminates.

\subsection{Stability of travelling periodic wave solutions}
\label{sub:Stability}

To determine stability of the different branches of travelling periodic
wave solutions of the granular dimer chains (\ref{eq:ADD}), we compute Floquet multipliers
of the monodromy matrix for the linearized system (\ref{eq:ADD-linear}).
To do this, we use the travelling wave solution obtained with the
shooting method and the MATLAB function \verb"ode113" to
compute the fundamental matrix solution of the linearized system (\ref{eq:ADD-linear})
on the interval $[0,2\pi]$.

\begin{figure}[t]
\begin{center}
\includegraphics[width=7cm,height=5cm]{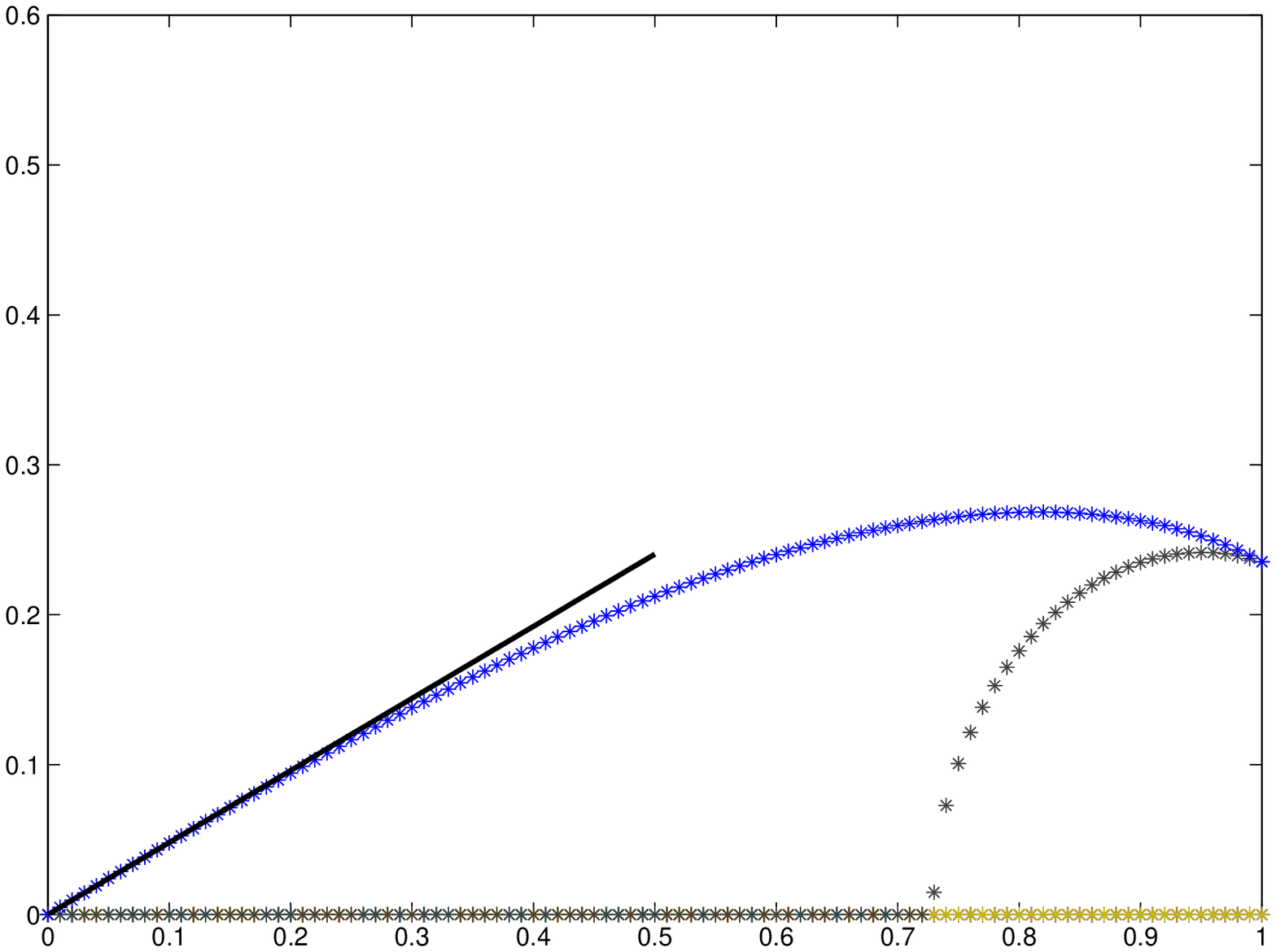} \hspace{0.1cm}
\includegraphics[width=7cm,height=5cm]{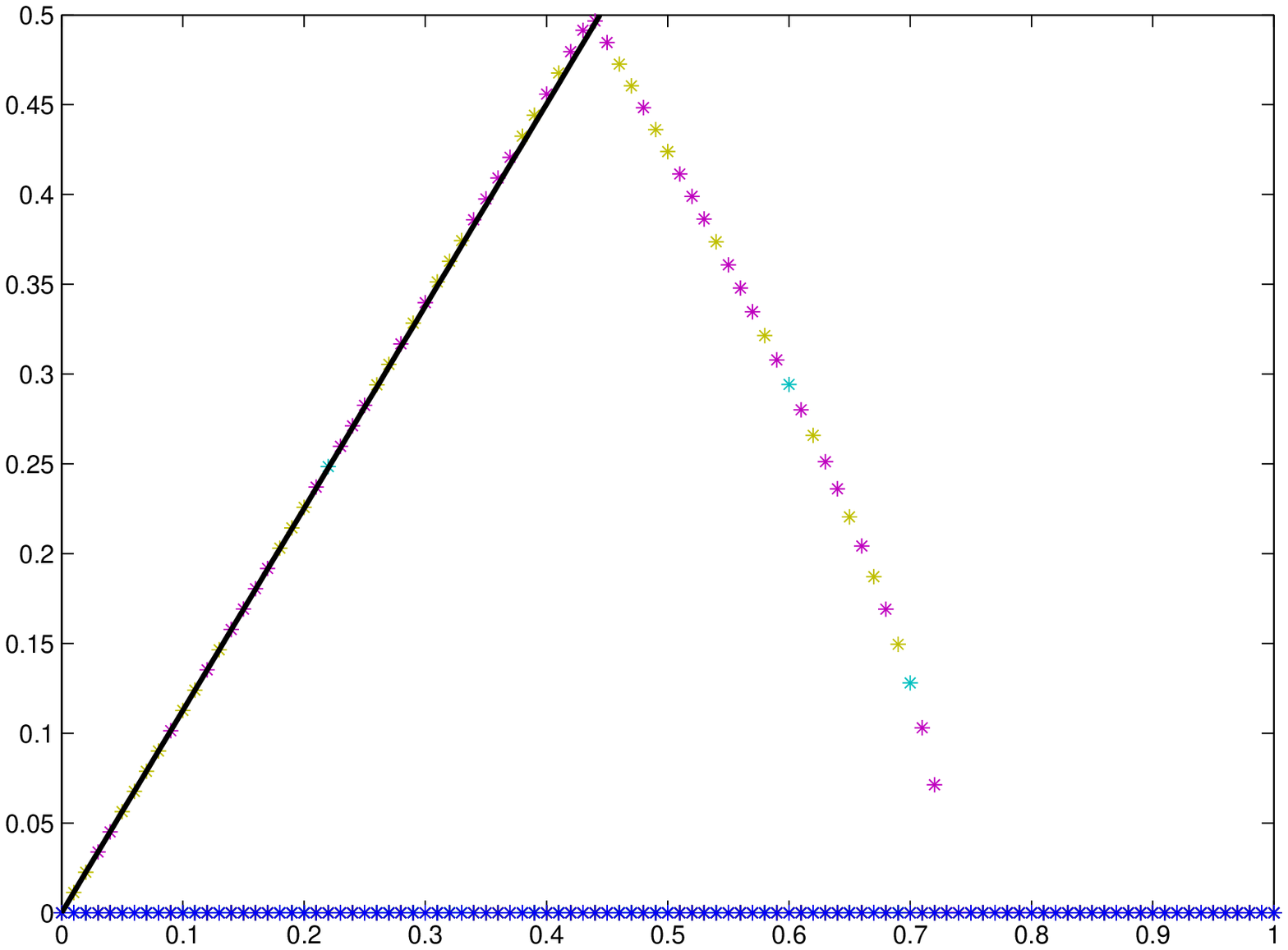} \\
\includegraphics[width=7cm,height=5cm]{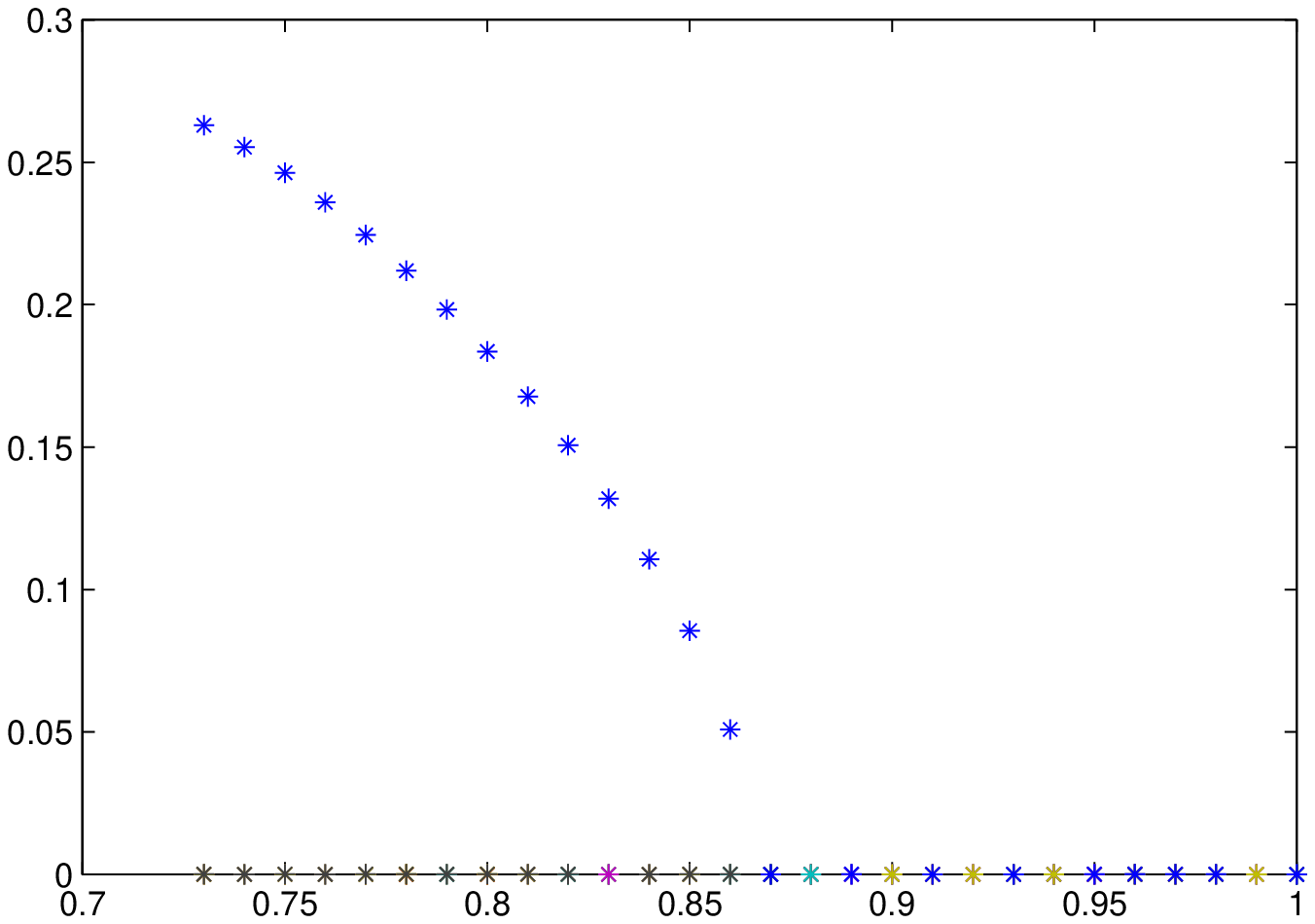} \hspace{0.1cm}
\includegraphics[width=7cm,height=5cm]{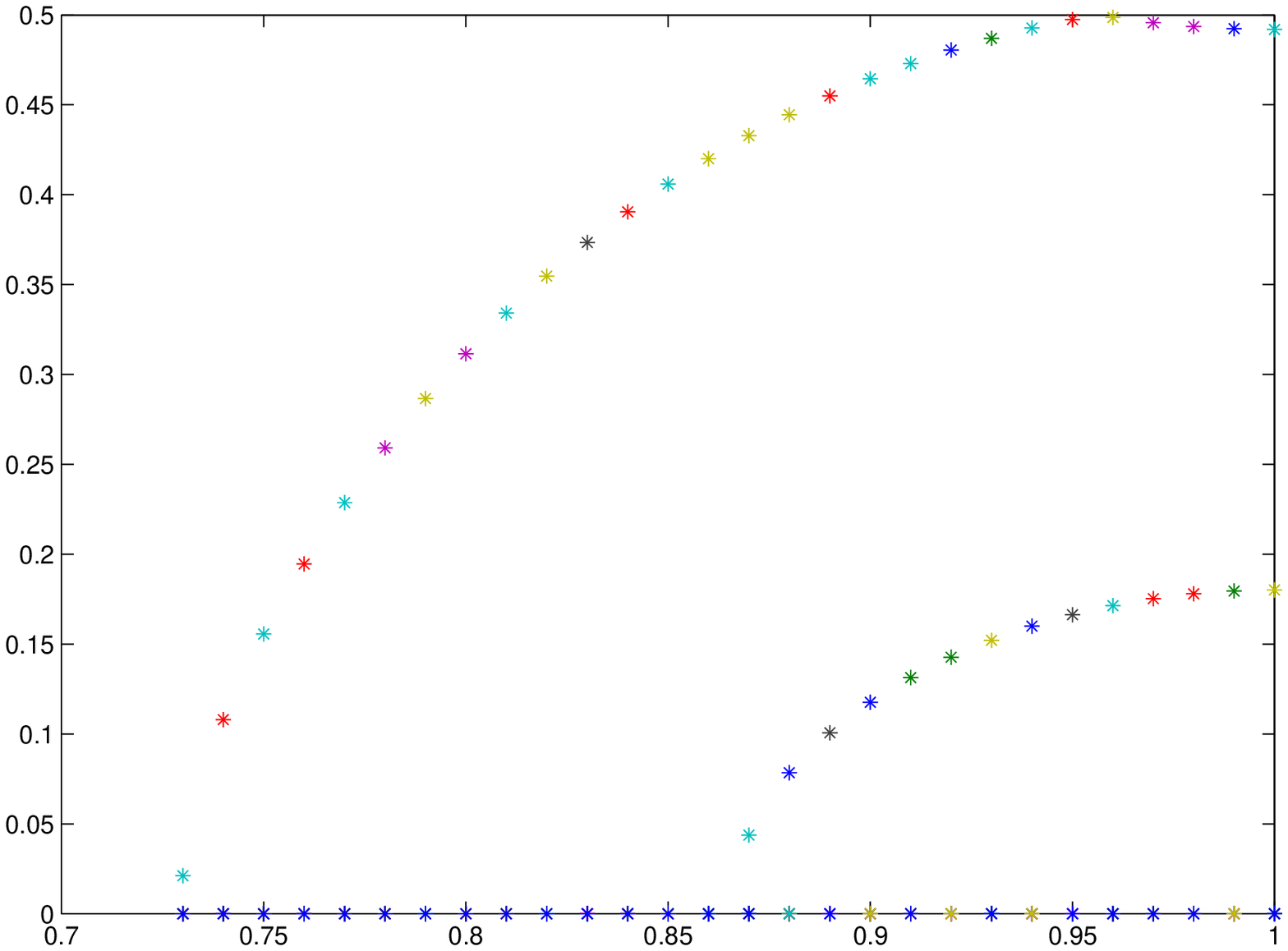}
\end{center}
\caption{Real (left) and imaginary (right) parts of the
characteristic exponents $\lambda$ versus $\varepsilon$ for $N = 2$ for branch 1
(top) and branch 2 (bottom).}
\label{fig:N4stability}
\end{figure}

By Theorem \ref{theorem-2}, the travelling waves of branch $1$ for $N = 2$
($q=\frac{\pi}{2}$) are unstable for small values of $\varepsilon$.
Figure \ref{fig:N4stability} (top) shows real and imaginary parts of
the characteristic exponents associated with branch $1$
for all values of $\varepsilon$ in $[0,1]$. Only positive values of ${\rm Re}(\lambda)$ and
${\rm Im}(\lambda)$ are shown, moreover, ${\rm Im}(\lambda) \in \left[0,\frac{1}{2}\right]$
because of $1$-periodicity of the characteristic exponents along the imaginary axis.

Thanks to the periodic boundary conditions, the system of linearized equations (\ref{eq:ADD-linear-1})
for $N = 2$ is closed at $4$ second-order linearized equations, which have
$8$ characteristic exponents as follows. The exponent $\lambda = 0$
has multiplicity $4$ for small positive $\varepsilon$, and two pairs of nonzero exponents
(one is real, the other one is purely imaginary) bifurcate according to the roots of
the characteristic equation (\ref{characteristic-eq}) for $\theta = \frac{\pi}{2}$.
These asymptotic approximations are shown on the top panels of Figure \ref{fig:N4stability} by solid curves,
in excellent agreement with the numerical data. We can see that
the unstable real $\lambda$ persist for all values of $\varepsilon$ in $[0,1]$.
The pitchfork bifurcation at $\varepsilon = \varepsilon_0 \approx 0.72$ in Figure \ref{fig:4soln} (top left)
corresponds to the coalescence of the pair of purely imaginary characteristic exponents on Figure
\ref{fig:N4stability} (top right) and appearance
of a new pair of real characteristic exponents
for $\varepsilon > \varepsilon_0$ on Figure \ref{fig:N4stability} (top left).
Therefore, the branch continued from $\varepsilon = 0$ is unstable for all $\varepsilon \in [0,1]$.

Bottom panels on Figure \ref{fig:N4stability} shows real and imaginary parts of the characteristic exponents
associated with branch $2$ (same for $2'$ by symmetry) for all values of $\varepsilon$ in $[\varepsilon_0,1]$.
We can see that these travelling waves are spectrally stable near $\varepsilon = 1$ in agreement with
the numerical results of James \cite{James2}. When $\varepsilon$ is decreased, these travelling waves lose
spectral stability near $\varepsilon = \varepsilon_1 \approx 0.86$ because of coallescence of
the pair of purely imaginary characteristic exponents and appearance
of a new pair of real characteristic exponents for $\varepsilon < \varepsilon_1$.
The two solution branches disappear as a result of the pitchfork bifurcation at $\varepsilon = \varepsilon_0
\approx 0.72$, which is again induced by the coalescence of the second pair of purely imaginary
characteristic exponents.

\begin{figure}[t]
\begin{center}
\includegraphics[width=7cm,height=5cm]{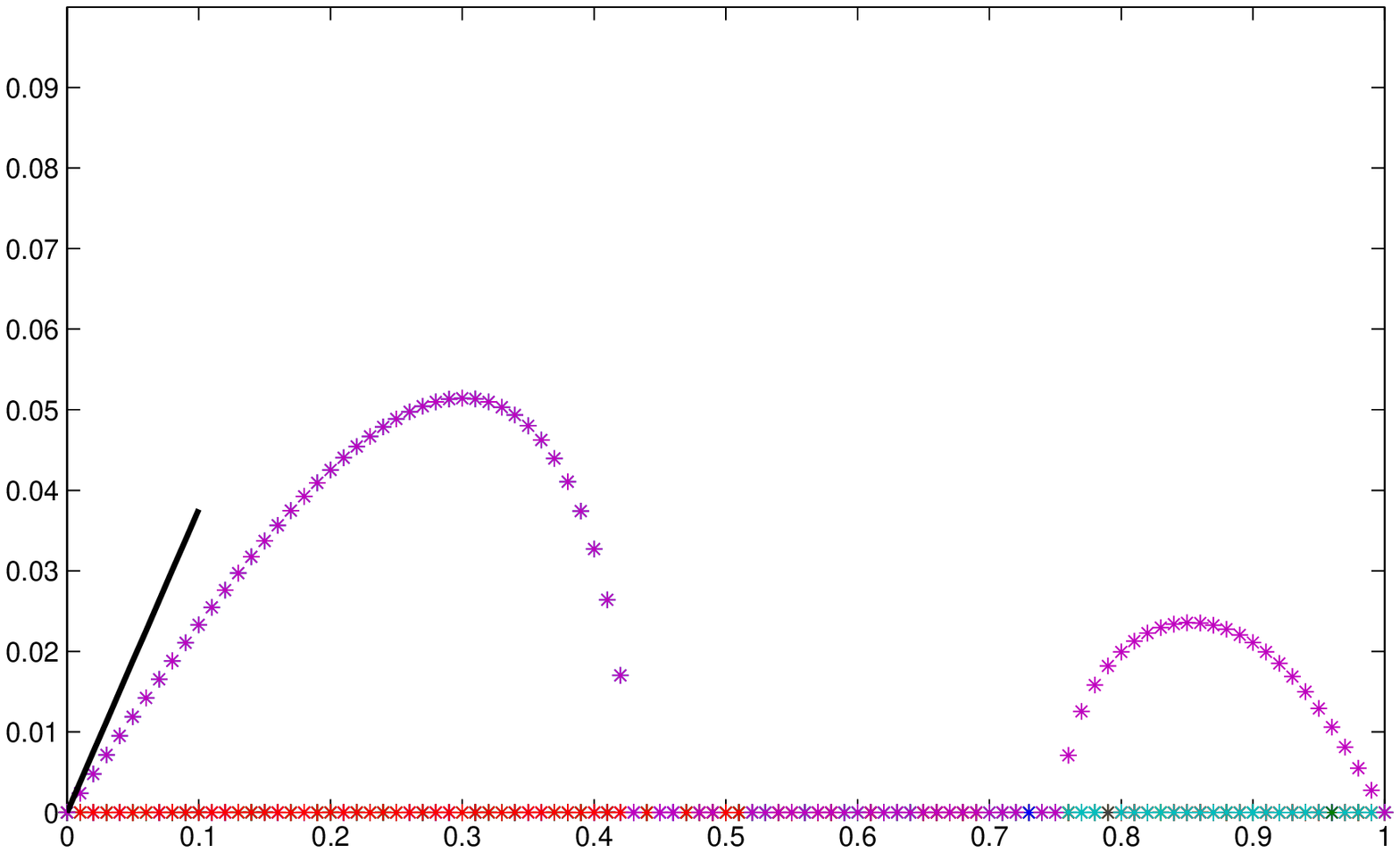} \hspace{0.1cm}
\includegraphics[width=7cm,height=5cm]{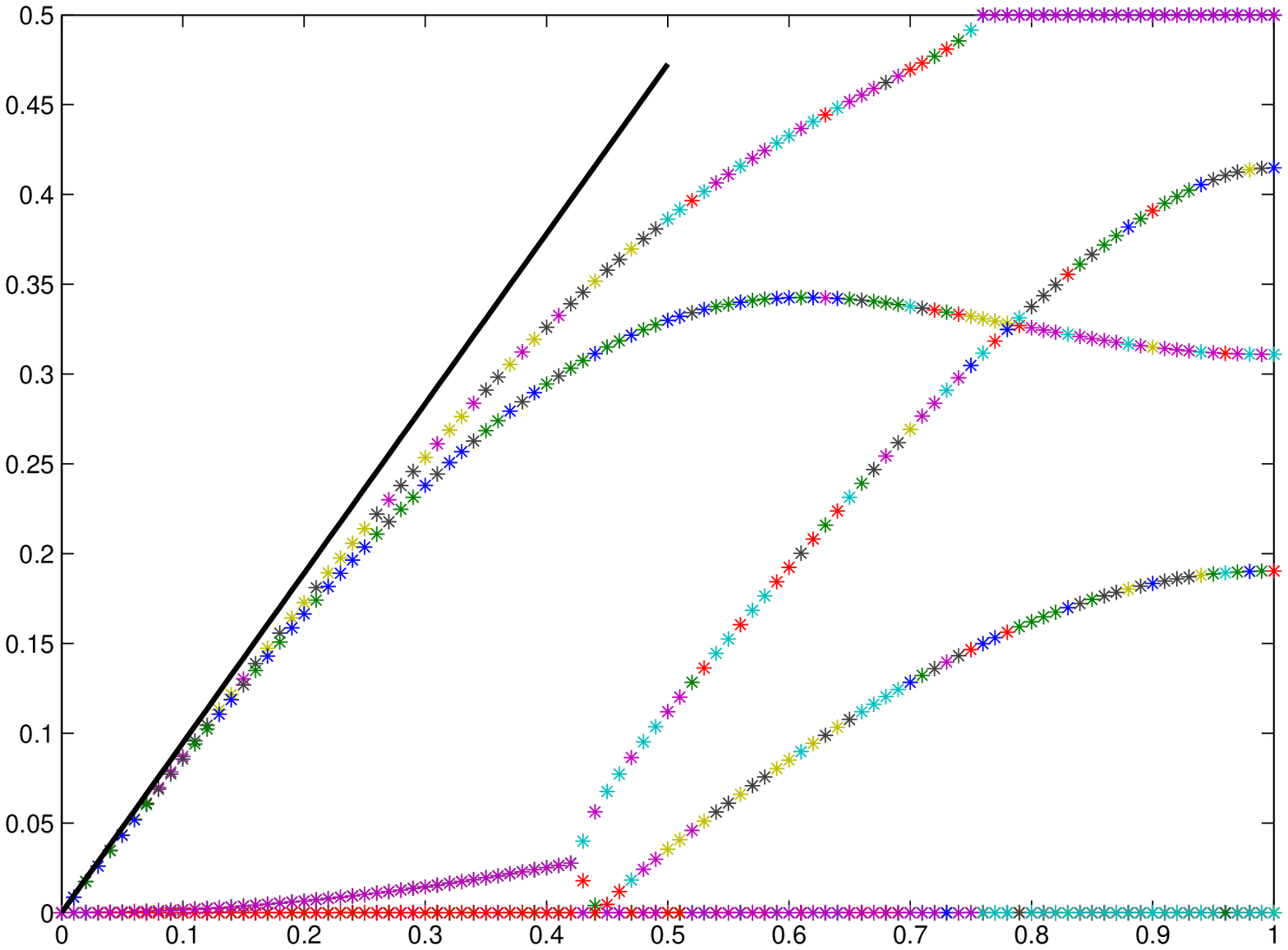} \\
\includegraphics[width=7cm,height=5cm]{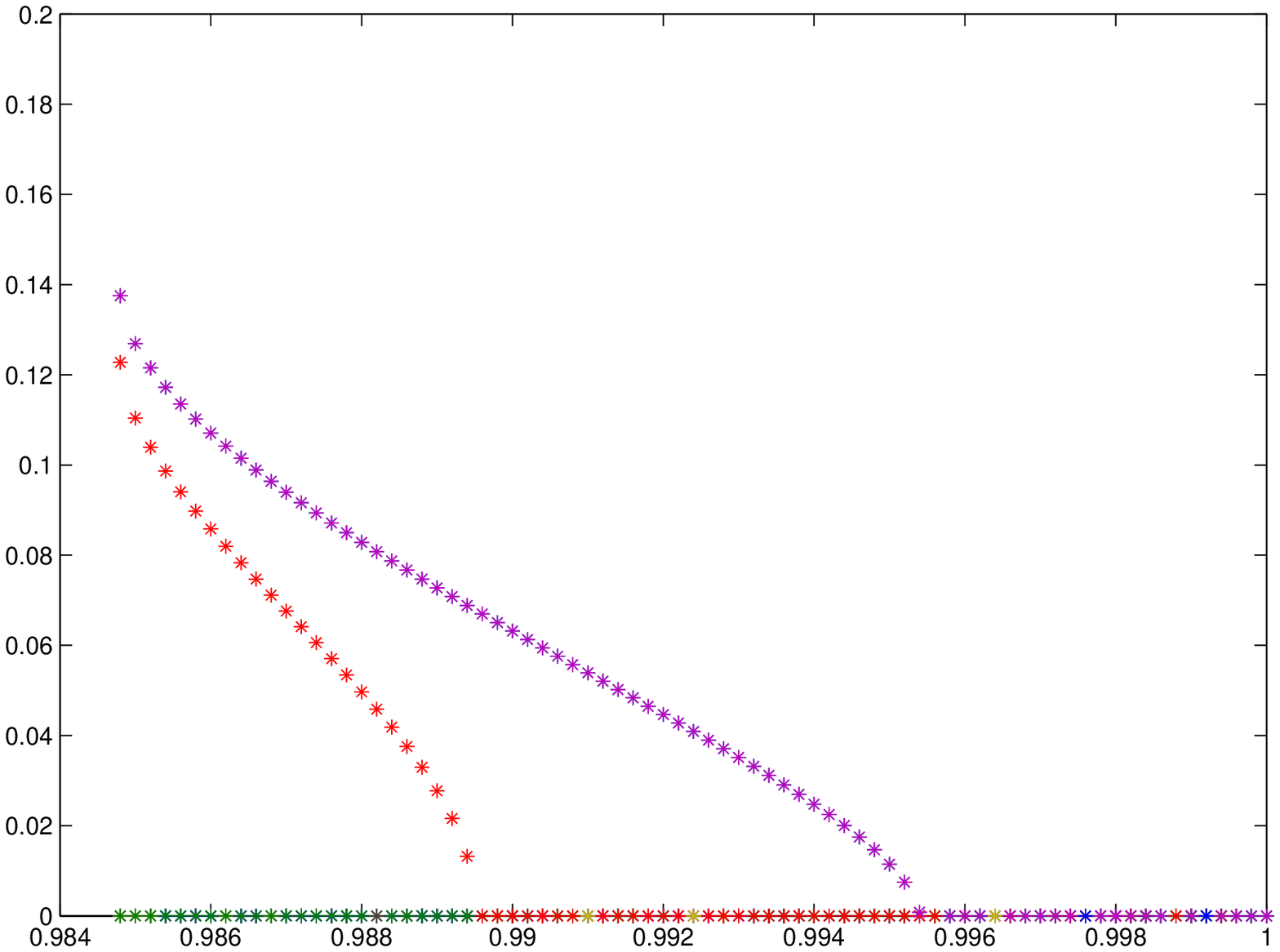} \hspace{0.1cm}
\includegraphics[width=7cm,height=5cm]{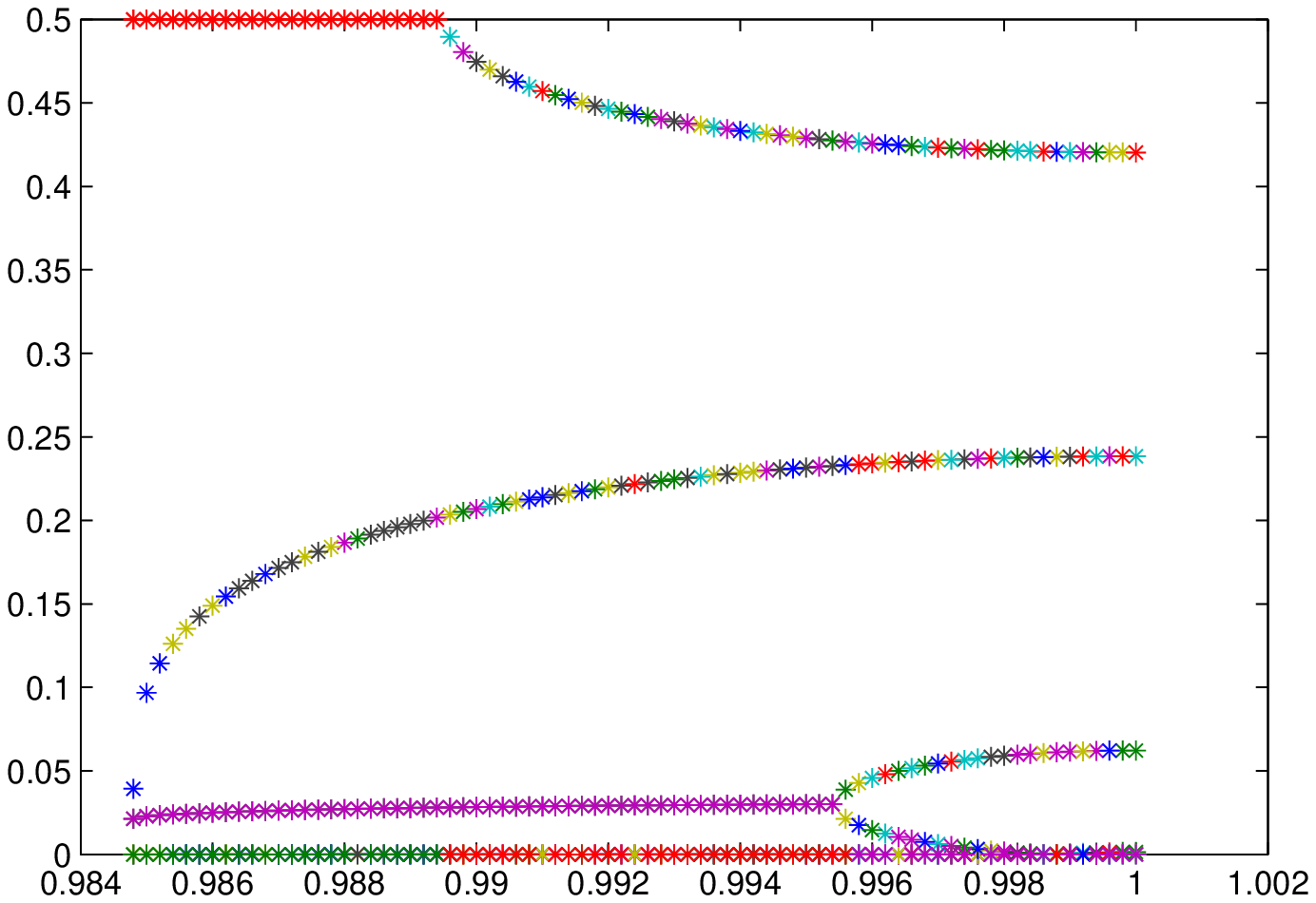}
\end{center}
\caption{Real (left) and imaginary (right) parts of the characteristic exponents 
$\lambda$ versus $\varepsilon$ for $N = 3$ for branch 1 (top) and branch 2 (bottom).}
\label{fig:N6stability1}
\end{figure}

For $N = 3$ ($q=\frac{\pi}{3}$), the system of linearized equations (\ref{eq:ADD-linear-1})
is closed at $6$ second-order linearized equations. Besides
the characteristic exponent $\lambda = 0$ of multiplicity four,
we have $8$ nonzero characteristic exponents $\lambda$.
The characteristic equation (\ref{characteristic-eq})
with $\theta = \frac{\pi}{3}$ and $\theta = \frac{2\pi}{3}$
predicts a double pair of real $\lambda$ and a double pair of purely imaginary $\lambda$.
Figure \ref{fig:N6stability1} (top)
shows ${\rm Re}(\lambda)$ (left) and ${\rm Im}(\lambda)$ (right)
for solutions of branch 1.
The double pair of purely imaginary $\lambda$ split
along the imaginary axis for small $\varepsilon > 0$. On the other hand,
the double pair of real $\lambda$ splits along the transverse direction
and results in occurrence of a quartet of complex-valued $\lambda$ for small $\varepsilon > 0$.
These complex characteristic exponents approach the imaginary axis at
$\varepsilon = \varepsilon_1 \approx 0.43$ (Neimark--Sacker bifurcation)
and then split along the imaginary axis as two pairs of
purely imaginary $\lambda$ for $\varepsilon > \varepsilon_1$. At the same time, one pair of
of the purely imaginary $\lambda$ continued from $\varepsilon = 0$ approaches the line
$\pm \frac{i}{2}$ (corresponding to the Floquet multiplier at $-1$)
at $\varepsilon = \varepsilon_2 \approx 0.72$  (period-doubling bifurcation)
and splits along the negative real axis. In summary, the periodic travelling
wave of branch 1 for $N = 3$ is stable for $\varepsilon \in (\varepsilon_1,\varepsilon_2)$
but unstable near $\varepsilon = 0$ and $\varepsilon = 1$.

Figure \ref{fig:N6stability1} (bottom)
shows ${\rm Re}(\lambda)$ (left) and ${\rm Im}(\lambda)$ (right)
for solutions of branch 2 that exists only for $\varepsilon \in [\varepsilon_*,1]$,
where $\varepsilon_* \approx 0.985$. All four pairs of the characteristic exponents
$\lambda$ are purely imaginary near $\varepsilon = 1$ that corresponds to the
numerical results for stability of travelling waves in monomer chains in \cite{James2}.
Two pairs coalesce at $\varepsilon \approx 0.995$ resulting in the
complex characteristic exponents (Neimark--Sacker bifurcation).
One more pair crosses the line
$\pm \frac{i}{2}$ for $\varepsilon \approx 0.989$ resulting in the negative
characteristic exponents (period-doubling bifurcation). The last remaining
pair of purely imaginary $\lambda$ crosses zero near $\varepsilon = \varepsilon_* \approx 0.985$
that indicates that termination of branch 2 is related to a local bifurcation.
However, we are not able to identify numerically any other branch of travelling wave solutions
in the neighborhood of branch 2 for $\varepsilon \approx \varepsilon_*$.

\begin{figure}[t]
\begin{center}
\includegraphics[width=7cm,height=5cm]{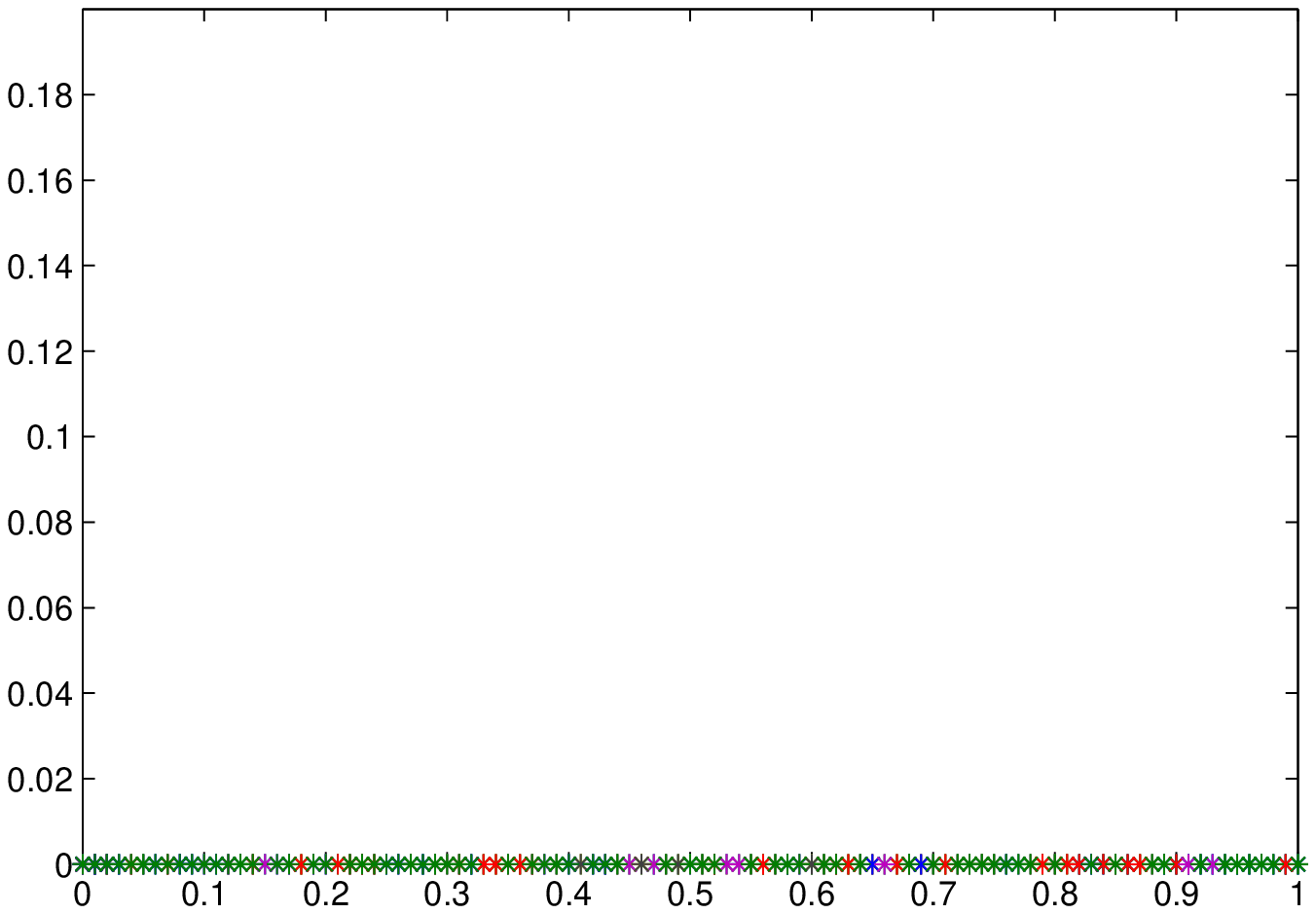}\hspace{0.1cm}
\includegraphics[width=7cm,height=5cm]{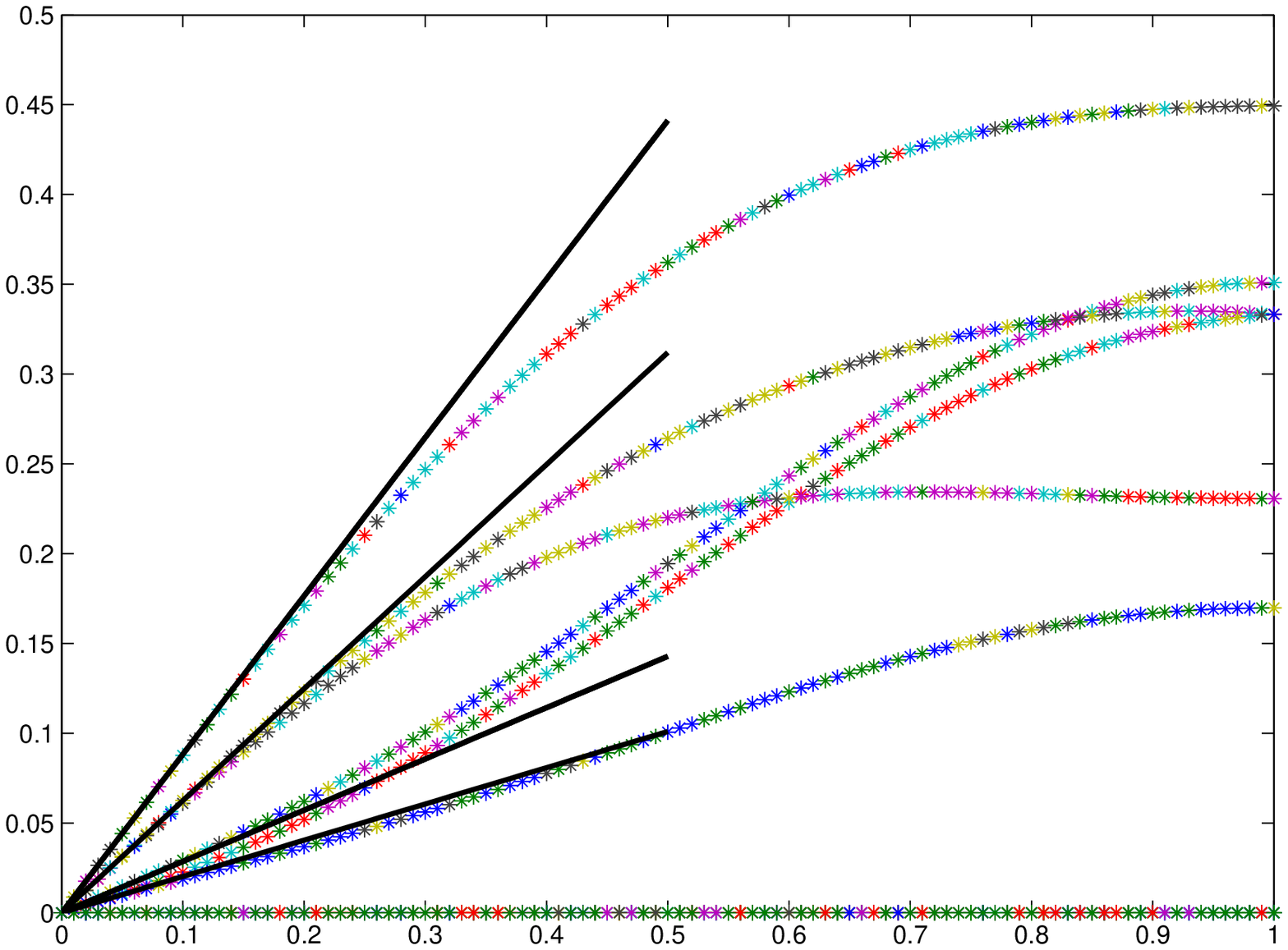} \\
\includegraphics[width=7cm,height=5cm]{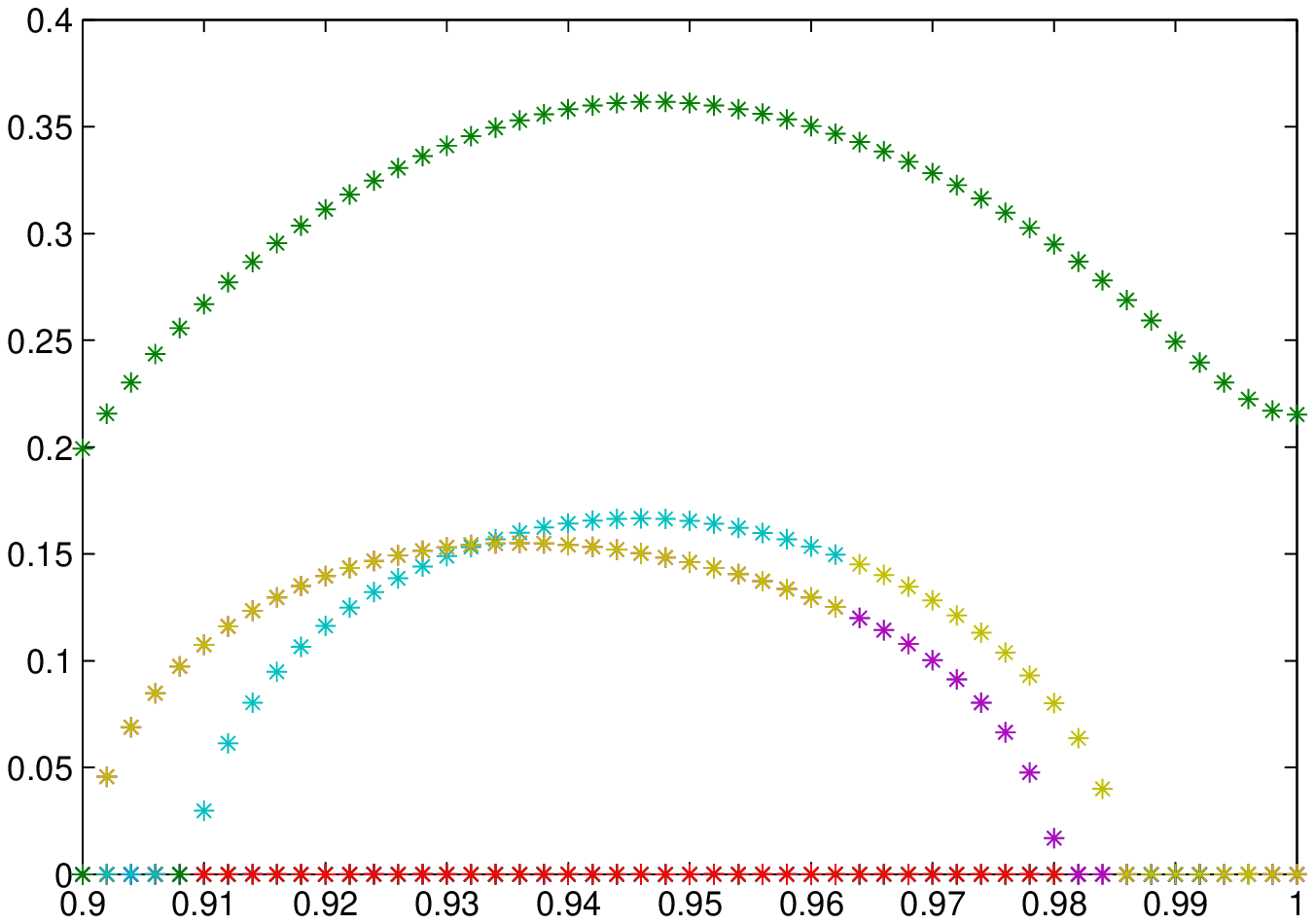} \hspace{0.1cm}
\includegraphics[width=7cm,height=5cm]{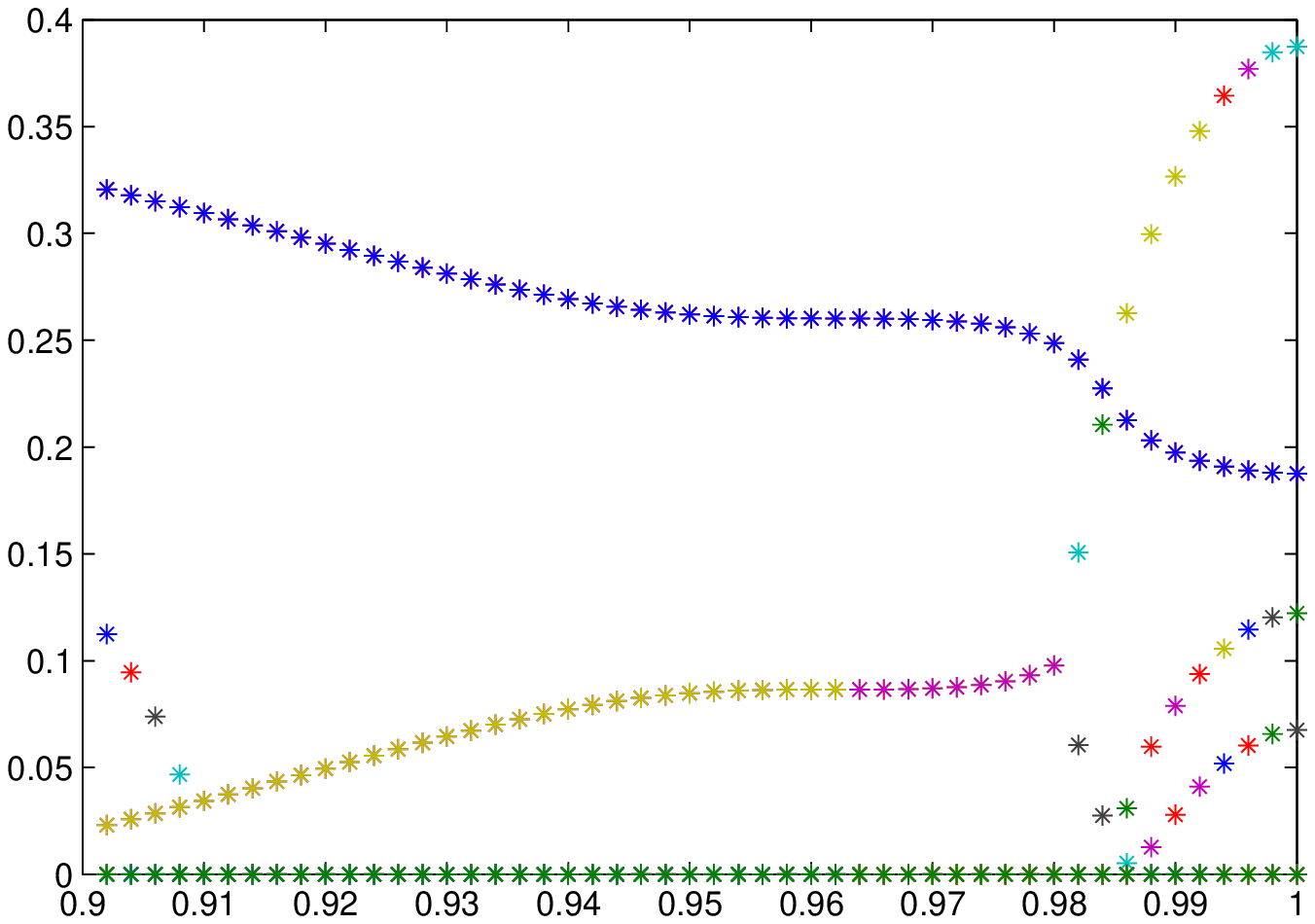}
\end{center}
\caption{Real (left) and imaginary (right) parts of the
characteristic exponents $\lambda$ versus $\varepsilon$ for $N = 4$ for branch 1 (top)
and branch 2 (bottom).}
\label{fig:N8stability}
\end{figure}

Recall that the coefficient $M_1$ changes sign at $q \approx 0.915$, as seen in Figure \ref{fig:coeff}.
Therefore, for $N \geq 4$, the characteristic equation (\ref{characteristic-eq})
for any values of $\theta$ predicts pairs of purely imaginary $\lambda$ only.
This is illustrated on the top panel of Figure \ref{fig:N8stability} for $N = 4$ ($q=\frac{\pi}{4}$).
We can see that all double pairs of purely imaginary $\lambda$ split along the imaginary axis
for small $\varepsilon > 0$ and that the periodic travelling waves of branch 1
remain stable for all $\varepsilon \in [0,1]$. The figure also illustrate
the validity of asymptotic approximations obtained from
roots of the characteristic equation (\ref{characteristic-eq}).

It is interesting that Figure \ref{fig:N8stability} shows safe coalescence
of characteristic exponents for larger
values of $\varepsilon$. Recall from Remark \ref{remark-Krein} that
the characteristic exponents have opposite
Krein signature for small values of $\varepsilon$ in such a way that
larger exponents on Figure  \ref{fig:N8stability}
have negative Krein signature $\sigma$ and smaller exponents have
positive Krein signature $\sigma$. It is typical
to observe instabilities after coalescence of two purely imaginary
eigenvalues of the opposite Krein signature \cite{MacKay-new}
but this only happens when the double eigenvalue at the coalescence
point is not semi-simple. When the double eigenvalue
is semi-simple, the coalescence does not introduce any instabilities \cite{Bridges}.
This is precisely what
we observe on Figure \ref{fig:N8stability}. After coalescence for
larger values of $\varepsilon$, the purely imaginary
characteristic exponents $\lambda$ reappear as simple exponents
of the opposite Krein signature and the exponents with positive Krein signatures
are now above the ones with negative Krein signatures.

Figure \ref{fig:N8stability} (bottom)
shows ${\rm Re}(\lambda)$ (left) and ${\rm Im}(\lambda)$ (right)
for solutions of branch 2 that exists only for $\varepsilon \in [\varepsilon_*,1]$,
where $\varepsilon_* \approx 0.90$. Besides the pairs of
purely imaginary characteristic exponents $\lambda$, there exists one
pair of real exponents $\lambda$ near $\varepsilon = 1$ that corresponds to the
numerical results for instability of travelling waves in monomer chains in \cite{James2}.
For smaller values of $\varepsilon$, more instabilities arise for the
solutions of branch 2 because of various bifurcations of pairs
of purely imaginary exponents $\lambda$.

\begin{figure}[t]
\begin{center}
\includegraphics[width=7cm,height=5cm]{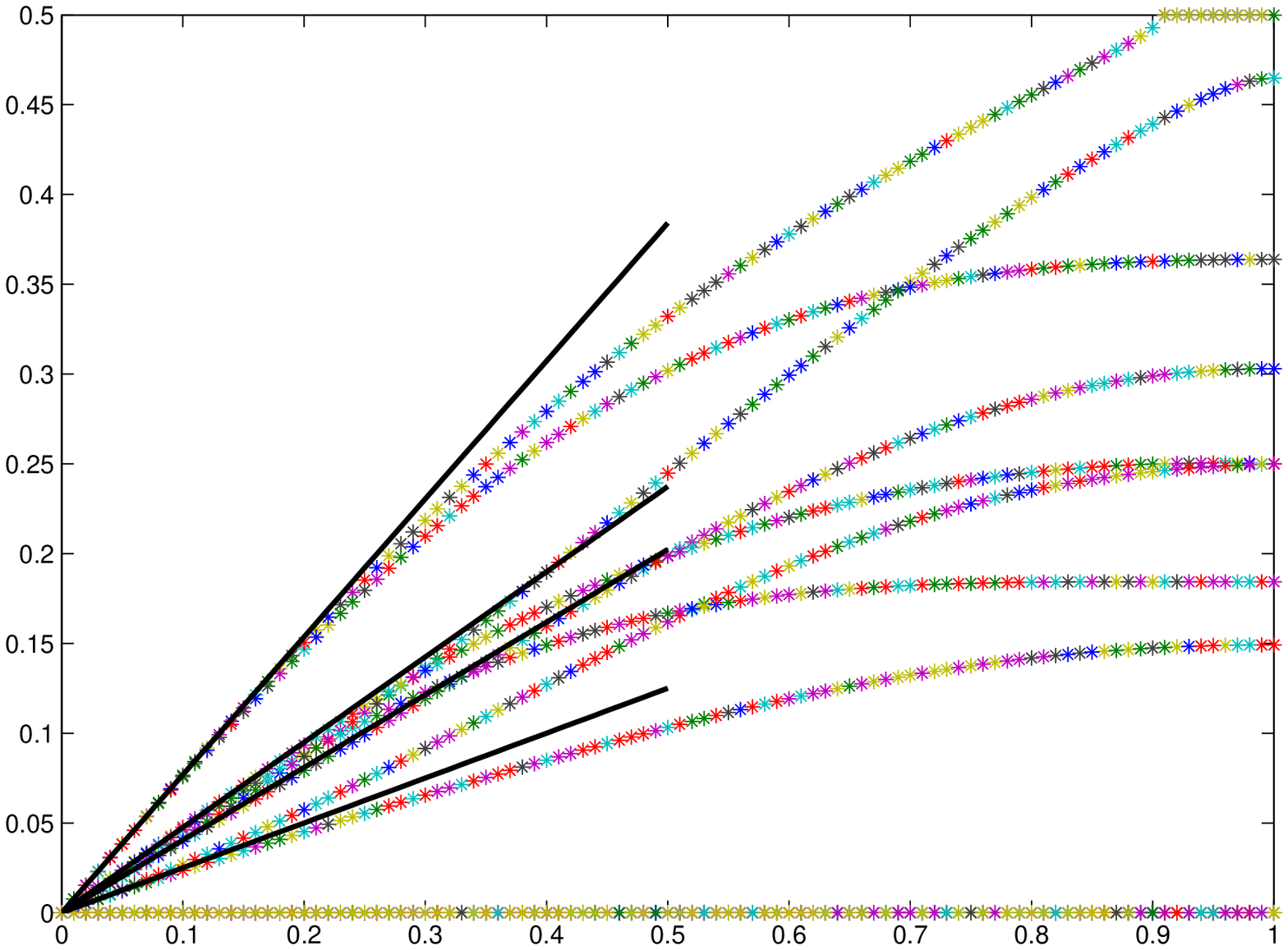} \hspace{0.1cm}
\includegraphics[width=7cm,height=5cm]{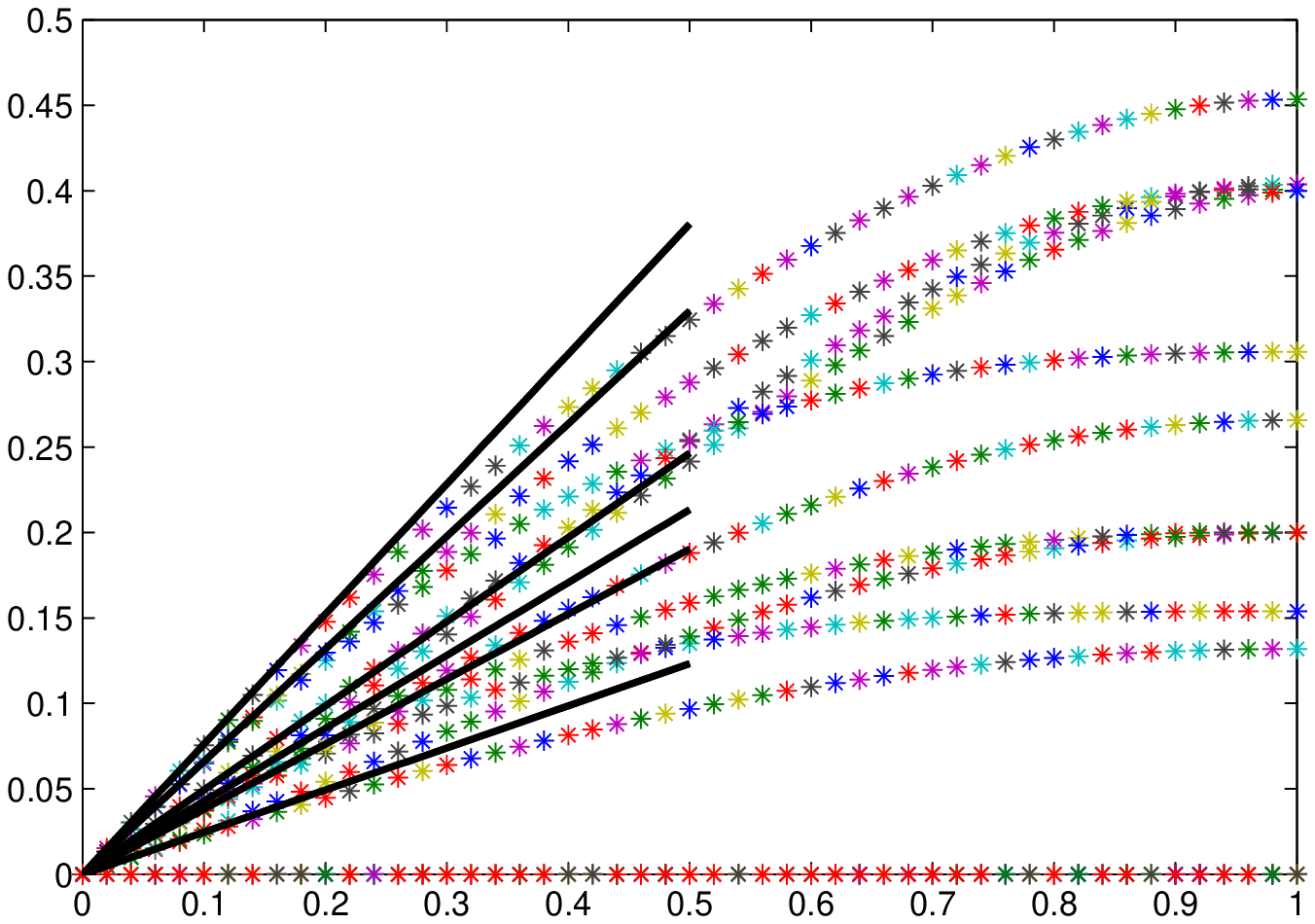}
\end{center}
\caption{Imaginary parts of the characteristic exponents $\lambda$ versus $\varepsilon$ 
for $N=5$ (left) and $N=6$ (right). The real part of all the exponents is zero.}
\label{fig:N10stability}
\end{figure}

Finally, Figure \ref{fig:N10stability} illustrate the stability of
solutions of branch 1 for $N = 5$ (left) and $N = 6$ (right). Not only
the double pairs of purely imaginary $\lambda$ split safely
along the imaginary axis for small $\varepsilon > 0$,
various coalescence of purely imaginary exponents $\lambda$ of opposite Krein
signature never result in occurrence of complex exponents $\lambda$.
The solutions of branch 1 remain stable for all $\varepsilon \in [0,1]$.

\subsection{Stability of the uniform periodic oscillations}

The periodic solution with $q=0$ (which is no longer a traveling wave but a uniform oscillation
of all sites of the dimer) is given by the exact solution (\ref{exact-2}).
Spectral stability of this solution is obtained from the system of linearized equations
(\ref{eq:ADD-linear-2}). Using the boundary conditions
$$
u_{2n+1} = e^{2i \theta} u_{2n-1}, \quad w_{2n+2} = e^{2i \theta} w_{2n}, \quad n \in \Z,
$$
where $\theta \in [0,\pi]$ is a continuous parameter, we obtain
the system of two closed second-order equations,
\begin{equation}
\label{eq:ADD-linear-3}
\left\{ \begin{array}{l}
\ddot{u} + \frac{\alpha}{1 + \varepsilon^2} |\varphi|^{\alpha - 1} u =
\frac{\varepsilon}{1 + \varepsilon^2}
\left( V''(- \varphi) + V''(\varphi) e^{-2i \theta} \right) w, \\
\ddot{w} + \frac{\alpha \varepsilon^2}{1 + \varepsilon^2} |\varphi|^{\alpha - 1} w
= \frac{\varepsilon}{1 + \varepsilon^2} \left( V''(-\varphi) + V''(\varphi) e^{2i \theta} \right) u.
\end{array} \right.
\end{equation}
The characteristic equation (\ref{characteristic-eq}) for $q = 0$
predicts a double pair (\ref{limiting-roots})
of purely imaginary $\Lambda$ for any $\theta \in (0,\pi)$.
We confirm here numerically that the double pair
is preserved for all $\varepsilon \in [0,1]$.

\begin{figure}[t]
\begin{center}
\includegraphics[width=7cm,height=5cm]{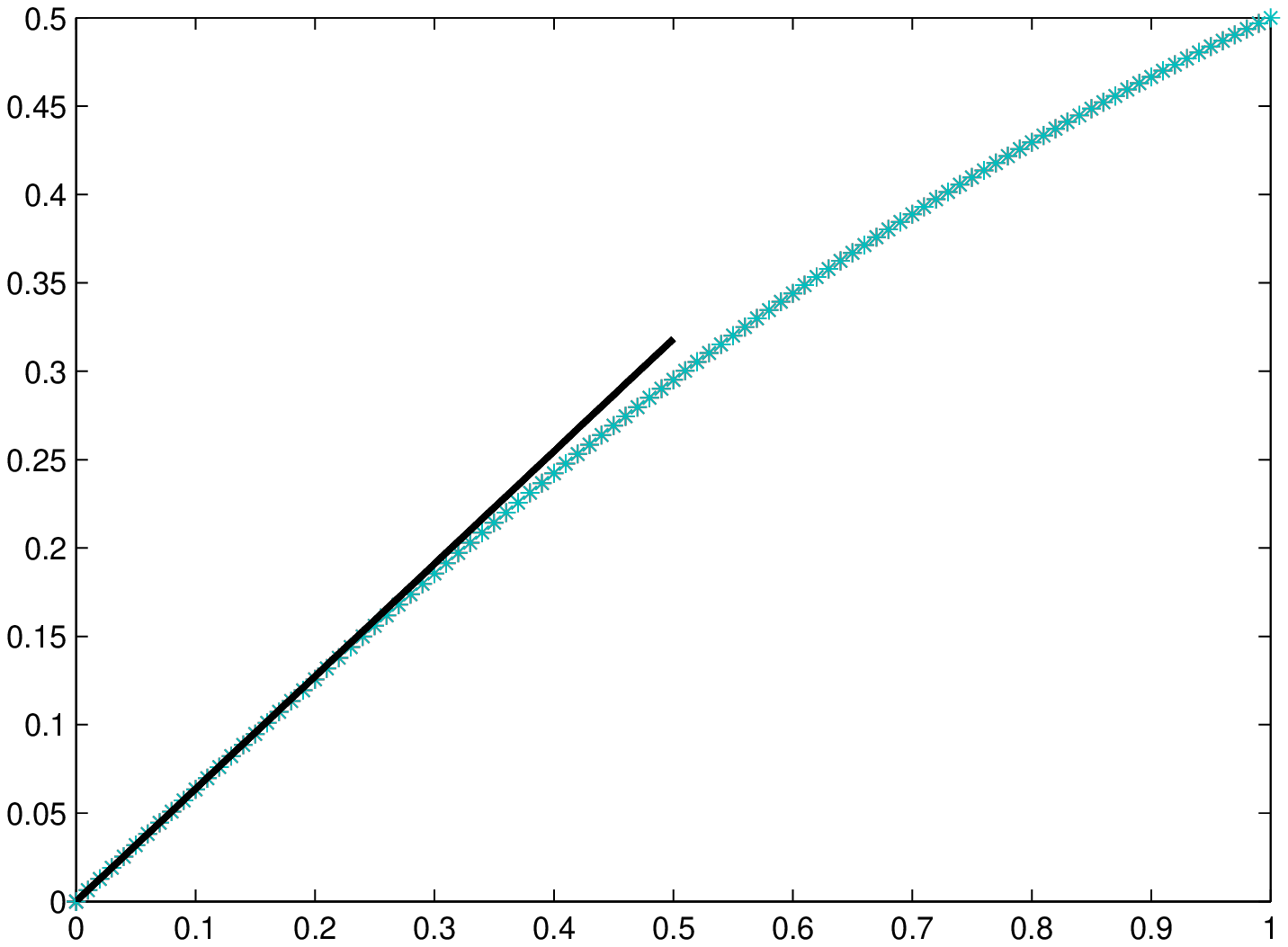} \hspace{0.1cm}
\includegraphics[width=7cm,height=5cm]{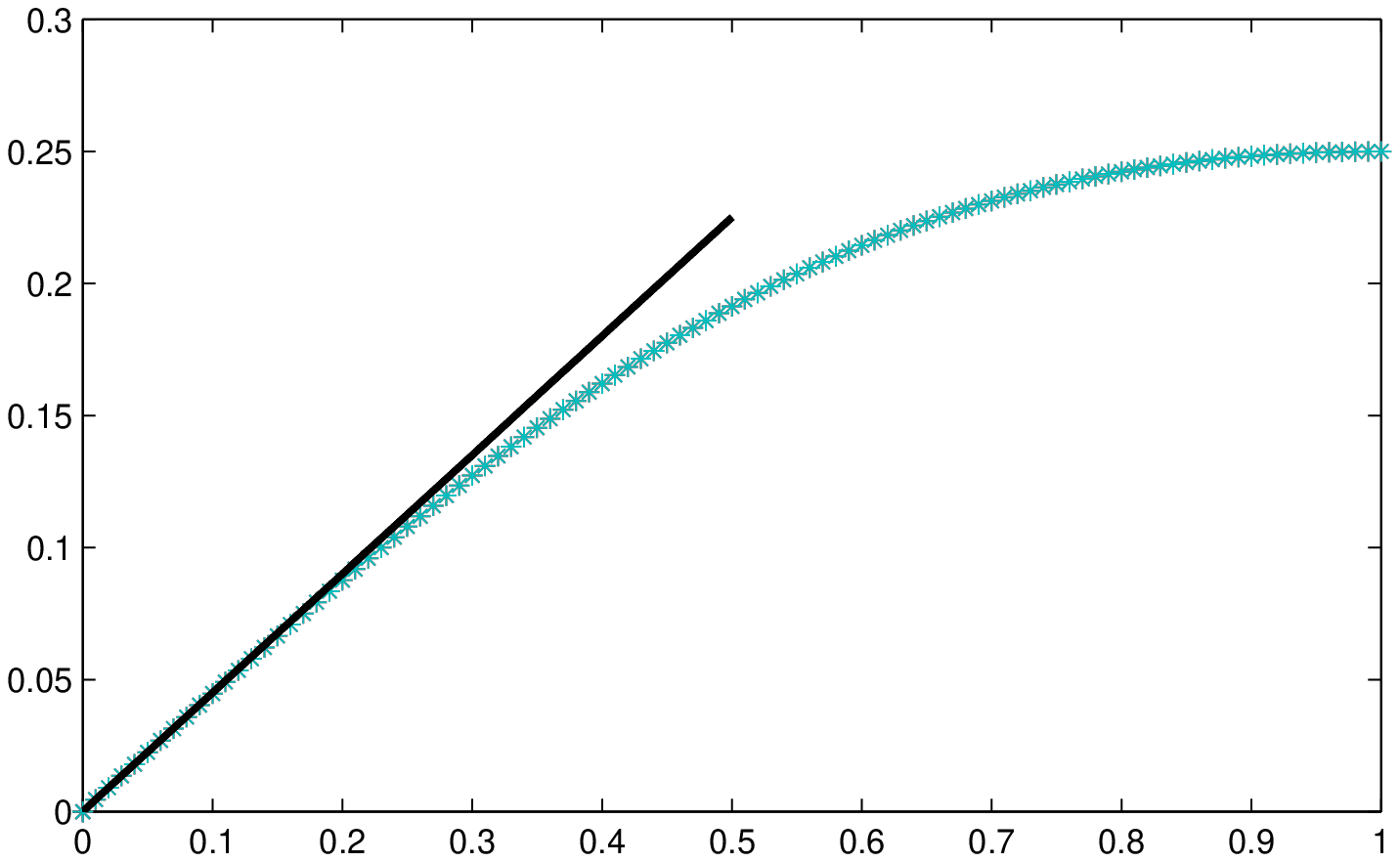}
\end{center}
\caption{Imaginary parts of the characteristic exponents $\lambda$ versus $\varepsilon$ 
for $\theta = \frac{\pi}{2}$ (left) and $\theta = \frac{\pi}{4}$ (right). 
The real part of all the exponents is zero.}
\label{fig:Q0stability}
\end{figure}

Figure \ref{fig:Q0stability} shows the imaginary part of the characteristic exponents
$\lambda$ of the linearized system (\ref{eq:ADD-linear-3}) for $\theta = \frac{\pi}{2}$ (left)
and $\theta = \frac{\pi}{4}$ (right). Similar results are obtained for
other values of $\theta$. Therefore, the periodic solution with $q = 0$
remains stable for all values of $\varepsilon \in [0,1]$.

The pattern on Figure \ref{fig:Q0stability} suggests a hidden symmetry in this case.
Suppose $\lambda_{\theta}$ is a characteristic exponent of the system (\ref{eq:ADD-linear-3})
for the eigenvector
\begin{equation}
\label{eigenvector-1-symmetry}
\left[ \begin{array}{c} u \\ w \end{array} \right] = 
\left[ \begin{array}{c} U_{\theta}(t) \\ W_{\theta}(t) \end{array} \right] e^{\lambda_{\theta} t},
\end{equation}
where $U_{\theta}(t)$ and $W_{\theta}(t)$ are $2\pi$-periodic and the subscript $\theta$ indicates that the system
(\ref{eq:ADD-linear-3}) depends explicitly on $\theta$. Recall that the unperturbed solution
satisfies the symmetry $\varphi(t + \pi) = -\varphi(t)$ for all $t$. Using this symmetry and
the trivial identity $e^{2 \pi i} = 1$, we can verify that there is another solution of the system
(\ref{eq:ADD-linear-3}) with the same $\theta$ for the characteristic exponent $\lambda_{\pi-\theta}$:
\begin{equation}
\label{eigenvector-2-symmetry}
\left[ \begin{array}{c} u \\ w \end{array} \right] = \left[ \begin{array}{c} U_{\pi - \theta}(t + \pi) \\
e^{2i \theta} W_{\pi - \theta}(t+\pi) \end{array} \right] e^{\lambda_{\pi-\theta} t}.
\end{equation}
From the symmetry of roots (\ref{limiting-roots}) and the corresponding characteristic exponents, 
we have $\lambda_{\theta} = \lambda_{\pi - \theta}$. 
The eigenvectors (\ref{eigenvector-1-symmetry}) and (\ref{eigenvector-2-symmetry}) 
are generally linearly independent and coexist for the same value of 
$\lambda = \lambda_{\theta} = \lambda_{\pi-\theta}$.
This argument explains the double degeneracy of characteristic exponents $\lambda$ 
for the case $q = 0$ for all values of $\varepsilon \in [0,1]$.

\section{Conclusion}

We have studied periodic travelling waves in granular dimer chains
by continuing these solutions from the anti-continuum limit, when
the mass ratio between the light and heavy beads
is zero. We have shown that every limiting periodic wave is uniquely
continued for small mass ratio parameters.
Although the vector fields of the granular dimer chain equations are not smooth, we can still use
the implicit function theorem to guarantee that the continuation is $C^1$ with respect to the mass ratio parameter.
We have also used rigorous perturbation theory to compute characteristic exponents in the linearized stability problem.
From this theory, we have seen that the periodic waves with the wavelength larger than
a certain critical value are spectrally stable for small mass ratios. 

Numerical computations are developed to show that the stability of these 
periodic waves with larger wavelengths extends all way to the limit of equal mass ratio.
On the other hand, we have also computed periodic travelling waves that are continued from solutions of
the granular monomer chains at the equal mass ratio, their spectral stability, and their terminations
for smaller mass ratios.

Among open problems, we have not clarified the nature of bifurcation, where the solutions of branch 2 
terminate at a $\varepsilon_* \in (0,1)$ for $N = 3, 4$. We have not been able to find 
another solution nearby for $\varepsilon \gtrapprox \varepsilon_*$. Safe coalescence of 
purely imaginary characteristic exponents $\lambda$ of opposite Krein signatures is also amazing
and we have not been able to explain the hidden symmetry that would explain 
why the eigenvalues at the coalescence point 
remain semi-simple. These problems as well as analysis of the periodic travelling wave solutions 
for other values of $q$ will wait for further studies. 

\vspace{0.25cm}

{\bf Acknowledgements:} The research of the authors was supported
in part by the NSERC Discovery grant. The authors thank
G. James for useful discussions.

\end{document}